%% file: ANA-STDM-2019-26-PAPER.tex
\pdfoutput=1
\documentclass[PAPER, atlasdraft=false, cernpreprint=true, texlive=2025, UKenglish,orcidlogo, texmf, orcidlogo]{atlasdoc}
\usepackage[full]{atlaspackage}
\usepackage{atlasbiblatex}

\usepackage{atlasphysics}
\graphicspath{{logos/}{figures/}{aux_figures/}}

\usepackage{ANA-STDM-2019-26-PAPER-defs}

\AtlasTitle{Measurements and interpretations of $W^{\pm}Z$ production cross-sections in $pp$ collisions at $\sqrt{s} = 13$~TeV with the ATLAS detector}

\AtlasAbstract{%
Measurements of integrated and differential cross-sections for $W^{\pm}Z$ production in proton--proton collisions are presented.
The data collected by the ATLAS detector at the Large Hadron Collider from $2015$ to $2018$ at a centre-of-mass energy of $\sqrt{s} = 13$~TeV are used, corresponding to an integrated luminosity of $140$~fb$^{-1}$.
The $W^{\pm}Z$ candidate events are reconstructed using leptonic decay modes of the gauge bosons into electrons or muons.
The integrated cross-section per lepton flavour for the production of $W^{\pm}Z$ is measured in the detector fiducial region with a relative precision of 4\%.
The measured value is compared with the Standard Model prediction at a precision of up to next-to-next-to-leading-order in QCD and next-to-leading-order in electroweak.
Cross-sections for $W^+Z$ and $W^-Z$ production and their ratio are presented.
The $W^{\pm}Z$ production is also measured differentially as functions of various kinematic variables, including new observables sensitive to CP-violation effects.
All measurements are compared with state-of-the-art Standard Model predictions from fixed-order calculations or Monte Carlo generators based on next-to-leading-order matrix elements interfaced with parton showers.
An effective field theory interpretation of the measurements is performed, considering both CP-conserving and CP-violating dimension-6 operators modifying the $W^{\pm}Z$ production.
In the absence of observed deviations from the Standard Model, limits on CP-conserving Wilson coefficients are extracted using the transverse mass of the $W^{\pm}Z$ system.
For CP-violating coefficients a machine learning approach is used to construct an observable with enhanced sensitivity to CP-violation effects.
}

\AtlasRefCode{STDM-2019-26}

\PreprintIdNumber{CERN-EP-2025-145}

\arXivId{2507.03500}

\AtlasJournalRef{\JHEP 11 (2025) 006}
\AtlasDOI{10.1007/JHEP11(2025)006}

\AtlasCoverEgroupAnalysisTeam{atlas-stdm-2019-26-analysis-team@cern.ch}

\hypersetup{pdftitle={ATLAS document},pdfauthor={The ATLAS Collaboration}}

\newcommand{\AtlasCoordFootnote}{%
ATLAS uses a right-handed coordinate system with its origin at the nominal interaction point (IP)
in the centre of the detector and the \(z\)-axis along the beam pipe.
The \(x\)-axis points from the IP to the centre of the LHC ring,
and the \(y\)-axis points upwards.
Polar coordinates \((r,\phi)\) are used in the transverse plane,
\(\phi\) being the azimuthal angle around the \(z\)-axis.
The pseudorapidity is defined in terms of the polar angle \(\theta\) as \(\eta = -\ln \tan(\theta/2)\).
Angular distance is measured in units of \(\Delta R \equiv \sqrt{(\Delta y)^{2} + (\Delta\phi)^{2}}\).}

\DeclareUnicodeCharacter{306}{\v{}}

\usepackage{rotating}

\begin{document}

\maketitle

\tableofcontents

\section{Introduction}
\label{sec:intro}
%

The study of  \wz diboson production is an important test of the Standard Model (SM) due to its sensitivity to gauge
boson self-interactions which are related to the non-Abelian structure of the electroweak interaction.
It provides the means to directly probe the triple gauge boson couplings (TGC), in particular the $WWZ$ gauge coupling, and to look for deviations from the SM.
Of particular interest is the exploration of potential deviations in the boson couplings with respect to the SM predictions, inducing a violation of the  combined charge conjugation and parity (CP) invariance because, according to the Sakharov condition~\cite{Sakharov:1967dj}, the observed baryon asymmetry of the universe requires additional sources of CP violation beyond those measured which are described in the SM by the Cabibbo–Kobayashi–Maskawa (CKM) complex phase.
Additionally, the \wz inclusive production enables the search for new physics in the couplings of quarks, leptons and bosonic fields. %
Improved constraints on these couplings can potentially probe scales of new physics in the multi-\TeV\ range and provide a way to look for signals of new physics in a model-independent way.
The SM Effective Field Theory (SMEFT)~\cite{Buchmuller:1985jz, Leung:1984ni,Degrande:2012wf,Brivio:2017vri,Isidori:2023pyp} offers a way to set these constraints by parameterising effects of Beyond Standard Model (BSM) physics at energies below the masses of the related new particles that would be responsible for new physics.
New terms are added to the SM Lagrangian without assuming the existence of any additional field.  Each term is proportional to a Wilson coefficient and is suppressed by increasing powers of $1/\Lambda$,
where $\Lambda$ represents the new physics scale.

Measurements of the \wz production cross-section in proton--antiproton collisions at a centre-of-mass energy of $\sqrt{s} = 1.96$~\TeV{} were published by the CDF and D\O~collaborations~\cite{CDF:2012WZ,D0:2012WZ} using integrated luminosities of  $7.1$~fb$^{-1}$ and $8.6$~fb$^{-1}$, respectively.
At the Large Hadron Collider (LHC)~\cite{Evans:2008zzb}, \wz measurements in proton--proton ($pp$) collisions, were performed at centre-of-mass energies of $5$~\TeV, $7$~\TeV\ and $8$~\TeV, by the ATLAS  and CMS Collaborations~\cite{CMS:2021pqj,Aad:2012twa,ATLASWZ_8TeV,Khachatryan:2016poo}.
Measurements of \wz production at $\sqrt{s}=13$~\TeV\ were reported by the ATLAS~\cite{Aaboud:2016yus,WZPol36fb,ATLAS:2022oge,STDM-2020-01} and CMS~\cite{CMS:2019efc,CMS:2021icx} Collaborations using integrated luminosities up to $137$~fb$^{-1}$.
The latest ATLAS analyses~\cite{ATLAS:2022oge,STDM-2020-01}, focused on polarisation measurements, provided the first observation of the joint polarisation of the two final state bosons.
Recently, the CMS experiment reported a measurement of the \wz production cross-section at a centre-of-mass energy of $\sqrt{s}=13.6$~\TeV~\cite{CMS:2024ild}.
Constraints on CP-conserving SMEFT terms including only gauge boson fields were considered in previous \wz measurements~\cite{Aad:2012twa,ATLASWZ_8TeV,Khachatryan:2016poo,CMS:2019efc,CMS:2021icx}.
Constraints on CP-violating terms were performed  in vector-boson fusion (VBF) production of $Z$~\cite{ATLAS:2020nzk}, $H \to \tau\tau$~\cite{ATLAS:2024wfv} or $H \to W W^*$~\cite{ATLAS:2023pwa,ATLAS:2025hki} using the azimuthal angle difference between jets, $\Delta \phi_{\mathrm{jj}}$.
Besides individual angular variables, analyses of VBF production of $H \to \gamma \gamma$~\cite{ATLAS:2022tan}, $H \to ZZ^*$~\cite{ATLAS:2023mqy} and the analysis of CP properties of $H \to \tau\tau$~\cite{ATLAS:2025bts} used Optimal Observables to derive limits on CP-violating terms.

This paper presents precise measurements of the integrated and differential \wz production cross-sections in $pp$ collisions using data recorded by the ATLAS detector from 2015 to 2018 (Run 2), at a centre-of-mass energy of $\sqrt{s} = 13$~\TeV, and corresponding to an integrated luminosity of 140~\ifb.
The $W$ and $Z$ bosons are reconstructed using their decays into electrons or muons and cross-sections are measured in a fiducial region close to the detector acceptance.
The ratio of the $W^+Z$ cross-section to the $W^-Z$ cross-section is also measured. %
The reported measurements are compared with predictions from state-of-the art Monte Carlo (MC) event generators
and with the most recent calculations at next-to-next-to-leading-order (NNLO) in QCD and next-to-leading-order (NLO) in electroweak (EW) from MATRIX~\cite{Grazzini:2016swo, Grazzini:2017ckn,Grazzini:2019jkl}.

With respect to the previous ATLAS results~\cite{WZPol36fb}, the measurements reported in this paper extend the analysed data to the full Run-2 data sample, allowing to do differential cross-section measurements with a higher granularity and a better precision reached in the tails of distributions.
The cross-section measurements of this analysis therefore supersede those of Ref.~\cite{WZPol36fb}.
In addition, the set of measured differential variables  related to the kinematics of the \wz system and to the jet activity in the event is significantly extended to include observables  sensitive to CP-violation effects.
The measured cross-sections and the Rivet~\cite{Bierlich:2019rhm} routine for this measurement are published on HEPData~\cite{Maguire:2017ypu}.
The precision achieved in the reported cross-section measurements is important to validate the theory predictions.

Limits on several CP-conserving and CP-violating BSM terms using detector-level data are extracted in the context of the SMEFT.
Constraints on CP-conserving terms including bosonic fields or describing potential new fermion-boson couplings are set using the distribution of the transverse mass of the \wz pair.
For constraining the CP-violating terms a machine learning approach is used to construct an observable with enhanced sensitivity to CP-violation effects.

%

%
%
%
%
%
%
%
%
%
%
%
%
%

%
%

%

%
%


%
\section{ATLAS detector}
\label{sec:detector}
%

%
%

The ATLAS experiment~\cite{PERF-2007-01} at the LHC is a multipurpose particle detector
with a forward--backward symmetric cylindrical geometry and a near \(4\pi\) coverage in
solid angle.\footnote{\AtlasCoordFootnote}
It consists of an inner tracking detector (ID) surrounded by a thin superconducting solenoid
providing a \SI{2}{\tesla} axial magnetic field, electromagnetic and hadron calorimeters and a muon spectrometer (MS).
The inner tracking detector covers the pseudorapidity range \(|\eta| < 2.5\).
It consists of silicon pixel, silicon microstrip and transition radiation tracking detectors.
Lead/liquid-argon (LAr) sampling calorimeters provide electromagnetic (EM) energy measurements
with high granularity.
A steel/scintillator-tile hadron calorimeter covers the central pseudorapidity range (\(|\eta| < 1.7\)).
The endcap and forward regions are instrumented with LAr calorimeters
for both the EM and hadronic energy measurements up to \(|\eta| = 4.9\).
The muon spectrometer surrounds the calorimeters and comprises separate trigger and
high-precision tracking chambers measuring the deflection of muons in a magnetic field generated by the superconducting air-core toroidal magnets.
The field integral of the toroids ranges between \num{2.0} and \qty{6.0}{\tesla\metre}
across most of the detector.
Three layers of precision chambers, each consisting of layers of monitored drift tubes, cover the region \(|\eta| < 2.7\),
complemented by cathode-strip chambers in the forward region, where the background is highest.
The muon trigger system covers the range \(|\eta| < 2.4\) with resistive-plate chambers in the barrel, and thin-gap chambers in the endcap regions.
A two-level trigger system is used to select events.
The first-level trigger is implemented in hardware and uses a subset of the detector information
to accept events at a rate below \SI{100}{\kHz}.
This is followed by a software-based trigger that
reduces the accepted event rate to about \qty{1.25}{\kHz} on average
depending on the data-taking conditions.
An extensive software suite~\cite{SOFT-2022-02} is used in the reconstruction and analysis of real
and simulated data, in detector operations, and in the trigger and data acquisition systems of the experiment.


%
%
%
%
%

%
\section{Signal and background simulation}
\label{sec:samples}
%


%
%
%
%
%
A sample of simulated \wz\ events is used to correct the signal yield for detector effects and to compare the measurements with the theory predictions.
The production of \wz\ pairs and the subsequent leptonic decays of the vector bosons were simulated with the \SHERPA[2.2.12]~\cite{Bothmann:2019yzt}  MC event generator.
The \SHERPA[2.2.12] calculation includes all diagrams with four electroweak vertices. The
matrix elements were computed for up to one additional parton at NLO in QCD and up to three partons at leading-order (LO) using \textsc{Comix}~\cite{Gleisberg:2008fv} and \OPENLOOPS~\cite{Cascioli:2011va}, and merged with the \Sherpa parton shower (PS)~\cite{Schumann:2007mg} according to the ME+PS$@$NLO prescription~\cite{Hoeche:2012yf}.
The \NNPDF[3.0nnlo] set of parton distribution functions (PDF) was used along with the associated set of tuned parton-shower parameters developed by the \Sherpa authors.
An alternative signal sample was generated at NLO in QCD using the \POWHEGBOX[v2]~\cite{powheg1:2004, powheg, Alioli:2010xd, Melia:2011tj} generator, interfaced to \PYTHIA[8.210]~\cite{Sjostrand:2014zea} for simulation of parton showering, hadronisation and the underlying event.
Final-state radiation resulting from QED interactions is simulated using  \PYTHIA[8.210] and the \AZNLO~\cite{AZNLO:2014}
set of tuned parameters.
The \CT[10nlo]~\cite{CT10} PDF set was used for the hard-scattering process, while the \textsc{CTEQ6L1}~\cite{Pumplin:2002vw} PDF set was used for the parton shower.
The sample was generated with dynamic renormalisation and factorisation QCD scales, $\mu_\textrm{R}$ and $\mu_\textrm{F}$, equal to  half of $m_{WZ}$, where $m_{WZ}$ is the invariant mass of the $WZ$ system.

A third \wz\  sample with on-shell $W$ and $Z$ is also considered for comparisons to measured jet observables.
It uses the \MGNLO[2.6.5] MC event generator, interfaced with \PYTHIA[8.240] using the A14 set of tuned parameters~\cite{ATL-PHYS-PUB-2014-021} for the modelling of the PS, hadronisation and underlying event.
Matrix elements containing three leptons, one neutrino and up to two jets in the final state were calculated at \NLO in QCD and merged with the PS from  \PYTHIA[8.210] using the \FXFX scheme~\cite{Frederix:2012ps}.
The \NNPDF[3.0nlo]~\cite{Ball:2014uwa} PDF set was used for the hard-scattering process,  while the \NNPDF[2.3lo]~\cite{Ball:2012cx}  PDF set was used for the PS.
The default dynamic renormalisation and factorisation scales set by \MGNLO~\cite{Hirschi:2015iia} were used.

The measured fiducial cross-sections and the differential distributions are compared with predictions provided by the state-of-the-art MATRIX 2.1 program~\cite{Grazzini:2016swo, Grazzini:2017ckn, Grazzini:2019jkl,Cascioli:2011va, Denner:2016kdg,Gehrmann:2015ora,Catani:2012qa,Catani:2007vq},  which implements fixed-order calculations for the full $pp \rightarrow \ell^{'} \nu \ell \ell$ process.
The calculations are performed in the fiducial phase space of the measurement defined in Section~\ref{sec:phase_space} at \NNLO in QCD including \NLO EW corrections combined with the additive or the multiplicative prescriptions restricted to the $q\bar{q}$ channel, while photon-induced corrections are treated in an additive way.
The two combination schemes are denoted by \nnlopew and \nnloxew, respectively.
The $G_\mu$ scheme~\cite{Dittmaier:2009cr} is used for the input EW parameters.
Charged leptons are considered at dressed level as defined in Section~\ref{sec:phase_space}.
Dynamic renormalisation and factorisation QCD scales are implemented equal to half the sum of the $W$ and $Z$ bosons transverse masses.
The \NNPDF[3.1nnlo\_as\_118\_luxqed]~\cite{Bertone:2017bme} set, which accounts for QED effects in the DGLAP evolution, the momentum sum rule, and deep-inelastic scattering coefficient functions, is used for the PDFs.

The predictions from the \powhegpythia sample were rescaled by a global factor of $1.18$ to match the \NNLO QCD cross-section predicted by \MATRIX.

The background sources include processes with two or more electroweak gauge bosons, namely $ZZ$, $WW$ and $VVV$ ($V=W,Z$); processes with top quarks, such as \tt and \ttV, single top and \tzj; and processes with gauge bosons associated with jets or photons ($Z+j$ and $Z\gamma$).
Electroweak \wz production, \wzjjew, arising at the order $\alpha_{\mathrm{EW}}^6$ is also considered as a background and not part of the measured signal.
MC simulation is used to estimate the contribution from  background processes with three or more prompt leptons\footnote{A prompt lepton is a lepton that is not produced in a photon conversion or in the decay of a hadron, a $\tau$-lepton, or their descendants.}.
Background processes with at least one misidentified lepton are evaluated using data-driven techniques and simulated events are used to assess the systematic uncertainties of these backgrounds.

The \SHERPA[2.2.2] event generator was used to simulate both the $q\bar{q}$ and $gg$-initiated $ZZ$ and $WW$ processes, including $H \rightarrow VV$ production, using the \NNPDF[3.0nnlo] PDF set.
It provides a matrix element calculation accurate at NLO in $\alpha_{\mathrm{s}}$ for 0- and 1-jet final states, and LO accuracy for 2- and 3-jet final states.
The production of \ttV\ events was modelled using the \MGNLO[2.3.3] generator at \NLO with the \NNPDF[3.0nlo] PDF\@. The events were interfaced to \PYTHIA[8.210] using the A14 set of tuned parameters and the \NNPDF[2.3lo] PDF set.
The \tzj process was modelled using the \MGNLO[2.3.3] generator at \NLO with the \NNPDF[3.0nlo] PDF. The events were interfaced with \PYTHIA[8.230] using the A14 set of tuned parameters and the \NNPDF[2.3lo]  PDF set.
The \wzjjew events, corresponding to processes of order six (zero) in $\alpha_\mathrm{EW}$ ($\alphas$), were generated at \LO by the  \MGNLO[2.6.5] MC generator interfaced with \PYTHIA[8.240] for the modelling of the parton shower, hadronisation and underlying event.
The parton distribution function set was \NNPDF[3.0nlo].
In the simulation of the parton shower the  \NNPDF[2.3lo] PDF set was used.
The \zzjjew production was modelled using the \MGNLO[2.6.7] generator with matrix elements calculated at \LO in QCD and with the \NNPDF[3.0nlo] PDF set.
The events were interfaced with \PYTHIA[8.230] using the A14 set of tuned parameters and the \NNPDF[2.3lo] PDF set.
The on-shell production of triboson events with fully leptonic decays was simulated by the \SHERPA[2.2.2] event generator at NLO accuracy with zero additional partons and at LO accuracy with one and two additional partons and using the \NNPDF[3.0nnlo]  PDF set.
The production of \tt events was modelled using the \POWHEGBOX[v2]
generator at NLO with the \NNPDF[3.0nlo] PDF set and the \hdamp parameter\footnote{The
\hdamp parameter is a resummation damping factor and one of the
parameters that controls the matching of \POWHEG matrix elements to
the parton shower and thus effectively regulates the
high-\pT radiation against which the \ttbar system recoils.} set to 1.5\,\mtop~\cite{ATL-PHYS-PUB-2016-020}.  The events were interfaced to \PYTHIA[8.230] to model the parton shower, hadronisation, and underlying event, with parameters set according to the A14 tune and using the \NNPDF[2.3lo] set of PDFs.
The production of $Z\gamma$  final states was simulated with the \SHERPA[2.2.4] event generator at \NLO QCD accuracy using the \NNPDF[3.0nnlo] PDF set.
Finally, the production of $Z+j$ events was modelled with the \POWHEGBOX[v1] MC generator interfaced to \PYTHIA[8.186]~\cite{Sjostrand:2007gs} for the modelling of the parton shower, hadronisation, and underlying event, with parameters set according to the \AZNLO set of tuned parameters.
The \CT[10nlo] PDF set was used for the hard-scattering processes, whereas the \CTEQ[6L1] PDF set was used for the parton shower.
The decays of bottom and charm hadrons were simulated using the \EVTGEN[1.2.0] program~\cite{Lange:2001uf}, except for processes modelled using \SHERPA.
All generated MC events were passed through the ATLAS detector simulation~\cite{Aad:2010ah}, based on
\GEANT~\cite{Agostinelli:2002hh}, and processed using the same reconstruction software as used for the data.
Multiple interactions in the same and neighbouring bunch crossings (pile-up) were modelled by overlaying the hard-scattering event with simulated inelastic $pp$ events generated with \PYTHIA[8.186] using the \NNPDF[2.3lo] PDF set and the A3~\cite{ATL-PHYS-PUB-2016-017} set of tuned parameters. %
The MC events were weighted to reproduce the distribution of the average number of interactions per bunch crossing
(\(\left<\mu \right>\)) observed in the data. The \(\left<\mu \right>\) value in data was rescaled by a factor of \(1.03 \pm 0.04\) to improve agreement between data and simulation in both charged-particle track distributions and  the visible inelastic \(pp\) cross-section~\cite{Aaboud:2016mmw}.

Furthermore, the lepton and jet momentum scale and resolution, the lepton reconstruction, identification, isolation and trigger efficiencies and the pile-up jets and $b$-jet veto efficiencies in the simulation were corrected to match those measured in data.

%
%
%
%
%
%


%
\section{Event reconstruction and selection}
\label{sec:selection}
%


Only data recorded with stable beam conditions and with all relevant detector subsystems operational are considered~\cite{DAPR-2018-01}.
Candidate events are selected using triggers that require at least one electron or muon~\cite{TRIG-2016-01,ATLAS:2019dpa,ATLAS:2020gty}.
These triggers require leptons to satisfy different transverse momentum thresholds and isolation criteria, which depend on the data-taking run period and the instantaneous luminosity.
The combined efficiency of these triggers is $99\%$ for \wz events satisfying the offline selection criteria.
Events are required to have a primary vertex compatible with the LHC luminous region inside the ATLAS detector.
The primary vertex is defined as the reconstructed vertex with at least two charged-particle tracks that has the largest sum $\pT^{2}$ of its associated tracks.

Muon candidates are identified by tracks reconstructed in the MS and matched to tracks
reconstructed in the ID\@. Muons are required to satisfy a ``medium'' identification selection~\cite{ATLAS:2020auj}.
The efficiency of this selection averaged over \pt and $\eta$ is larger than $98\%$.
The muon momentum is measured by combining the MS measurement, corrected for the energy deposited in
the calorimeters, and the ID measurement.
The $\pt$ of the muon must be greater than $15$~\GeV{} and its pseudorapidity must satisfy $|\eta|<2.5$.

Electron candidates are reconstructed from energy clusters in the electromagnetic calorimeter matched to ID tracks.
Electrons are identified using a discriminant that is the value of a likelihood function constructed with information about the shape of the electromagnetic showers in the calorimeter, the track properties and the quality of the track-to-cluster matching for the candidate~\cite{ATLAS:2019qmc, ATLAS:2023dxj}.
Electrons must satisfy a ``medium'' likelihood requirement, which provides an overall identification efficiency of $90\%$.
The electron momentum is computed from the cluster energy and the direction of the track.
The $\pt$ of the electron must be greater than $15$~\GeV{} and the pseudorapidity of the cluster must satisfy
$\abseta< 1.37$ or $1.52 < \abseta < 2.47$.

Electron and muon candidates are required to originate from the primary vertex. Thus, the significance of
the track's transverse impact parameter calculated relative to the beam line, $|{d_{0}/\sigma_{d_{0}}}|$,
must be smaller than $3.0$ for muons and less than $5.0$ for electrons. Furthermore, the longitudinal impact parameter,
$z_{0}$ (the difference between the value of $z$ of the point on the track at which $d_{0}$ is defined and the
longitudinal position of the primary vertex), is required to satisfy $|z_0\cdot \sin(\theta)|<0.5$~mm.

Electrons and muons are required to be isolated from other particles using both calorimeter-cluster and ID-track information.
The isolation requirement for electrons is tuned for an efficiency of  at least $88\%$ for $\pT>15$~\GeV\ and
at least $99\%$ for $\pT>60$~\GeV~\cite{ATLAS:2019qmc}, while particle-flow-based isolation variables are used for muons, providing an efficiency above $90\%$ for $\pT>15$~\GeV\ and at least $99\%$ for $\pT>60$~\GeV~\cite{ATLAS:2020auj}.

Jets of hadrons are reconstructed using a particle-flow algorithm based on noise-suppressed positive-energy topological clusters in the calorimeters~\cite{ATLAS:2017ghe}.
Energy deposited in the calorimeters by charged particles is subtracted and replaced by the momenta of tracks which are matched to those topological clusters.
Tracks which are not matched to the primary vertex are not used in jet reconstruction, which effectively removes contribution from charged particle pile-up.
The jets are clustered using the anti-$k_{t}$ algorithm~\cite{antikt,Fastjet} with a radius parameter $R = 0.4$.
They are  calibrated according to in situ measurements of the jet energy scale~\cite{JETM-2018-05}.
All jets must have $\pt>25$~\GeV{} and be reconstructed in the pseudorapidity range $|\eta|<4.5$.
To reduce the impact of jets originating from pile-up interactions, jets with
$|\eta|<2.4$ and $\pt < 60$~\GeV, or with $2.5 < |\eta| < 4.5$ and $\pt < 120$~\GeV, are required to satisfy the ``tight'' and ``loose'' working points of the jet vertex~\cite{Aad:2015ina} and forward jet vertex~\cite{ATL-PHYS-PUB-2019-026} tagging algorithms, respectively.
Jets with $|\eta|<2.5$ containing a $b$-hadron are identified with a deep-learning neural network (NN)~\cite{FTAG-2019-07} which uses distinctive features of $b$-hadron decays in terms of the impact parameters of the tracks and the displaced vertices reconstructed in the ID.
Jets initiated by $b$-quarks are selected by setting the algorithm's output threshold such that a $b$-jet selection efficiency of $70\%$ is achieved in simulated $t\bar{t}$ events, with a rejection factor of $500$ against light-flavour jets~\cite{FTAG-2019-07}.

To avoid cases where the detector response to a single physical object is reconstructed as  two different final-state objects,  e.g. an electron reconstructed as both an electron and a jet, several steps are followed to remove such overlaps, as described in Ref.~\cite{Aaboud:2016lpj}.

The transverse momentum of the neutrino is estimated from the missing transverse momentum in the event, \met,
calculated as the negative vector sum of the transverse momentum of all identified hard physics objects
(electrons, muons, jets), with a contribution from an additional soft term.
This soft term is calculated from ID tracks matched to the primary vertex and not assigned to any of the hard objects~\cite{JETM-2020-03}.

Events are required to contain exactly three lepton candidates satisfying the selection criteria described above.
To ensure that the trigger efficiency is well determined, at least one of the candidate leptons is required to have
$\pt > 25$~\GeV\ for $2015$ and $\pt > 27$~\GeV\ for $2016$--$2018$ data, and being  geometrically matched to a lepton that was selected by the trigger.
No requirement on the number of jets is applied.

To suppress background processes with at least four prompt leptons, events with a fourth lepton candidate
satisfying looser selection criteria are rejected. For this looser selection, the lepton $\pt$ requirement is
lowered to $\pt>5$~\GeV{}, electrons are allowed to be reconstructed in the region $1.37< \abseta< 1.52$ and ``loose'' identification requirements~\cite{ATLAS:2019qmc,ATLAS:2020auj} are used for both the electrons and muons.
A less stringent requirement is applied for electron isolation and is based only on ID track information.
No dedicated identification algorithm is used to suppress events with electrons and muons originating from the decay of $\tau$-leptons.

Candidate events are required to have at least one pair of leptons with the same flavour and opposite charge,
with an invariant mass that is consistent with $m_Z^{\textrm{PDG}}$ to within $10$~\gev, where $m_Z^{\textrm{PDG}}$ is the world average mass of the $Z$ boson, as reported by the Particle Data Group~\cite{PhysRevD.98.030001}.
This pair is considered to be the $Z$ boson candidate. If more than one pair can be formed, the pair whose invariant mass is closest to the world-average measured $Z$ boson mass is taken as the \Zboson\ boson candidate.
The remaining third lepton is assigned to the $W$ boson decay.
The transverse mass of the $W$ candidate, computed using \met and the \pt of the associated lepton, is required to be greater than $30$~\GeV.

Backgrounds originating from candidate leptons that are not prompt leptons, also called  ``misidentified'' leptons, are suppressed by requiring the lepton associated with the $W$ boson to satisfy more stringent selection criteria.
The transverse momentum of this lepton is therefore required to be greater than $20$~\GeV. This lepton is also required to satisfy the ``tight'' identification requirements, which results in an efficiency between $90\%$ and $98\%$ for muons and an overall efficiency of $85\%$ for electrons.
Finally, leptons associated with the $W$ boson must also pass a tighter isolation requirement, tuned for an efficiency of at least 65\% and 75\% for electrons and muons, respectively.

%

%
%
%
%
%
%
%
%
%
%
%
%
%
%
%
%
%
%
%
%
%
%
%
%
%

%
%
%
%
%
%
%
%

%

%

%

%
%


%
\section{Background estimation}
\label{sec:background}
%

%
The background sources are classified into two groups: events where at least one of the candidate leptons is not a prompt lepton (reducible background)
and events where all candidates are prompt leptons or are produced in the decay of a $\tau$-lepton (irreducible background).
Background events amount to 19\% of all selected events.

Events of reducible backgrounds mostly originate from  $Z+j$, $Z\gamma$, $t\bar{t}$, and $WW$ production processes and constitute about $5\%$ of all selected events.
This reducible background is  estimated with a data-driven method based on the inversion of a matrix containing the efficiencies and  misidentification probabilities for prompt and misidentified leptons~\cite{ATLASWZ_8TeV,Aad:2014pda,WZPol36fb,ATLAS:2022swp}. %
The method exploits  the classification of the leptons as  loose  or tight  candidates and  the probability that a non-prompt lepton is misidentified as a loose or tight lepton.
Tight leptons are leptons satisfying the selection criteria described in Section~\ref{sec:selection}.
Loose leptons are leptons that do not meet the isolation and identification criteria for signal leptons but satisfy only looser criteria.
The misidentification probabilities for leptons are determined from data using dedicated control samples each enriched  in misidentified leptons from light- or heavy-flavour jets and from photon conversions, respectively.
The lepton efficiencies for prompt leptons are determined as detailed in Refs.~\cite{ATLAS:2019qmc,ATLAS:2020auj}.
The lepton efficiencies and misidentification probabilities are combined with event rates in data samples of \wz candidate events where at least one and up to three of the leptons is loose.
Then, solving a system of linear equations, the number of events with at least one misidentified lepton  is obtained, which represents the amount of reducible background in the \wz sample.
About $2\%$ of this background contribution arises from events with two misidentified leptons.
The background from events with three misidentified leptons, e.g. from multijet processes, is negligible.
The method allows the shape of any kinematic distribution of reducible background events to be estimated.
Another independent method of assessing the reducible background was also considered.
This method estimates the amount of reducible background using MC simulations scaled to data by process-dependent factors determined from data-to-MC comparisons in dedicated control regions.
Agreement within $20\%$ with the matrix method estimate is obtained in both yield and shape of the distributions of irreducible background events.
Systematic uncertainties affecting the  matrix method estimate are detailed in Section~\ref{sec:systematics}.

The events contributing to irreducible background processes represent $\sim14$\% of all selected events.
They originate from $ZZ$, $t\bar{t} +V$, $VVV$ (where $V$ = $Z$ or $W$) and  \tz events.
Events from electroweak \wzjjew production are not part of the measured signal and are therefore considered as backgrounds.
The amount of irreducible  background is estimated by using MC simulations. %
The dominant contribution in this second group is from $ZZ$ production, where one of the leptons from the $ZZ$ decay falls outside the detector acceptance. It represents about $7\%$ of all selected events.
The MC-based estimates of the $ZZ$ and $t\bar{t} +V$ backgrounds are validated by comparing the MC expectations with the event yield and several kinematic distributions of data samples enriched in $ZZ$ and $t\bar{t} +V $ events, respectively.
The $ZZ$ control sample is selected from events with a $Z$ candidate which meets all of the analysis selection criteria and which is accompanied by two additional leptons, satisfying the lepton criteria described in Section~\ref{sec:selection}.
The $ZZ$ MC expectation is rescaled by a factor of $1.13$ to match the observed event yield of data in this control region.
This scaling factor relative to \sherpa predictions is in agreement with the $ZZ$ cross-section measurements performed at $\sqrt{s} = 13$~\TeV~\cite{Aaboud:2017rwm}.
Good agreement is observed between the shapes of the main kinematic distributions and the MC predictions.
The $m_{Z}$ distribution of events in the $ZZ$ control sample is presented in Figure~\ref{fig:ZZCR_ttVCR_controlplot}(a).

The $t\bar{t} +V$ control sample is defined as a sub-sample of \wz\ candidate events by requiring the presence of at least two reconstructed $b$-jets.
The observed data event yield in this validation region is matching the $t\bar{t} +V$ MC prediction rescaled by a factor of $1.3$.
This scaling factor relative to predictions is inline with the $t\bar{t} + Z$ cross-section measurements performed at $\sqrt{s} = 13$~\TeV~\cite{ATLAS:2021fzm}. %
The contribution of $t\bar{t} +V$ events amounts to $\sim 4\%$ of all selected events.
The distributions of the main kinematic variables are found to be well described by the MC predictions.
The \mtw distribution of events in the $t\bar{t}+V$ control sample is presented in Figure~\ref{fig:ZZCR_ttVCR_controlplot}(b).
Systematic uncertainties affecting the rescaling of $ZZ$ and $t\bar{t} +V $ MC predictions are discussed in Section~\ref{sec:systematics}.

\begin{figure}[!htbp]
\centering
\subfloat[]{\includegraphics[width=.45\textwidth]{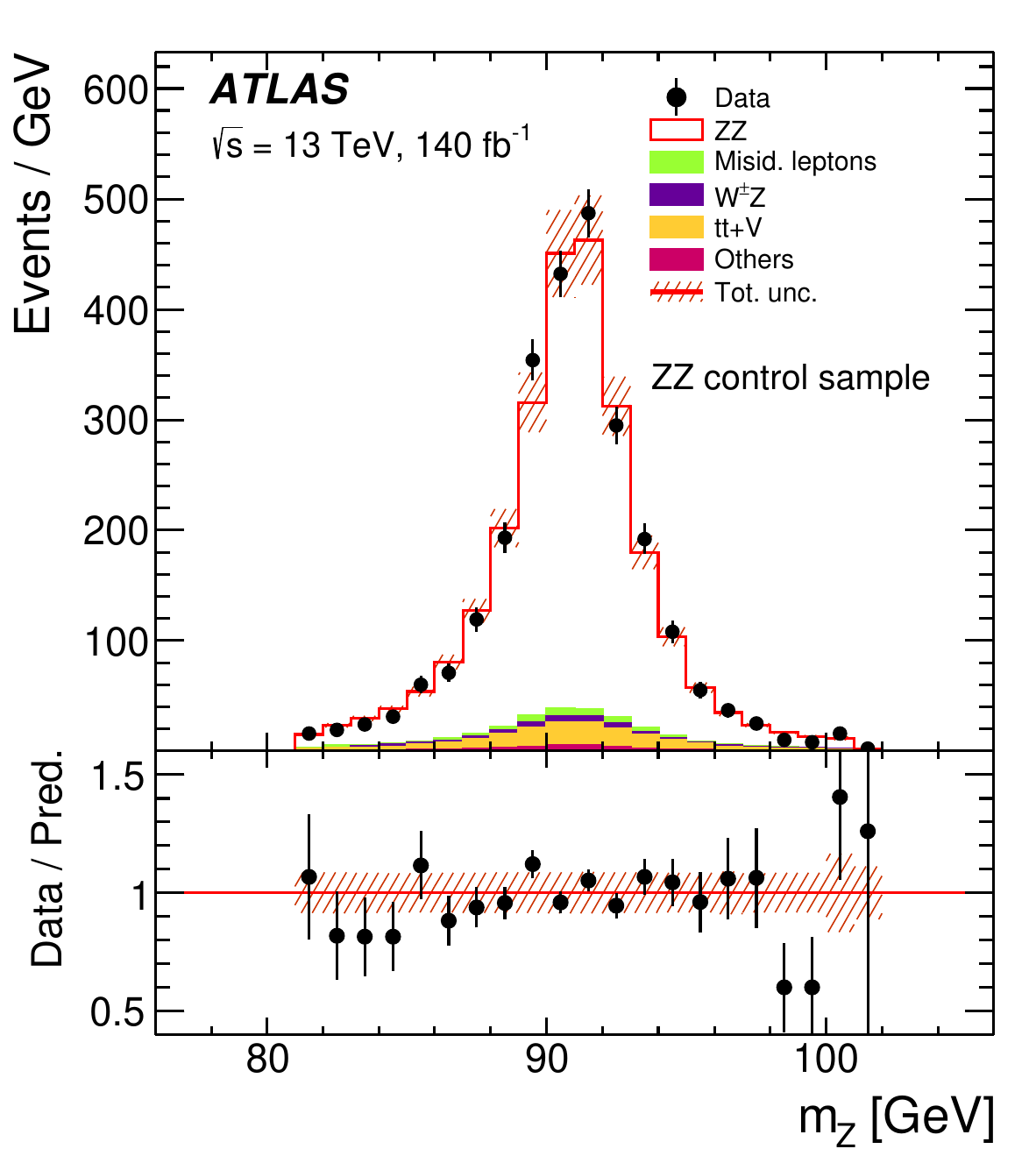}}
\subfloat[]{\includegraphics[width=.45\textwidth]{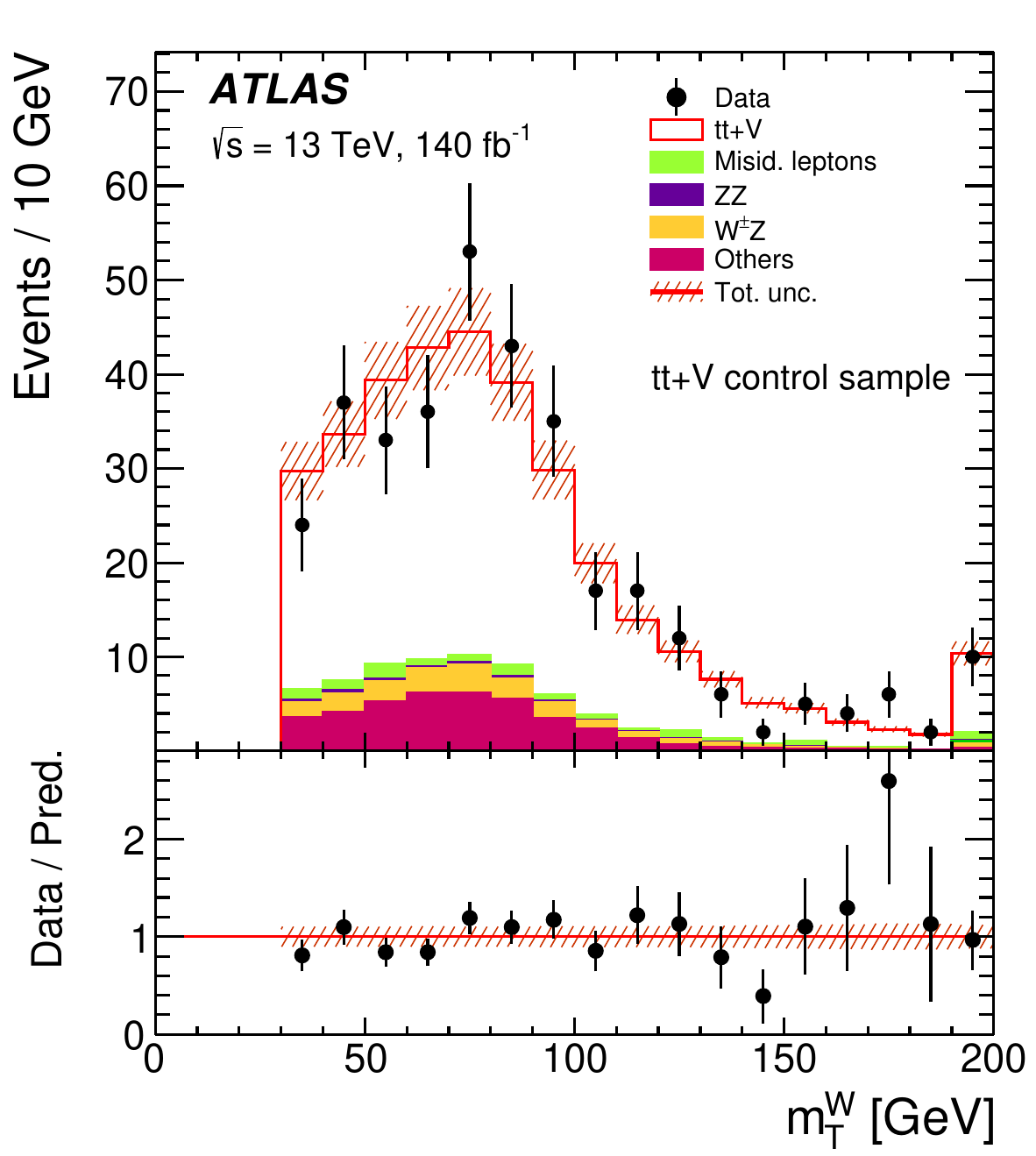}}
\caption{
Distribution of (a) $m_{Z}$ for events in the $ZZ$ control sample and (b) \mtw for events in the $t\bar{t}+V$ control sample.
The points correspond to the data with the error bars representing the statistical uncertainties, and the histograms correspond to the predictions of the various SM processes.
The sum of the background processes with misidentified leptons is labelled ``Misid. leptons''.
The red histogram shows the total prediction; the shaded band is the total uncertainty of this prediction.
The last bin in (b) contains the overflow.
The lower panels in each figure show the ratio of the data points to the open red histogram with their respective uncertainties.
}
\label{fig:ZZCR_ttVCR_controlplot}
\end{figure}


%
\section{Measurements methodology}
\label{sec:MeasMethodology}
\subsection{Phase space for cross-section measurements} \label{sec:phase_space}

The fiducial \wz\ cross-section is measured in a phase space chosen to closely follow the event selection criteria described in Section~\ref{sec:selection}.
It is based on the kinematics of particle-level objects as defined in Ref.~\cite{TruthWS}.
These are final-state prompt leptons associated with the $W$ and $Z$ boson decays.
Charged leptons after QED final-state radiation are ``dressed'' by adding to the lepton four-momentum the contributions from photons with an angular distance $\Delta R < 0.1$ from the lepton.
Dressed leptons, and final-state neutrinos that do not originate from hadron or $\tau$-lepton decays,  are matched to the $W$ and $Z$ boson decay products using an algorithm that does not depend on details of the MC generator, called the ``resonant shape'' algorithm~\cite{ATLASWZ_8TeV}.

The reported cross-sections are measured in a fiducial phase space defined at particle level as follows.
The dressed leptons from $Z$ and $W$ boson decays must have $|\eta| < 2.5$ and transverse momentum \pt\ above $15$~\GeV\ and  $20$~\GeV, respectively;
the invariant mass of the two leptons from the $Z$ boson decay must be within $10$~\GeV{}
from the world average value of the $Z$ boson mass $m_Z^{\textrm{PDG}}$. %
The $W$ transverse mass, defined as $m_{\textrm{T}}^{W} = \sqrt{2 \cdot p_{\mathrm{T}}^\nu \cdot p_{\mathrm{T}}^\ell \cdot[1 -\cos{\Delta \phi(\ell, \nu)}]}$, where $\Delta \phi(\ell, \nu)$
is the angle between the lepton and the neutrino in the transverse plane, and $\pt^\ell$ and $\pt^\nu$ are the transverse momenta of the lepton from $W$ boson decay and of the neutrino, respectively, must be greater than $30$~\GeV.
In addition, to match isolation criteria used in the detector-level event selection, it is required that the angular distance $\Delta R$
between the charged leptons from the $W$ and $Z$ decay is larger than $0.3$, and that
$\Delta R$ between the two leptons from the $Z$ decay is larger than $0.2$.
A requirement that the transverse momentum of the leading lepton be above $27$~\GeV\ reduces the acceptance of the fiducial phase space by less than $0.5\%$.
This criterion is therefore not added to the definition of the fiducial phase space, while it is present in the selection at the detector level.

For the differential measurements related to jets, particle-level jets are reconstructed from stable particles
with a lifetime of $\tau>30$~ps in the simulation after parton showering, hadronisation, and the decay of particles with $\tau<30$~ps.
Muons, electrons, neutrinos and photons associated with $W$ and $Z$ decays are excluded from the jet collection. The particle-level jets are reconstructed with the anti-$k_{t}$
algorithm with a radius parameter $R=0.4$ and are required to have a \pt above $25$~\GeV{} and an absolute value of the pseudorapidity below $4.5$.
The angular distance between all selected leptons and jets is required to be $\Delta R (j,\ell)>0.3$.
If this requirement is not satisfied, the jet is discarded.


%
\subsection{Integrated cross-section measurements}\label{sec:intXsectionMethod}

For a given channel \wz\ $\rightarrow \ell^{'\pm} \nu \ell^+ \ell^-$, where $\ell$ and $\ell^{'}$ indicate each type of lepton ($e$ or $\mu$), the integrated fiducial cross-section that includes the leptonic branching fractions of the $W$ and $Z$ bosons is calculated as
\begin{equation*}\label{eq:sigma_fid}
\sigma^{\mathrm{fid.}}_{W^\pm Z \rightarrow \ell^{'} \nu \ell \ell}  = \frac{  N_{\textrm{data}} - N_{\textrm{bkg}} }{   \mathcal{L}  \cdot  C_{WZ} }  \times  \left( 1 - \frac{N_\tau}{N_{\textrm{all}} } \right) \, ,
\end{equation*}
where  $N_{\textrm{data}}$ and  $N_{\textrm{bkg}}$ are the number of observed events and the estimated number of background events, respectively, $\mathcal{L}$ is the integrated luminosity, and $C_{WZ}$, obtained from simulation, is the ratio of the number of selected signal events at detector level to the number of events at particle level in the fiducial phase space.
This factor corrects for detector efficiencies and for QED final-state radiation effects.
The contribution from $\tau$-lepton decays, amounting approximately to $4\%$, is removed from the cross-section definition by introducing the term in parentheses.
This term is computed using simulation, where $N_\tau$ is the number of selected events at detector level in which at least one of the bosons decays into a $\tau$-lepton and $N_{\textrm{all}}$ is the number of selected  $WZ$ events with decays into any lepton.

The  $C_{WZ}$ factors for $W^- Z$,  $W^+ Z$, and $W^\pm Z$ inclusive processes computed with \sherpa for each of the four leptonic channels are shown in Table~\ref{tab:cwz}.

\begin{table}[tb]
\caption{The $C_{WZ}$ factors for each of the $eee$, $\mu{ee}$, ${e}\mu\mu$, and $\mu\mu\mu$ inclusive channels.
The  \sherpa MC event sample with the ``resonant shape'' lepton assignment algorithm at particle level is used. Only statistical uncertainties are reported. }
\label{tab:cwz}
\begin{center}
\begin{tabular}{cccc}
\hline
Channel &   $C_{W^-Z}$ &  $C_{W^{+}Z}$  & $C_{W^\pm Z}$\\
\hline
$eee$ 		 	 & 0.353 $\pm$ 0.002  & 0.343 $\pm$ 0.001 & 0.347 $\pm$ 0.001 \\
$\mu{ee}$ 	 	 & 0.453 $\pm$ 0.002  & 0.450 $\pm$ 0.002 & 0.451 $\pm$ 0.001 \\
${e}\mu\mu$      & 0.525 $\pm$ 0.002  & 0.510 $\pm$ 0.002 & 0.516 $\pm$ 0.001 \\
$\mu \mu\mu$     & 0.694 $\pm$ 0.003  & 0.695 $\pm$ 0.002 & 0.695 $\pm$ 0.002 \\
\hline
\end{tabular}
\end{center}

\end{table}

The measured fiducial  cross-sections for the four channels are combined using a $\chi^2$ minimisation method that accounts for correlations between the sources of systematic uncertainty affecting each channel~\cite{Glazov:2005rn,Aaron:2009bp,Aaron:2009wt}.

\subsection{Differential cross-section measurements}\label{sec:diffXsectionMethod}

The differential detector-level distributions within the fiducial phase space are corrected for detector resolution and for QED final-state radiation effects using simulated signal events and an iterative Bayesian unfolding method~\cite{DAgostini:1994zf}, as implemented in the RooUnfold toolkit~\cite{RooUnfold}.
Background events are subtracted from data following the estimates of Section~\ref{sec:background}.
Events from \wz production with at least one $\tau$-lepton decay are also subtracted from data.
A fiducial correction is applied to take into account events that are selected at detector-level but originate from outside of the fiducial phase space at particle-level.
The unfolding procedure corrects for migrations of the events  between bins in the distributions from the particle to the detector-level reconstruction.
Finally, efficiency corrections take into account events selected in the fiducial phase space at particle-level that are not reconstructed at detector-level due to detector inefficiencies.
To reduce the bias arising from using a chosen MC prediction to model the ‘true’ particle-level distribution, the method can be applied with multiple iterations, at the cost of an increased statistical uncertainty.

For the measurements of the differential distributions, all four decay channels, $eee$, $e\mu\mu$, $\mu ee$,
and $\mu\mu\mu$, are added together.
The resulting distributions are unfolded with a response matrix computed using a \sherpa MC signal sample that includes all four decay channels and is divided by four such that cross-sections refer to final states where the $W$ and $Z$ bosons decay in a single leptonic channel with muons or electrons.
The \sherpa signal sample is used for unfolding, since it provides a fair description of the data distributions.
The number of iterations was tuned to obtain the best trade-off in minimising both the unfolding bias and the unfolding statistical uncertainty.
For each observable, the number of iterations ranges from two to five, depending on the resolution in the unfolded variable.
The width of the bins in each distribution is chosen according to the experimental resolution and to the statistical significance of the expected number of events in each bin.
The fraction of signal MC events generated in a bin which are reconstructed in the same bin is always greater than $40\%$ and around $70\%$ on average.
%
%
%
%
%


%
\section{Systematic uncertainties}
\label{sec:systematics}
%


%

Experimental and theory sources of systematic uncertainties in the measured cross-sections and limits on BSM effects are considered, including uncertainties in the correction procedure  for detector effects, in the background estimate and in the luminosity.

The systematic uncertainties  in the $C_{WZ}$ factors  related to the theory modelling are due to the choice of PDF set, QCD renormalisation  $\mu_R$ and factorisation $\mu_F$ scales, and MC modelling of higher-order QCD effects simulated by PS and of the underlying event.
The uncertainties due to the PDF and the $\alphas$ value used in the PDF determination are evaluated using the \texttt{PDF4LHC} prescription~\cite{Butterworth:2015oua}.
Uncertainties due to missing higher-order QCD corrections are evaluated by varying the  $\mu_R$ and $\mu_F$ scales independently by factors of two and one-half, removing combinations where the variations differ by a factor of four.
The impact of these variations in the $C_{WZ}$ factors is of maximally 0.4\%.
A global MC modelling uncertainty, that includes effects of the parton shower model, is estimated by evaluating the $C_{WZ}$ factors using the \sherpa and \powhegpythia MC event generators.
The difference between the $C_{WZ}$ factors from the two generators amounts to 3\%, independent of the decay channel and is considered as an uncertainty.
This discrepancy arises mostly from differences in the isolation of decay leptons from hadronic activity from the PS and the underlying event and, to a lower extent, from a slightly harder \pt spectrum of decay leptons in one MC generator compared to the other.

Uncertainties in differential cross-section measurements arising from the imperfect description of the data by the \sherpa MC simulation are evaluated using a data-driven method~\cite{Malaescu:2011yg}.
Each MC differential distribution is corrected to match the data distribution and the resulting weighted MC distribution at detector level is unfolded with the response matrix used in the actual data unfolding.
The new unfolded distribution is compared with the weighted MC distribution at particle level and the difference is taken as the systematic uncertainty.
This procedure results in an uncertainty in the measured cross-sections of the order of 0.2\% on average, with a maximal value of 2\% depending on the observable.
An additional uncertainty is attributed to the measurements to account for differences between the \sherpa and \powhegpythia generators.
The \powhegpythia generator is alternatively used to unfold the data and the deviation from the nominal result is taken as the uncertainty.
To remove effects already accounted for in the data-driven method, the \powhegpythia distributions were first reweighted to match \sherpa distributions at particle level.
The resulting uncertainty in the measured differential cross-section is 3\% on average.

Systematic uncertainties affecting the reconstruction and energy calibration of electrons, muons and jets are propagated through the analysis.
The uncertainties due to lepton reconstruction, identification, isolation requirements and trigger efficiencies as well as in the lepton momentum scale and resolution are assessed using tag-and-probe methods in $Z \to \ell \ell$ events~\cite{ATLAS:2019qmc,ATLAS:2020auj,ATLAS:2023mnw}.
The uncertainties in the jet energy scale are obtained from $\sqrt{s} = 13$~\TeV\ simulations and {\textit{in situ}} measurements~\cite{JETM-2018-05}.
The uncertainties in the jet energy resolution~\cite{JETM-2018-05}, in the suppression of jets originating from pile-up~\cite{Aad:2015ina}, in the $b$-tagging efficiency and in the mistag rate~\cite{FTAG-2019-07} are also considered.
The uncertainty in \MET is estimated by propagating the uncertainties in the transverse momenta of reconstructed objects and by applying momentum scale and resolution uncertainties to the track-based soft term~\cite{JETM-2020-03}. %
A variation in the pile-up reweighting of MC events is included to cover the uncertainty in the ratio of the predicted and measured $pp$ inelastic cross-sections~\cite{Aaboud:2016mmw}.
For the measurements of the $W$ charge-dependent cross-sections, an uncertainty arising from the charge misidentification of electrons is also considered~\cite{ATLAS:2019qmc}.
It affects only electrons and leads to an uncertainty of less than $0.15\%$ in the ratio of $W^+Z$ to $W^- Z$ integrated cross-sections determined by combining the four decay channels.

The dominant contribution among the experimental systematic uncertainties in the $eee$ and $\mu ee$  channels is due to the uncertainty in the electron reconstruction and identification efficiencies, contributing at most a $1.7\%$ uncertainty to the integrated cross-section, while in the $e \mu\mu$ and $\mu\mu\mu$ channels it originates from the muon reconstruction and isolation efficiencies  and is also at most  $1.7\%$.
The effect of pile-up uncertainties is of the order of 0.5\% to 1\% for muon and electron channels, respectively.

The uncertainty in the amount of background from misidentified leptons takes into account the limited number of events in the control regions as well as the difference in background composition between the control region used to determine the lepton misidentification rate and the control regions used to estimate the yield in the signal region.
This results in an uncertainty of about $25\%$ in the total  misidentified-lepton background yield and in the shape of the differential distributions of the reducible background events.

Uncertainties due to the theory modelling of \zz generated events are considered.
They arise from higher-order QCD corrections and the PDFs and are evaluated in the same way as for \wz\ events.
The uncertainty due to irreducible background sources other than \zz is evaluated by propagating the uncertainty in their MC cross-sections.
These are $30\%$ for $VVV$~\cite{ATL-PHYS-PUB-2016-002}, $25$\% for \wzjjew~\cite{WZVBS36fb} and $15\%$ for $tZj$~\cite{Aad:2020wog} and $t\bar{t}+V$~\cite{ATLAS:2021fzm,ATL-PHYS-PUB-2016-005}.
A normalisation uncertainty of 10\% is attributed to the contribution of events from \wz\ production with at least one $\tau$-lepton decay, corresponding to the QCD scale uncertainty of the Sherpa MC prediction.

\begin{table}[!htbp]
\caption{Summary of the relative uncertainties in the measured fiducial cross-section $\sigma^{\textrm{fid.}}_{W^\pm Z}$ for each channel and for their combination.
The uncertainties are reported as percentages.
The first rows indicate the main sources of systematic uncertainty for each channel and their combination, which are treated as correlated between channels.
A row with uncorrelated uncertainties follows, which comprise all uncertainties of statistical origin including MC statistical uncertainty and statistical uncertainties in the reducible background estimate, which are uncorrelated between channels.
}
\label{tab:SystematicSummary}
\begin{center}

\begin{tabular}{l r r r r r}
\hline
& $e e e$ & $\mu e e$  & $e \mu\mu$ & $\mu\mu\mu$ & Combined \\
\hline
\multicolumn{6}{c}{Relative uncertainties [\%]}\\
\hline
$e$ energy scale & $0.3$ & $< 0.1$ & $0.2$ & $< 0.1$ & $0.1$ \\
$e$ efficiency & $1.7$ & $1.1$ & $0.6$ & $< 0.1$ & $0.6$ \\
$\mu$ momentum scale & $< 0.1$ & $< 0.1$ & $< 0.1$ & $< 0.1$ & $< 0.1$ \\
$\mu$ efficiency & $< 0.1$ & $1.4$ & $0.5$ & $1.7$ & $0.9$ \\
$E_{\mathrm{T}}^{\mathrm{miss}}$ and jets & $0.4$ & $0.4$ & $0.5$ & $0.5$ & $0.4$ \\
Trigger & $< 0.1$ & $< 0.1$ & $< 0.1$ & $0.2$ & $0.1$ \\
Pile-up & $1.0$ & $0.9$ & $0.6$ & $0.4$ & $0.6$ \\
Misid. leptons background & $2.3$ & $1.1$ & $1.5$ & $1.4$ & $1.4$ \\
$ZZ$ background & $0.9$ & $0.8$ & $0.9$ & $0.8$ & $0.8$ \\
Other backgrounds & $0.9$ & $0.9$ & $0.9$ & $0.9$ & $0.9$ \\
\hline
Uncorrelated & $0.4$ & $0.3$ & $0.3$ & $0.3$ & $0.2$ \\
\hline
Total experimental uncertainty & $3.4$ & $2.6$ & $2.3$ & $2.6$ & $2.3$ \\
Luminosity & $0.9$ & $0.9$ & $0.9$ & $0.9$ & $0.9$ \\
Theoretical modelling & $3.0$ & $3.0$ & $3.0$ & $3.0$ & $3.0$ \\
Data statistics & $2.0$ & $1.8$ & $1.6$ & $1.4$ & $0.8$ \\
\hline
Total & $5.0$ & $4.5$ & $4.2$ & $4.3$ & $4.0$ \\
\hline
\end{tabular}


\end{center}
\end{table}

The uncertainty in the combined $2015$--$2018$ integrated luminosity is $0.83\%$~\cite{DAPR-2021-01}, obtained using the LUCID-2 detector~\cite{LUCID2} for the primary luminosity measurements, complemented by measurements using the inner detector and calorimeters.
It is applied to the signal normalisation as well as to all background contributions that are estimated
using only MC simulations. %

The total systematic uncertainty in the measured \wz fiducial cross-section, excluding the luminosity uncertainty, varies between  $4.2\%$ and $5\%$ for the four measurement channels, and is dominated by the theory modelling of the $C_{WZ}$ factors, arising from differences between the \sherpa and \powhegpythia  MC generators.
Table~\ref{tab:SystematicSummary} shows the statistical uncertainty and the main sources of systematic uncertainty in the \wz fiducial cross-section for each of the four channels and for their combination.
%
%
%
%


\clearpage
\section{Results}
\label{sec:CrossSections}

\subsection{Detector-level results}
\label{sec:RecoResults}


Table~\ref{tab:Results:YieldsSummary} summarises the predicted and observed number of events resulting from the \wz inclusive selection, together with the estimated background contributions.
Figure~\ref{fig:Results:WZControlPlots} shows the measured distributions of the transverse momentum and the invariant mass of the $Z$ candidate, the transverse mass of the $W$ candidate, and for the $WZ$ system the absolute difference between the azimuthal angle of the $Z$ boson and the charged lepton from the decay of the $W$ boson, \dPhiWlZ.
The \sherpa MC prediction is used for the \wz\ signal contribution.
Figure~\ref{fig:Results:WZControlPlots}  indicates that the MC predictions provide a fair description of the shapes of the data distributions.

\begin{table}[!htbp]
\centering
\caption{Total numbers of observed and expected events after the \wz inclusive selection described in Section~\ref{sec:selection} in each of the considered channels and for the sum of all channels.
The expected approximate size of \wz\ events from \sherpa and the estimated amount of background events from other processes are expressed as a fraction of the total number of predicted events.
The sum of background events containing misidentified leptons is labelled ``Misid. leptons''.
The total uncertainties in the total number of predicted events are reported.
}
\label{tab:Results:YieldsSummary}
%

%
%
%
%
%
%
%
%
%
%
%
%
%
%
%
%
%
%
%
%
%
%
%
%
%
%
%
%
%
%

\begin{tabular}{l rrrr r}
\toprule
Channel  & $\;\;\;\;\;e e e$ & $\;\;\;\;\;\mu e e$ & $\;\;\;\;\;e\mu\mu$ & $\;\;\;\;\;\mu\mu\mu$ & $\;\;\;\;\;\;\;$All \\
\midrule
Data 	& $3955$  	& $4600$  	& $5895$  	& $7486$  	& $21936$  	\\
\midrule
Total expected 	& $ 4000$ 	& $ 4900$ 	& $ 5900$ 	& $ 7600$ 	& $22400$ 	\\
& $\pm \, 500$ 	& $\pm \, 500$ 	& $\pm \, 700$ 	& $\pm \, 900$ 	& $\pm \, 2500$ 	\\
\midrule
$WZ$ 	& $78 \% $ 	& $83 \% $ 	& $79 \% $ 	& $82 \% $ 	& $81 \% $ 	\\
$ZZ$ 	& $ 8 \% $ 	& $ 7 \% $ 	& $ 8 \% $ 	& $ 7 \% $ 	& $ 7 \% $ 	\\
Misid. leptons 	& $ 7 \% $ 	& $ 4 \% $ 	& $ 6 \% $ 	& $ 5 \% $ 	& $ 5 \% $ 	\\
$t\bar{t}+V$ 	& $ 4 \% $ 	& $ 4 \% $ 	& $ 4 \% $ 	& $ 4 \% $ 	& $ 4 \% $ 	\\
$tZj$ 	& $ 2 \% $ 	& $ 2 \% $ 	& $ 2 \% $ 	& $ 1 \% $ 	& $ 2 \% $ 	\\
$WZjj\mathrm{-EW}$ 	& $ 1 \% $ 	& $ 1 \% $ 	& $ 1 \% $ 	& $ 1 \% $ 	& $ 1 \% $ 	\\
$VVV$ 	&  $< 1\% $ 	&  $< 1\% $ 	&  $< 1\% $ 	&  $< 1\% $ 	&  $< 1\% $ 	\\
\bottomrule
\end{tabular}

%
%
%
%
%
%
%
%
%
%
%
%
%
%
%
%
%
%
%
%
%
%


%
\end{table}

\begin{figure}[!htbp]
\centering
\subfloat[]{\includegraphics[width=.4\textwidth]{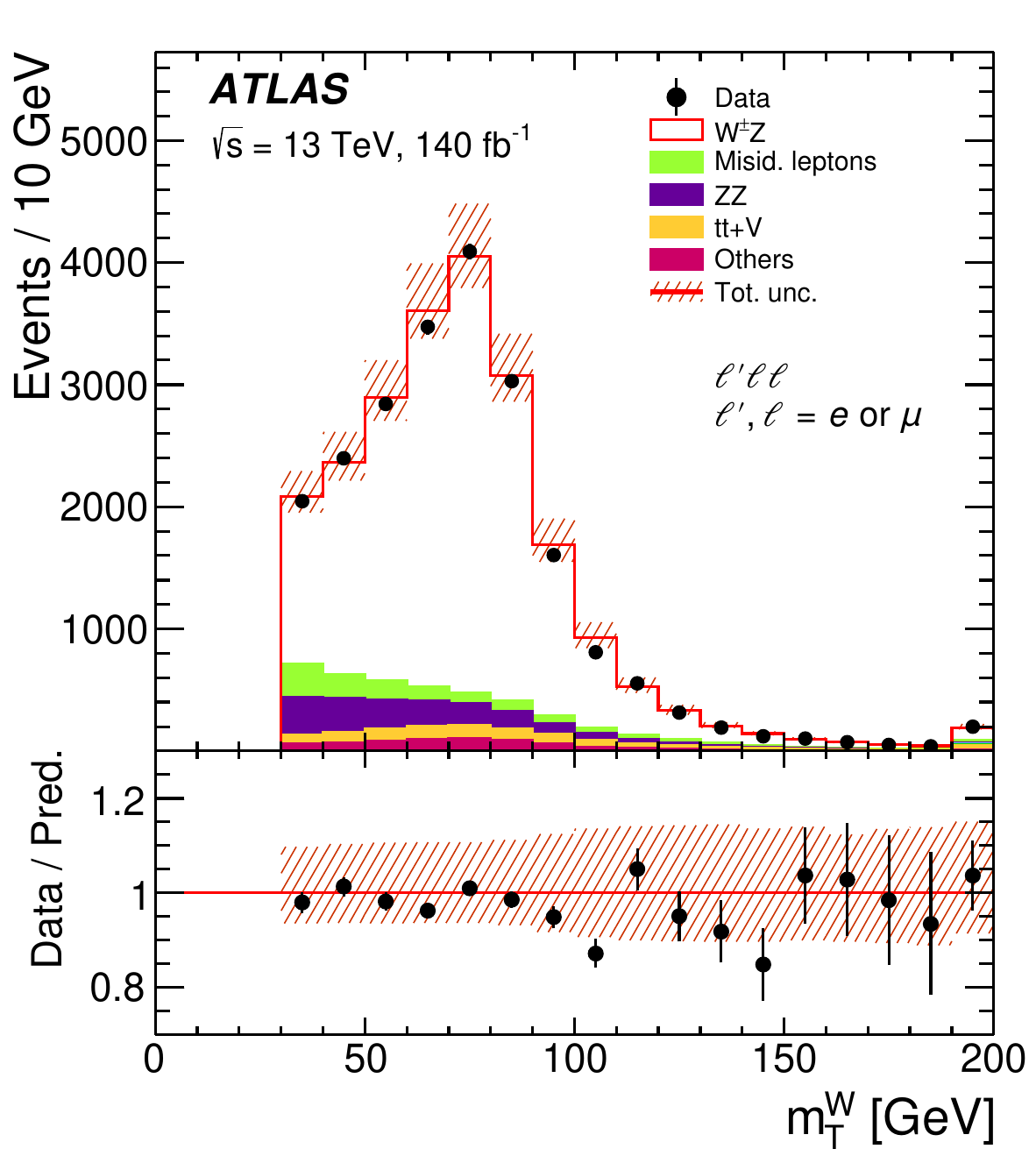}}
\subfloat[]{\includegraphics[width=.4\textwidth]{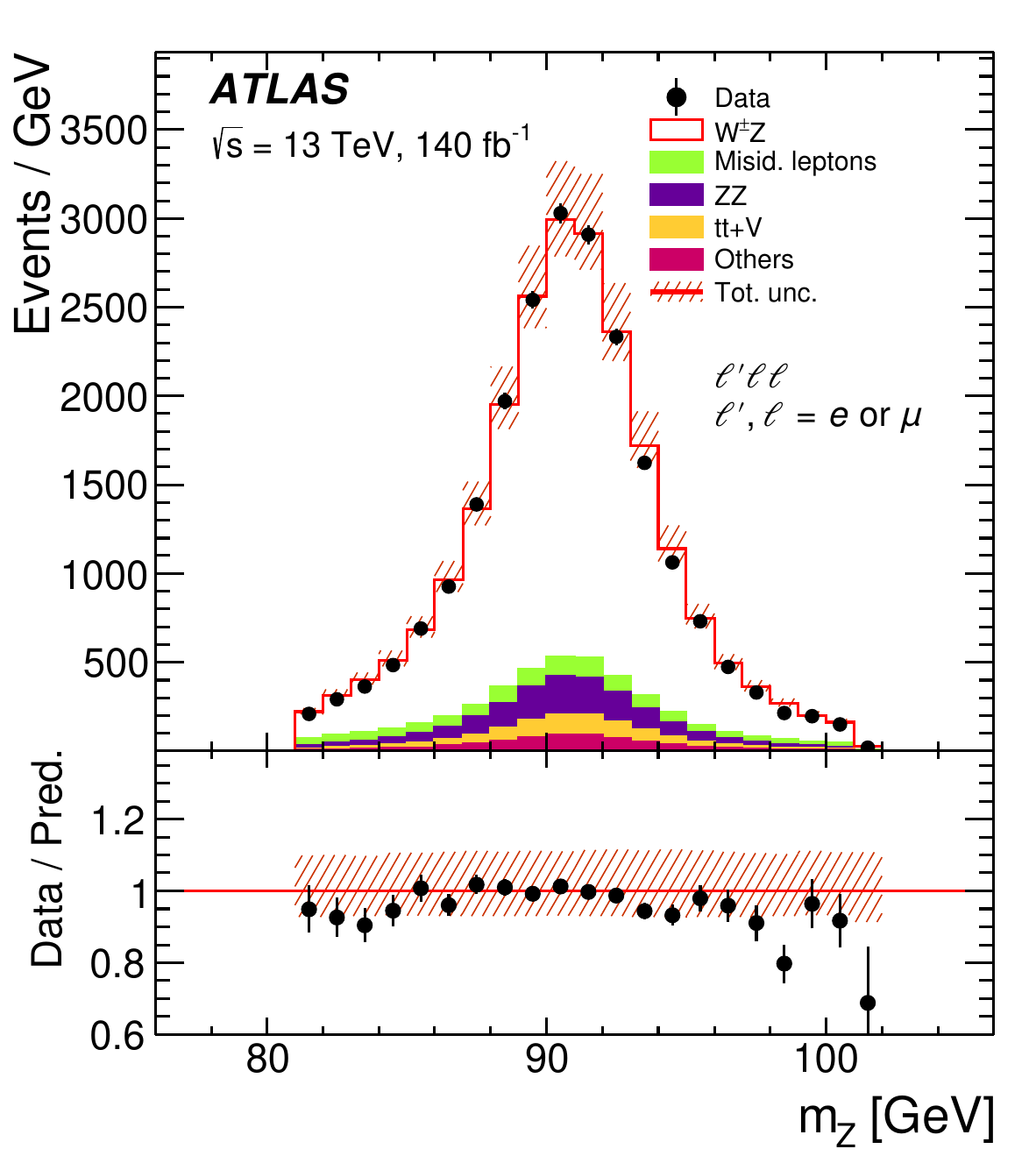}}\\
\subfloat[]{\includegraphics[width=.4\textwidth]{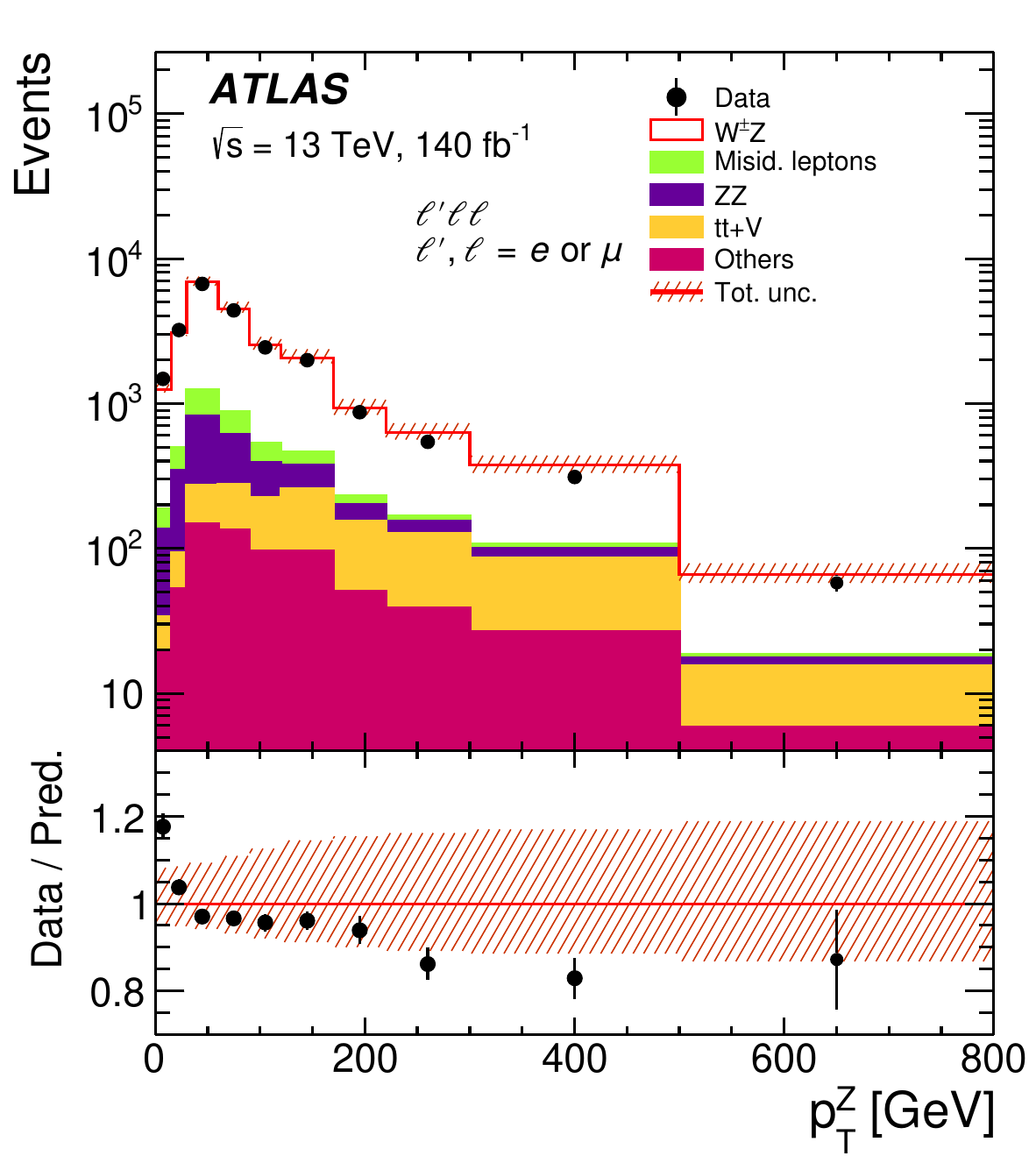}}
\subfloat[]{\includegraphics[width=.4\textwidth]{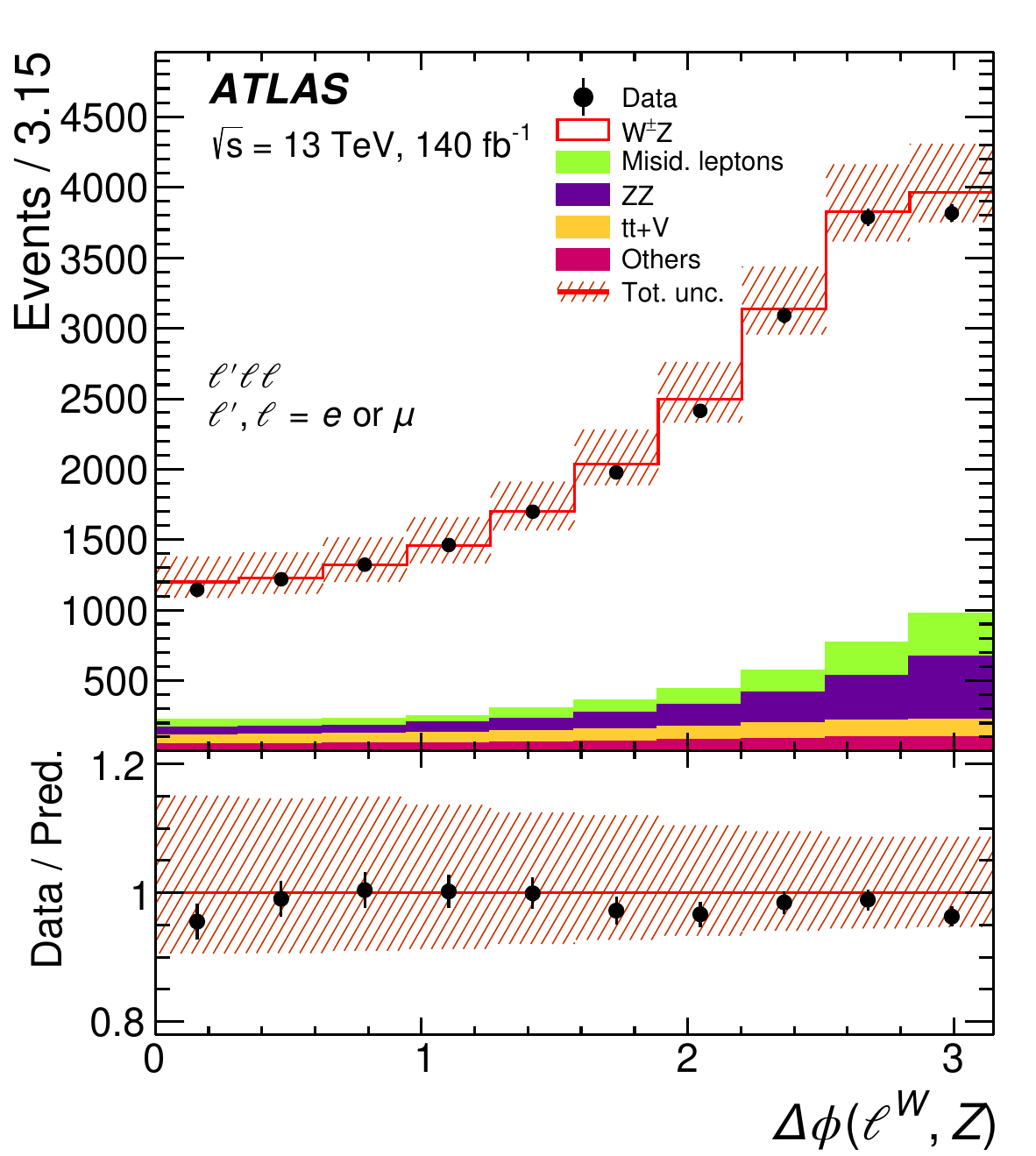}}
\caption{The distributions of observed and expected events after the \wz\ inclusive selection described in Section~\ref{sec:selection}, for the sum of all channels, of the kinematic variables (a) \mtw, (b) \mz, (c) \ptz and  (d) \dPhiWlZ.
The points correspond to the data with the error bars representing the statistical uncertainties, and the histograms correspond to the predictions of the various SM processes.
The sum of the background processes with misidentified leptons is labelled ``Misid. leptons''.
The \sherpa MC prediction is used for the \wz\ signal contribution.
The red histogram shows the total prediction; the shaded band is the total uncertainty in this prediction.
The last bin in (a), (c) and (d) contains the overflow.
The lower panels in each figure show the ratio of the data points to the open red histogram with their respective uncertainties.
}
\label{fig:Results:WZControlPlots}
\end{figure}


\clearpage
\subsection{Integrated cross-section measurements}
\label{sec:IntCrossSections}


%
%
%
The integrated \wz cross-sections in the fiducial phase space are measured in each of the four channels.
The combination of the \wz cross-sections for the four channels yields a $\chi^2$ per degree of freedom ($\mathrm{dof}$) of $\chi^2/n_{\mathrm{dof}} = 7.7/3$.
The combinations of the $W^+ Z$ and $W^- Z$ cross-sections separately yield $\chi^2/n_{\mathrm{dof}} = 1.9/3$ and $13.8/3$, respectively.
The \wz production cross-section in the fiducial phase space resulting from the combination of the four channels including the $W$ and $Z$ branching ratio in a single leptonic channel with muons or electrons is

\begin{equation*}
\sigma_{W^{\pm}Z \rightarrow \ell^{'} \nu \ell \ell}^{\mathrm{fid.}}  =  60.7  \pm 0.5 \,\mathrm{(stat.)} \pm 1.4 \,\mathrm{(exp. \, syst.)}  \pm  1.8 \,\mathrm{(mod. \, syst.)}   \pm  0.6 \,\mathrm{(lumi.)} \; \mathrm{fb},
\end{equation*}

where the uncertainties correspond to statistical, experimental systematic, modelling systematic and luminosity uncertainties, respectively.
The corresponding SM \nnloxew prediction from \MATRIX is $61.4 \pm 1.3 \, \mathrm{fb}$, where the uncertainty corresponds to the QCD scale uncertainty estimated conventionally by varying the scales $\mu_{\mathrm{R}}$ and $\mu_{\mathrm{F}}$ by factors of two around the nominal value with the constraint $0.5 \leq \mu_{\mathrm{R}} /\mu_{\mathrm{F}} \leq 2$.
Changing the scheme used to combine QCD and EW corrections to \nnlopew increases the \matrix prediction by $+3\%$.
Changing the PDF set used from \NNPDF[3.1nnlo\_as\_118\_luxqed] to \myMSHT[qed\_nnlo\_inelastic]~\cite{Bailey:2020ooq,Cridge:2021pxm} or \CT[18qed\_proton\_inelastic]~\cite{Hou:2019efy,Xie:2021equ} affects the \matrix prediction by  $-1\%$.
\begin{table}[t]
\caption{Fiducial integrated cross-section in fb, for $W^\pm Z$, $W^+ Z$ and $W^- Z$ production, measured in each of the channels $eee$, $\mu{ee}$, ${e}\mu\mu$, and $\mu\mu\mu$ and for all four channels combined.
The statistical ($\delta_{\mathrm{stat.}}$), experimental systematic ($\delta_{\mathrm{exp.\, syst.}}$), modelling systematic ($\delta_{\mathrm{mod.\, syst.}}$), luminosity ($\delta_{\mathrm{lumi.}}$) and total ($\delta_{\mathrm{tot.}}$) uncertainties are given in percent.
The \nnloxew SM predictions from \matrix using the \NNPDF[3.1nnlo\_as\_118\_luxqed] PDF set are also reported.
}
\label{tab:XSection:FiducialCSAll}
\begin{center}
%

\begin{tabular}{lcccccc}

\toprule
Channel   &  $\sigma^{\mathrm{fid.}}$ & $\delta_{\mathrm{stat.}}$ & $\delta_{\mathrm{exp.\, syst.}}$ & $\delta_{\mathrm{mod.\, syst.}}$& $\delta_{\mathrm{lumi.}}$ &$\delta_{\mathrm{tot.}}$\\
&  [fb] & [\%] &  [\%] & [\%]& [\%] &[\%]\\
\midrule
\multicolumn{6}{c}{~}\\ [-12.0pt]
\multicolumn{6}{c}{$\sigma_{W^{\pm}Z \rightarrow \ell^{'} \nu \ell \ell}^{\mathrm{fid.}}$} \\ [4.0pt]
\midrule
$e^\pm{ee}$ & 60.6 & 2.0 & 3.4 & 3.0 & 0.9 & 5.0 \\
$\mu^\pm{ee}$ & 57.5 & 1.8 & 2.6 & 3.0 & 0.9 & 4.5 \\
$e^\pm\mu\mu $ & 62.0 & 1.6 & 2.3 & 3.0 & 0.9 & 4.2 \\
$\mu^\pm\mu\mu$ & 60.7 & 1.4 & 2.6 & 3.0 & 0.9 & 4.3 \\
\midrule
Combined & 60.7 & 0.8 & 2.3 & 3.0 & 0.9 & 4.0 \\
SM prediction &  61.4 & --- & --- & --- & --- & 2.1 \\
\midrule
\multicolumn{6}{c}{~}\\ [-12.0pt]
\multicolumn{6}{c}{$\sigma_{W^{+}Z \rightarrow \ell^{'} \nu \ell \ell}^{\mathrm{fid.}}$} \\ [4.0pt]
\midrule
$e^+{ee}$ & 37.0 & 2.6 & 3.1 & 3.0 & 1.0 & 5.1 \\
$\mu^+{ee}$ & 35.0 & 2.3 & 2.6 & 3.0 & 1.0 & 4.7 \\
$e^+\mu\mu $ & 36.1 & 2.2 & 2.1 & 3.0 & 1.0 & 4.4 \\
$\mu^+\mu\mu$ & 35.5 & 1.8 & 2.5 & 3.0 & 1.0 & 4.4 \\
\midrule
Combined & 35.8 & 1.1 & 2.2 & 3.0 & 1.0 & 4.0 \\
SM prediction &  36.3 & --- & --- & --- & --- & 3.5 \\
\midrule
\multicolumn{6}{c}{~}\\ [-12.0pt]
\multicolumn{6}{c}{$\sigma_{W^{-}Z \rightarrow \ell^{'} \nu \ell \ell}^{\mathrm{fid.}}$} \\ [4.0pt]
\midrule
$e^-{ee}$ & 23.9 & 3.3 & 3.7 & 3.0 & 1.1 & 5.9 \\
$\mu^-{ee}$ & 22.5 & 2.9 & 2.8 & 3.0 & 1.1 & 5.1 \\
$e^-\mu\mu $ & 25.9 & 2.6 & 2.5 & 3.0 & 1.1 & 4.8 \\
$\mu^-\mu\mu$ & 25.2 & 2.2 & 2.8 & 3.0 & 1.0 & 4.8 \\
\midrule
Combined & 24.8 & 1.3 & 2.6 & 3.0 & 1.1 & 4.3 \\
SM prediction &  25.1 & --- & --- & --- & --- & 5.1 \\
\bottomrule
\end{tabular}


%
\end{center}
\end{table}

In Figure~\ref{fig:xsectionperchannel}, the measured \wz production cross-sections are compared with the SM \nnloxew or \nnlopew predictions from \MATRIX using the \NNPDF[3.1nnlo] PDF set.
The \matrix prediction using the \myMSHT[nnlo] PDF set is also represented.
All results for $W^\pm Z$, $W^+Z$ and $W^-Z$ final states are reported in Table~\ref{tab:XSection:FiducialCSAll}.
Individual cross-section measurements in each channel are in agreement within their respective uncertainties.
The MATRIX calculation at \nnloxew reproduces well the measured cross-sections.
The total uncertainty of 4\% on the combined measurement is of the same order as the QCD scale uncertainty of 2\% on the \matrix prediction and as the difference  of 3\% in the predictions from the \nnloxew or \nnlopew schemes.

\begin{figure}[t]
\begin{center}
\includegraphics[width=9cm]{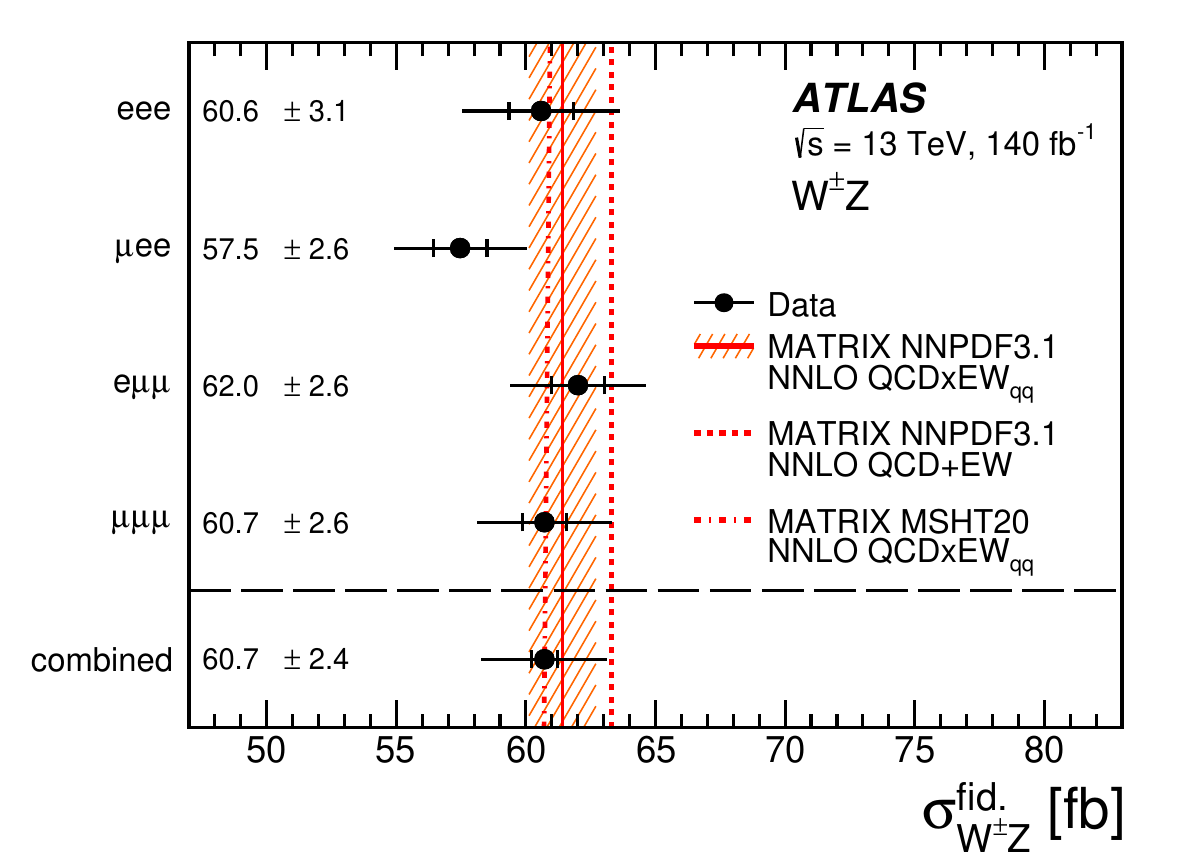}
\caption{The measured \wz\ integrated cross-sections in the fiducial phase space in each of the four channels and for their combination.
The inner and outer error bars on the data points represent the statistical and total uncertainties, respectively.
The \nnloxew SM prediction from \MATRIX using the \NNPDF[3.1nnlo\_as\_118\_luxqed] PDF set is shown as the red line;
the shaded band shows the effect of QCD scale uncertainties on this prediction.
The \nnlopew predictions from \matrix is also represented by the dashed-red line, as well as the \nnloxew \matrix prediction using the \myMSHT[nnlo] PDF set represented by the dotted-dashed red line.
}
\label{fig:xsectionperchannel}
\end{center}
\end{figure}

The ratio of the $ W^+Z$ to $W^-Z$ production cross-sections is

\begin{equation*}
\frac{\sigma_{W^{+}Z \rightarrow \ell^{'} \nu \ell \ell}^{\textrm{fid.}}}{\sigma_{W^{-}Z \rightarrow \ell^{'} \nu \ell \ell}^{\textrm{fid.}}}  =  1.456\pm 0.025 \,\textrm{(stat.)} \pm 0.010 \,\textrm{(syst.)}.
\end{equation*}

Most of the systematic uncertainties almost cancel out in the ratio, so that the measurement is dominated by the statistical uncertainty.
The measured cross-section ratios, for each channel and for their combination, are compared in Figure~\ref{fig:WPMXSection:WpWmRatio}
with the SM prediction of $1.435^{+0.013}_{-0.014}$, calculated with \MATRIX and the \CT[18nnlo] PDF set.
The same value of the cross-section ratio is predicted by the \sherpa MC simulation using the \NNPDF[3.0nnlo] PDF set.
The measured combined cross-section ratio is in agreement with the predictions.
Measurements in the $eee$ and $\mu ee$ channels are in agreement within $2\sigma$ with the predictions.
The uncertainties in the predictions correspond to PDF uncertainties estimated at NLO with \sherpa using the \NNPDF[3.0nnlo] eigenvectors and the envelope of the differences between the \CT[18nnlo],  \myMSHT[nnlo] and \NNPDF[3.1nnlo] PDF sets.
The effects of QCD scale uncertainties in the predicted cross-section ratio are negligible.
The predicted cross-section ratio is also calculated with \matrix using the \NNPDF[3.1nnlo] PDF set, yielding a value of  $1.447$  as shown in Figure~\ref{fig:WPMXSection:WpWmRatio}.
The total PDF uncertainties in the SM predictions from most recent PDF sets are 2\% while the total measurement uncertainty in this cross-section ratio is 1.8\%.
This comparison indicates that more statistics will set significant constraints.

\begin{figure}[t]
\begin{center}
\includegraphics[width=9cm]{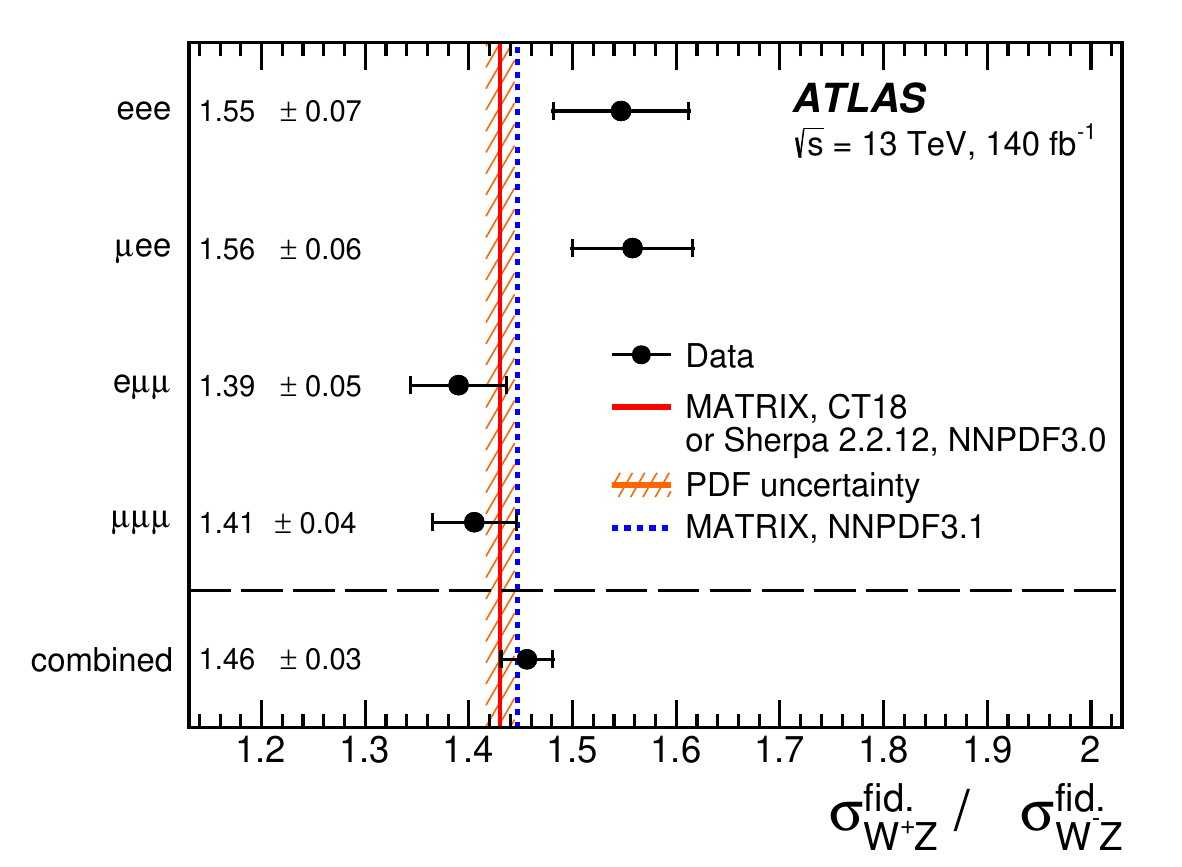}
\caption{Measured ratio $\sigma^{\mathrm{fid.}}_{W^{+}Z} / \sigma^{\mathrm{fid.}}_{W^{-}Z}$ of $W^{+}Z$ and $W^{-}Z$ integrated cross-sections in the fiducial phase space in each of the four channels and for their combination.
The error bars on the data points represent the total uncertainties, which are dominated by the statistical uncertainties.
The \nnloxew SM predictions from \MATRIX using the \CT[18nnlo] PDF set is represented as a single red line.
It is equal to the prediction from \sherpa using the \NNPDF[3.0nnlo] PDF set.
The dashed band represents the effect of PDF uncertainties estimated by using the \sherpa prediction from the \NNPDF[3.0nnlo] eigenvectors and the envelope of the differences between the \CT[18nnlo],  \myMSHT[nnlo], \NNPDF[3.1nnlo] and \PDFforLHC[15nnlo] PDF sets.
The \matrix prediction displayed by the dashed-blue line is using the  \NNPDF[3.1nnlo\_as\_118\_luxqed] PDF set.
}

\label{fig:WPMXSection:WpWmRatio}
\end{center}
\end{figure}

%
%
%
%
%

%
%
%
%

%
%
%
%
%


\clearpage
\subsection{Differential cross-section measurements}
\label{sec:DiffCrossSections}


%
\begin{figure}[t]
\centering
\subfloat[]{\includegraphics[width=.4\textwidth]{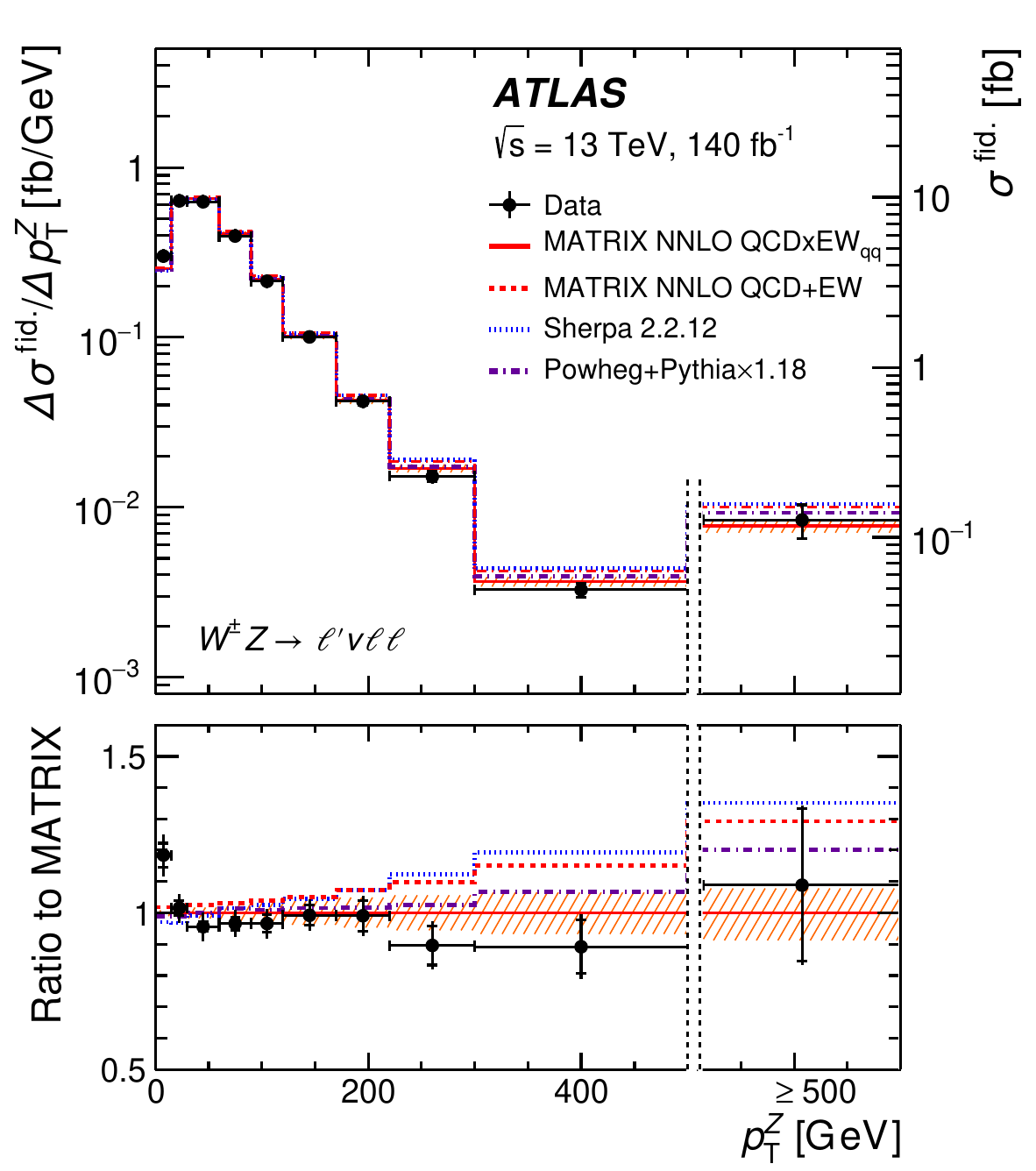}}
\subfloat[]{\includegraphics[width=.4\textwidth]{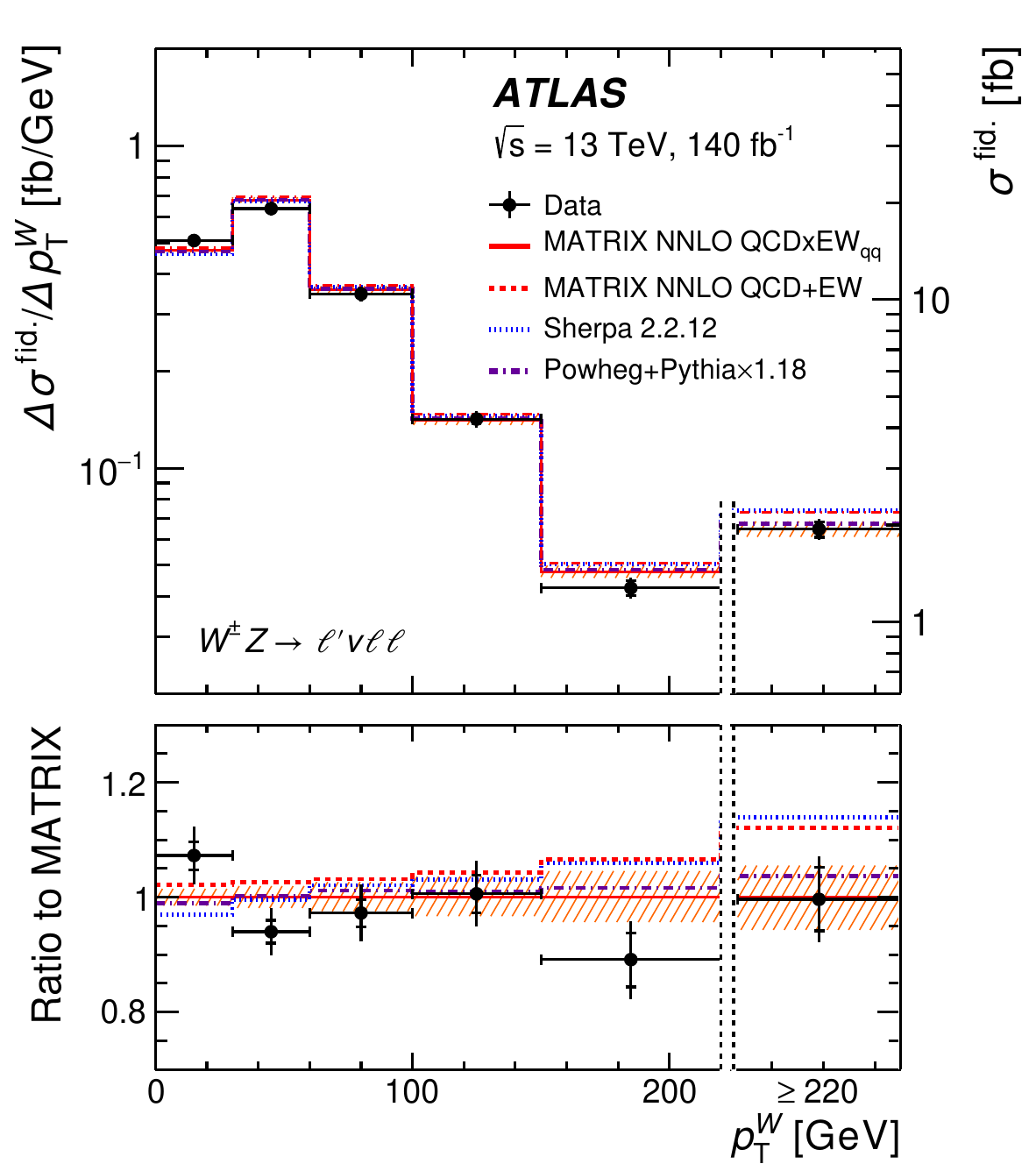}}\\
\subfloat[]{\includegraphics[width=0.4\textwidth]{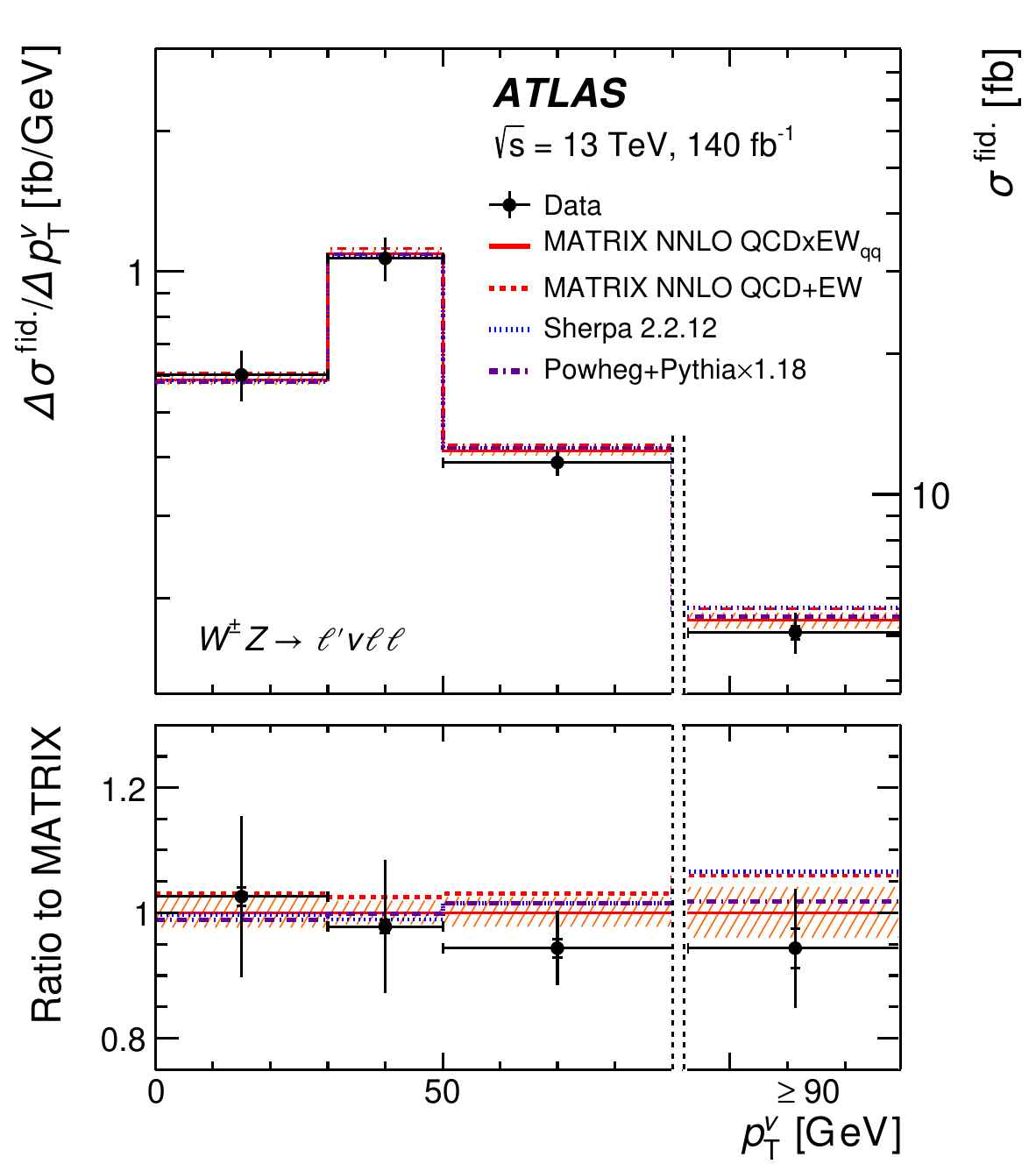}}
\subfloat[]{\includegraphics[width=0.4\textwidth]{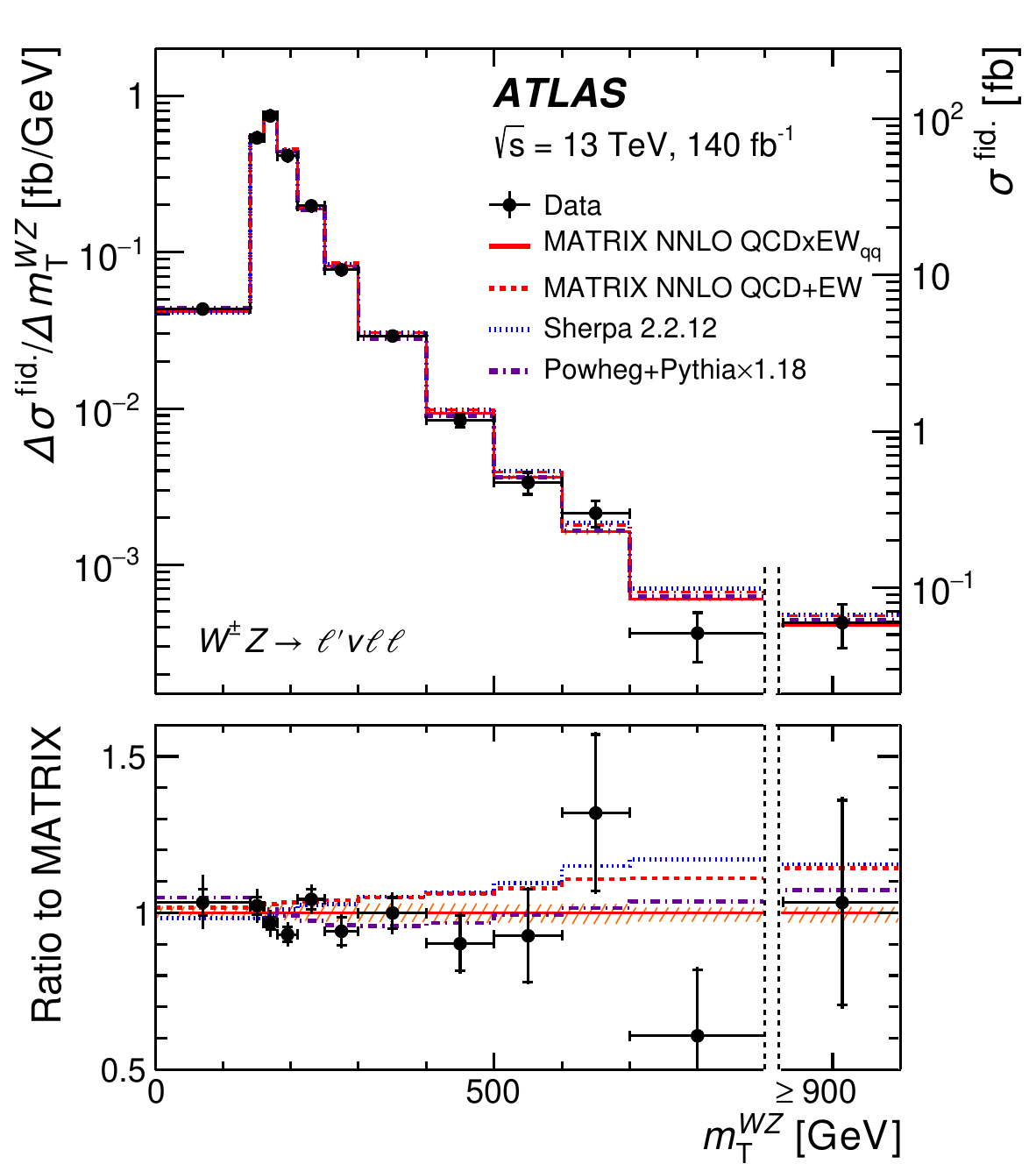}}
\caption{The measured $W^{\pm}Z$ differential cross-section in the fiducial phase space as a function of  (a)  $p_\textrm{T}^Z$,  (b) $p_\textrm{T}^W$, (c) \ptnu  and (d) \mtwz.
The inner and outer error bars on the data points represent the statistical and total uncertainties, respectively.
The measurements are compared with the prediction from \matrix using either the \nnloxew (red line)  or the \nnlopew  (dashed-red) line schemes.
The shaded band shows how the QCD scale uncertainties affect the \matrix predictions.
The predictions from the \sherpa and \powhegpythia MC generators are also indicated by dotted-blue and dotted-dashed violet lines, respectively.
The right vertical axis refers to the last cross-section point, separated from the others by vertical dashed lines,
as this last bin is integrated up to the maximum value reached in the phase space and the cross-section is not divided by the bin width. }
\label{fig:DiffXSection:pTZW_mTWZ}
\end{figure}

The \wz production cross-section is  measured as a function of several variables related to the energy of the \wz system produced: the transverse momenta of the $Z$ and $W$ bosons, \ptz and \ptw, the \pt of the neutrino associated with the decay of the $W$ boson, \ptnu and the transverse mass of the \wz system \mtwz, as presented in Figure~\ref{fig:DiffXSection:pTZW_mTWZ}.
The variable \mtwz is defined as follows: %
\begin{equation*}
m_{\mathrm{T}}^{WZ} = \sqrt{ \left( \sum_{\ell = 1}^3 p_{\mathrm{T}}^\ell + E_{\mathrm{T}}^{\mathrm{miss}} \right)^2
- \left[ \left(\sum_{\ell = 1}^3 p_x^\ell + E_{x}^{\mathrm{miss}} \right)^2 + \left(\sum_{\ell = 1}^3 p_y^\ell + E_{y}^{\mathrm{miss}} \right)^2 \right]} \, .
\end{equation*}
To derive \ptw and \ptnu from data events, it is assumed that the whole $E_\textrm{T}^\textrm{miss}$ of each event arises from the neutrino of the $W$ boson decay.
The validity of this assumption was verified for SM $WZ$ events using MC samples at the level of precision of the present results.
The measured  differential cross-sections in Figure~\ref{fig:DiffXSection:pTZW_mTWZ} are compared with the predictions from \MATRIX.
The predicted and measured cross-sections are in good agreement.
The measurements are also compared with NLO MC predictions from \powhegpythia, after a rescaling of its predicted integrated fiducial cross-section to the NNLO cross-section, and to \sherpa without rescaling its prediction.
Good agreement of the shapes of the measured distributions with the predictions of \powhegpythia and \sherpa is observed.
As shown in previous publications, the high energy tails of the \ptz~\cite{Aad:2012twa} and \mtwz~\cite{ATLASWZ_8TeV} observables are sensitive to anomalous triple gauge couplings (aTGC), \ptz having the disadvantage of being more subject to higher-order perturbative effects in QCD~\cite{Sapeta:2012} and electroweak theory~\cite{Biedermann:2017oae}.
This is also seen here with larger NNLO QCD scale uncertainties predicted by \matrix for \ptz\ than for \mtwz.
The difference between  \matrix predictions using the multiplicative or additive combination scheme for NLO EW corrections is also larger for \ptz than for \mtwz.
No excess of data events in the tails of these distributions is observed.
Deviations between data measurements and the theory predictions used are observed at low \ptz or \ptw values.
This is the manifestation in this kinematical regime of the domination of soft and collinear QCD radiations that cannot be described by the perturbative expansion in the strong coupling constant \alphas.
At low transverse momenta, a resummation of logarithmically-enhanced terms is required to obtain physical results~\cite{Grazzini:2015wpa,Kallweit:2020gva}.
The description of resummation effects with PS algorithms depends on the set of tuned parameters used, as it can be observed  in measurements of single $Z$ or $W$ production~\cite{STDM-2018-17}.

\begin{figure}[t]
\centering

\subfloat[]{\includegraphics[width=0.4\textwidth]{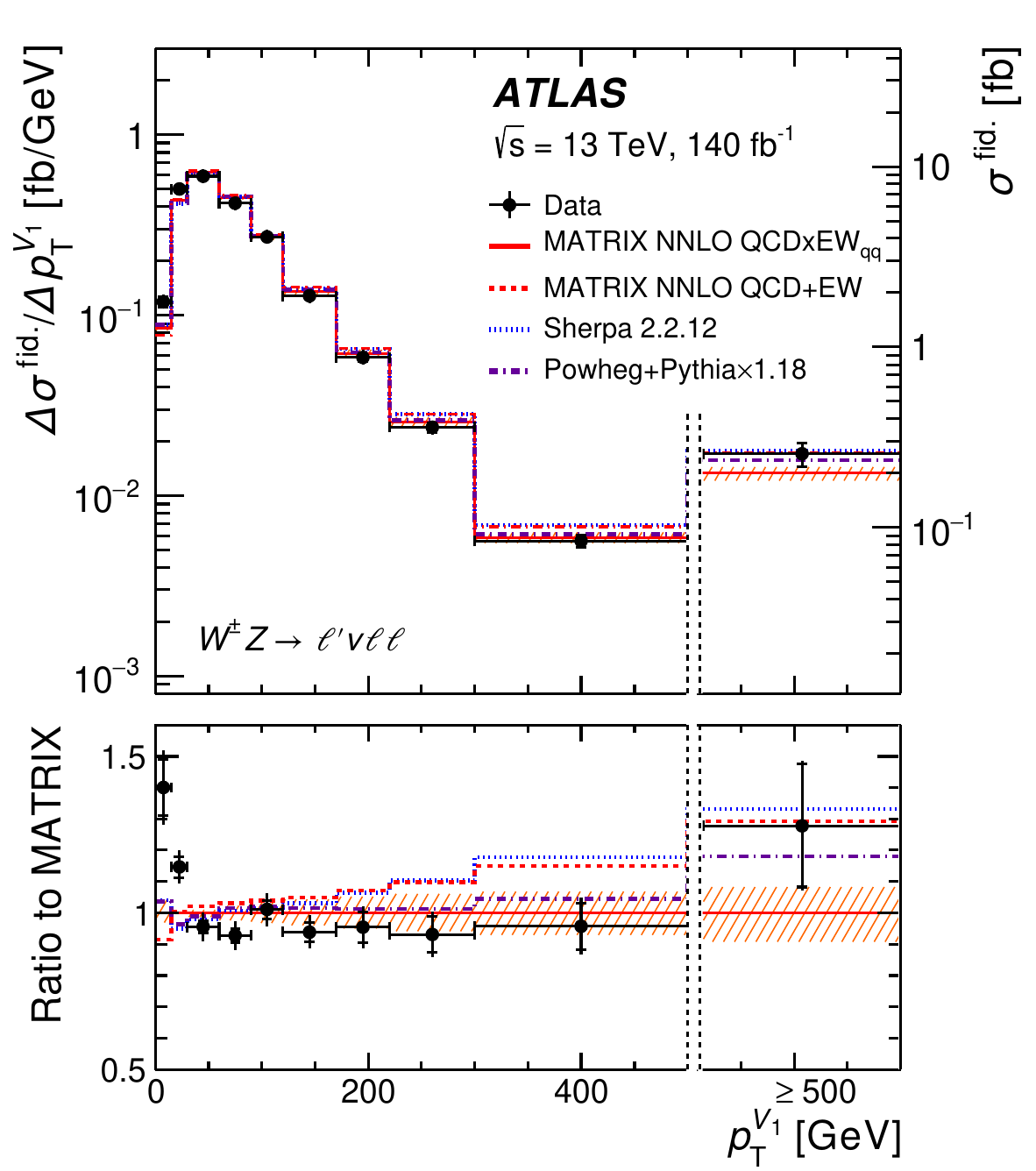}}
\subfloat[]{\includegraphics[width=0.4\textwidth]{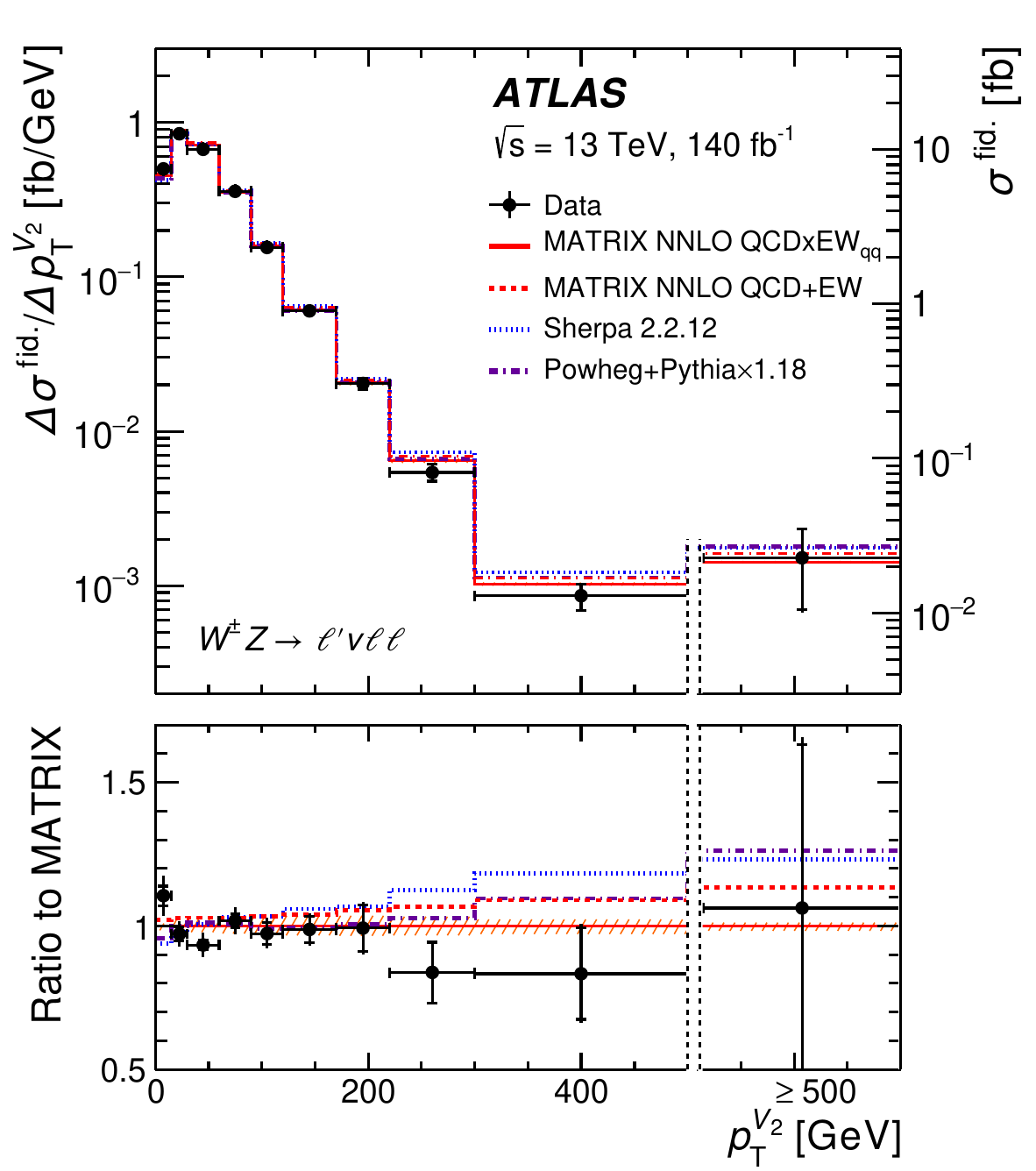}}
\caption{
The measured $W^{\pm}Z$ differential cross-section in the fiducial phase space as a function of (a) \ptvone  and (b) \ptvtwo.
The inner and outer error bars on the data points represent the statistical and total uncertainties, respectively.
The measurements are compared with the prediction from \matrix using either the \nnloxew (red line)  or the \nnlopew  (dashed-red) line schemes.
The shaded band shows how the QCD scale uncertainties affect the \matrix predictions.
The predictions from the \sherpa and \powhegpythia MC generators are also indicated by dotted-blue and dotted-dashed violet lines, respectively.
The right vertical axis refers to the last cross-section point, separated from the others by vertical dashed lines,
as this last bin is integrated up to the maximum value reached in the phase space and the cross-section is not divided by the bin width.
}
\label{fig:DiffXSection:pTV1_pTV2}
\end{figure}

Differential cross-sections as a function of the transverse momenta of the harder boson, \ptvone, and of the softer boson, \ptvtwo, are presented in Figure~\ref{fig:DiffXSection:pTV1_pTV2}.
The tail of the \ptvone distribution is especially sensitive to effects of NLO EW corrections~\cite{Grazzini:2019jkl}.
This is exemplified in Figure~\ref{fig:DiffXSection:pTV1_pTV2}(a) by the difference of 30\% observed in the last bin of the differential \ptvone cross-section between an additive combination of \NNLO QCD and \NLO EW correction,  \nnlopew, and  a factorised, \nnloxew, combination prescription, as calculated using \matrix~\cite{Grazzini:2019jkl}.
The data measurements of both \ptvone and \ptvtwo have a slope falling more rapidly than predicted by \sherpa and \matrix with the \nnlopew prescription, favouring therefore the \nnloxew scheme.
At very low \pt, the same mismodelling of data by present calculations as seen for \ptz and \ptw observables is observed.

Angular observables sensitive to QCD higher-order perturbative effects are presented in Figure~\ref{fig:DiffXSection:yZlW_all}: the absolute difference between the rapidities of the $Z$ boson and the charged lepton from the decay of the $W$ boson, \ayzlw, the difference between the azimuthal angle of the $Z$ boson and the charged lepton from the decay of the $W$ boson, \dPhiWlZ, and the azimuthal angle difference between the $W$ and $Z$ bosons, \dPhiWZ.
The rapidity correlations between the $W$ and $Z$ decay products have been found to be useful tools in searching for the approximately zero $WZ$ helicity amplitudes expected at LO in the SM, or for aTGCs~\cite{Baur:1994aj,Accomando:2005xp}.
These rapidity correlations are also sensitive to QCD corrections, PDF effects, and polarisation effects of the $W$ and $Z$ bosons.
The rapidity difference between the $W$ and $Z$ bosons, $|y_Z - y_W|$, is a boost-invariant substitute for the centre-of-mass scattering angle $\theta$ of the $W$ relative to the direction of the incoming quark.
Since the rapidity of the $W$ boson cannot be uniquely reconstructed due to the presence of the neutrino, the rapidity of the lepton from the $W$ boson decay is used instead.
The \dPhiWlZ and \dPhiWZ observables are also sensitive to higher-order QCD scale corrections.
An increase of the QCD scale uncertainties affecting the \matrix prediction is observed at low values of \dPhiWlZ and \dPhiWZ.
In this kinematical region, the prediction from \sherpa is seen to be very close to the \matrix prediction, while the \powheg prediction deviates more from \matrix.
Some difference between data and the prediction from \matrix or \sherpa is observed.
At low values of \ayzlw and \dPhiWZ the predictions have a tendency to overestimate the measured cross-sections.
The difference between data and predictions in these regions is not covered by QCD uncertainties alone, as estimated by using \matrix.
The difference observed could however arise from parton shower effects, as the difference of 3\% between \sherpa and \powheg MC predictions in the first bin in \ayzlw seems to indicate.
For \dPhiWlZ the difference at low values is covered by the QCD scale uncertainty derived for the \matrix prediction.
For these three observables, the difference in the treatment of NLO EW corrections is almost flat.

\begin{figure}[t]
\centering
\subfloat[]{\includegraphics[width=0.4\textwidth]{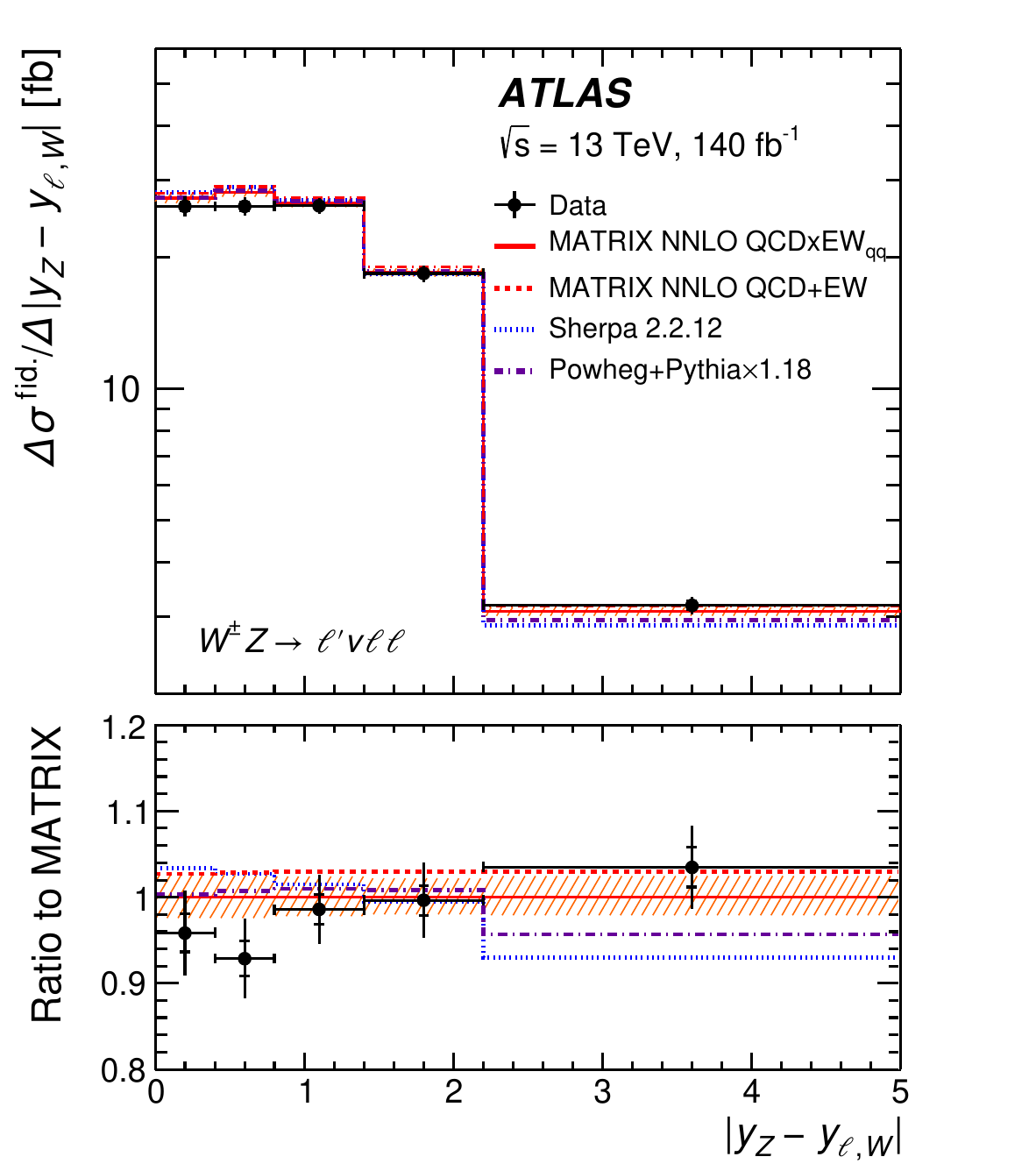}}
\subfloat[]{\includegraphics[width=0.4\textwidth]{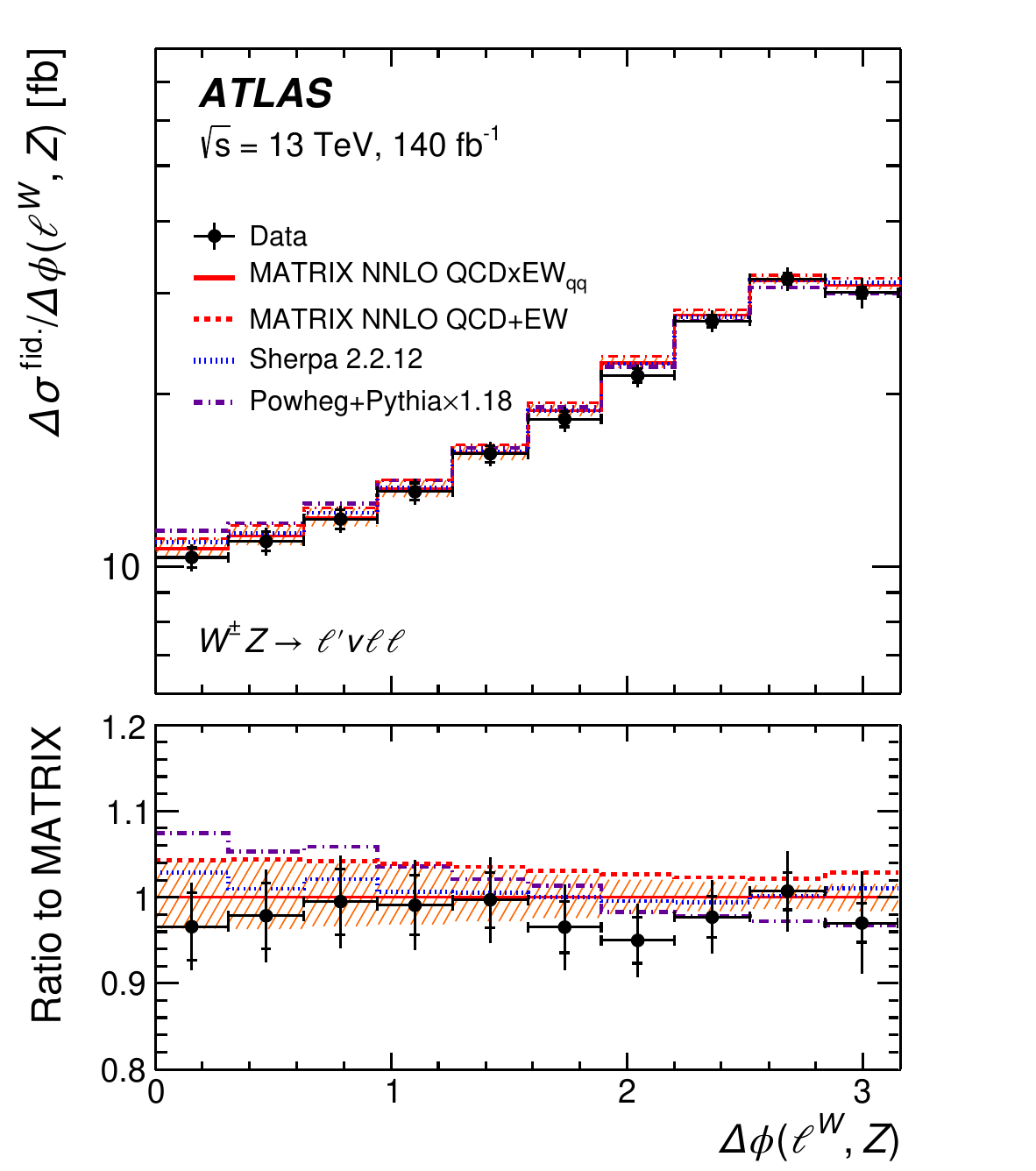}}\\
\subfloat[]{\includegraphics[width=0.4\textwidth]{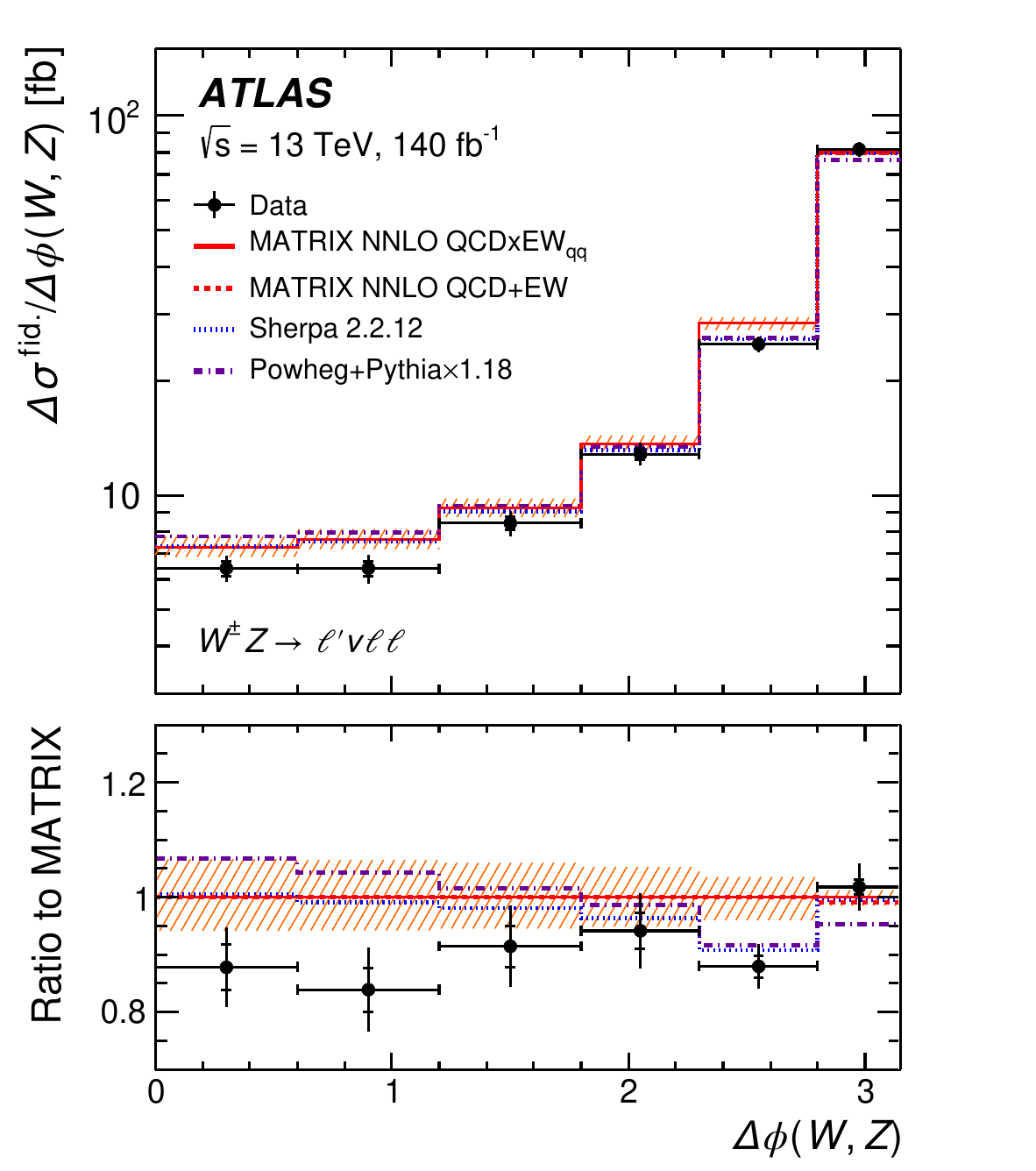}}
\caption{
The measured $W^{\pm}Z$ differential cross-section in the fiducial phase space as a function of (a) $|y_Z - y_{\ell,W}|$, (b) \dPhiWlZ and (c) \dPhiWZ.
The inner and outer error bars on the data points represent the statistical and total uncertainties, respectively.
The measurements are compared with the prediction from \matrix using either the \nnloxew (red line)  or the \nnlopew  (dashed-red) line schemes.
The shaded band shows how the QCD scale uncertainties affect the \matrix predictions.
The predictions from the \sherpa and \powhegpythia MC generators are also indicated by dotted-blue and dotted-dashed violet lines, respectively.
}
\label{fig:DiffXSection:yZlW_all}
\end{figure}

Differential cross-sections are also measured as functions of observables sensitive to possible CP-violation effects and presented in Figure~\ref{fig:DiffXSection:phiStar_and_TP}.
These variables are also further used to look for effects that would arise from CP-odd EFT operators.
The \phistarw (\phistarz) angle is defined following Ref.~\cite{Panico:2017frx} as the azimuthal angle of the charged (negatively charged) lepton resulting from the $W$ ($Z$) boson decay between each boson decay plane and the $x-z$ plane in the centre-of-mass of the diboson system.
The pseudorapidity of the $W$ boson is reconstructed using an estimate of the longitudinal momentum of the neutrino obtained using the neural-network based method developed in Ref.~\cite{ATLAS:2022oge}.
The triple product $p_{\perp}$ of three momenta, proposed in~\cite{Degrande:2021zpv} for CP-violation studies, is defined as:
\begin{equation}\label{eq:tp_1}
p_{\perp}(A,B,C) = \vec{p}_A\cdot \frac{\vec{p}_B \times \vec{p}_C}{|\vec{p}_B \times \vec{p}_C|} \, .
\end{equation}
In all triple products the positively (negatively) charged lepton from the $Z$ decay is denoted $\ell^{+}$ ($\ell^{-}$) while the lepton from the $W$ decay is denoted $\ell^{W}$ regardless of its charge.
The $\sum p^z$ notation represents  the $z$ components of the momenta resulting from the sum over the three final states leptons and the final state neutrino.
Following this definition, two triple product observables are measured differentially: \TPzZWLep and \TPZLeppzZLepmWLep.
Good agreement is observed between the measured cross-sections and the predictions for these four observables, as shown in Figure~\ref{fig:DiffXSection:phiStar_and_TP}.
Any asymmetry observed around $\pi/2$ in the \phistarw (\phistarz) distribution or around zero in the triple products distributions would be a hint for CP-violation effects in the weak-boson self-interactions.

\begin{figure}[t]
\centering
\subfloat[]{\includegraphics[width=0.4\textwidth]{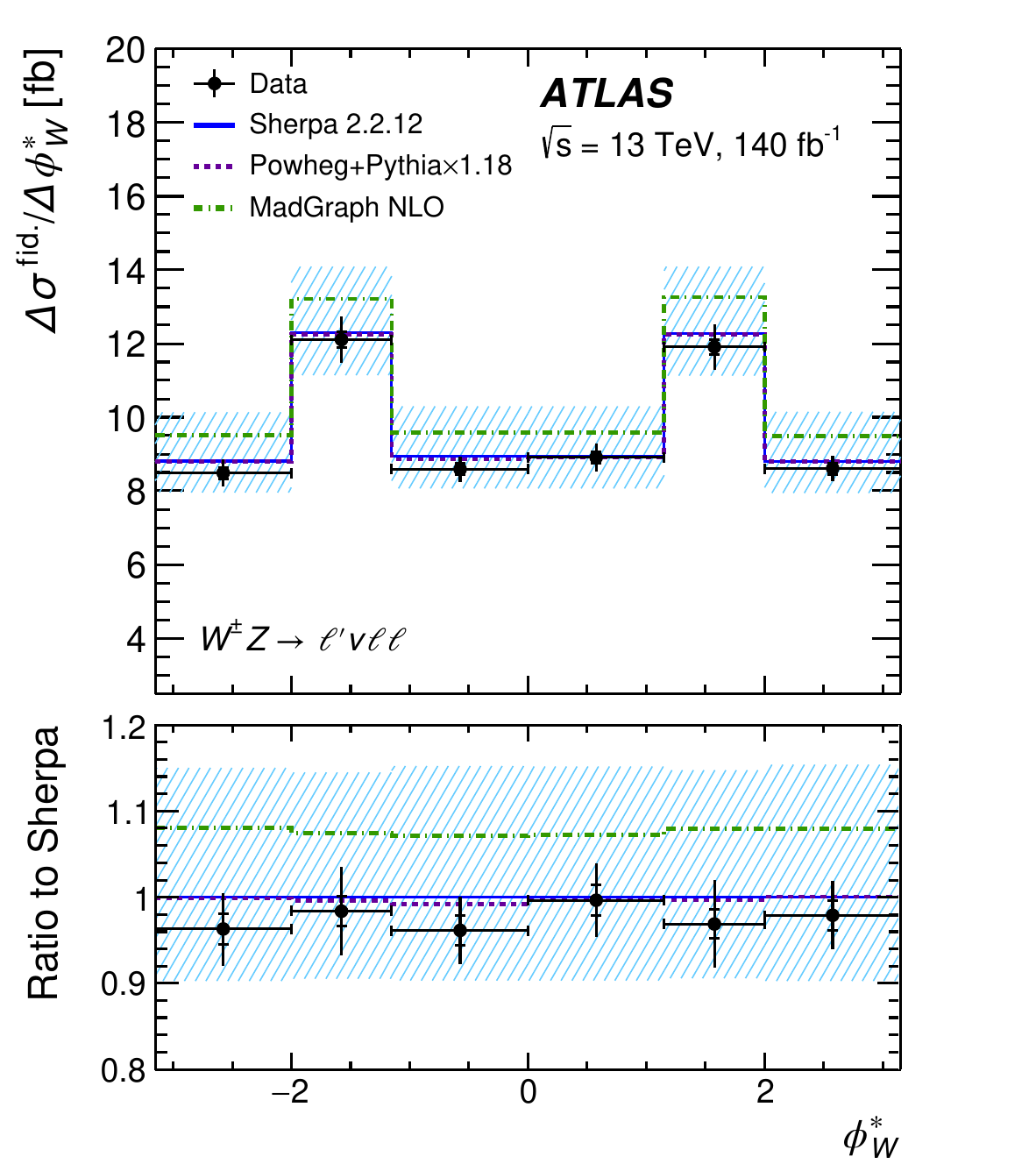}}
\subfloat[]{\includegraphics[width=0.4\textwidth]{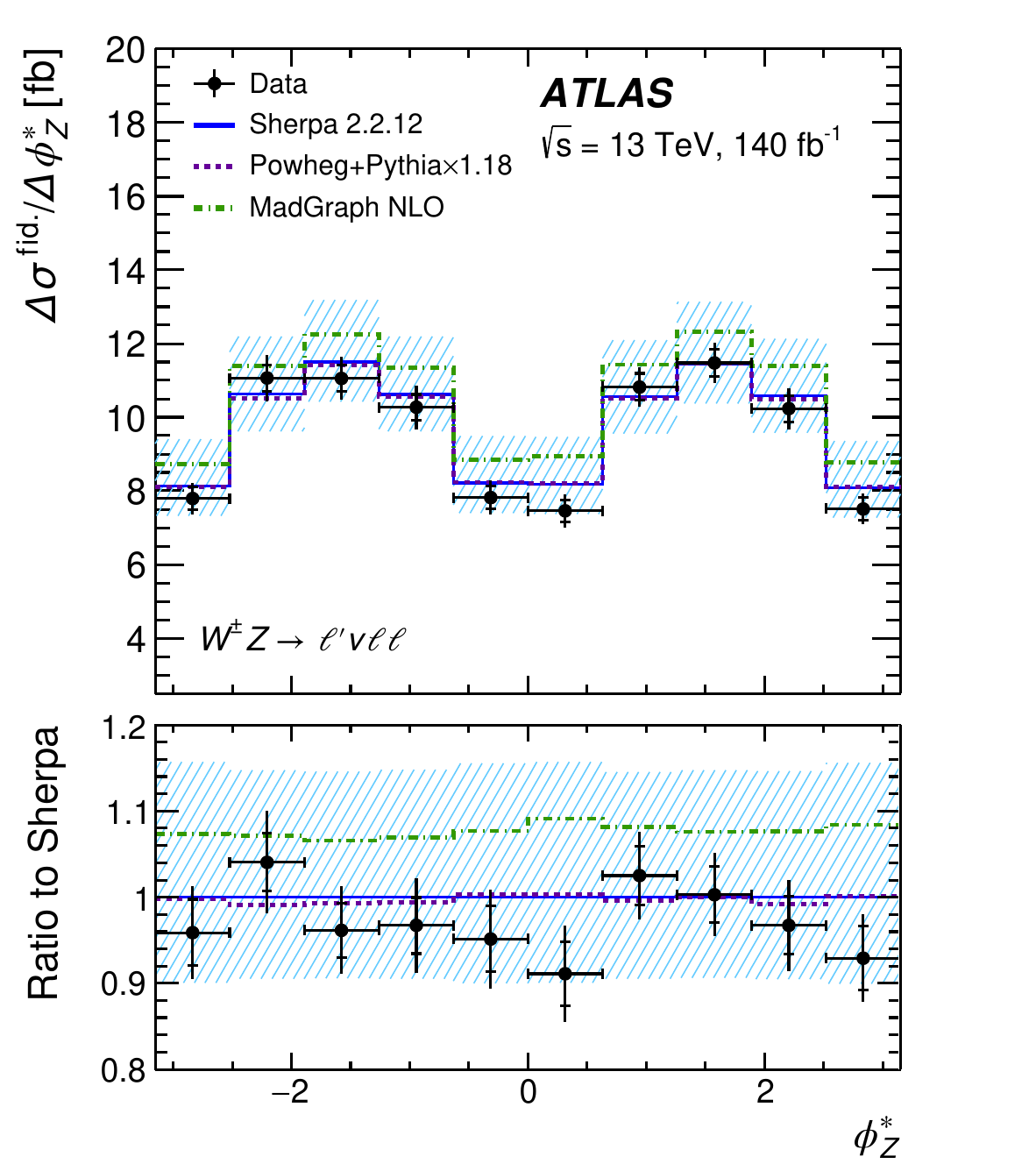}}\\
\subfloat[]{\includegraphics[width=0.4\textwidth]{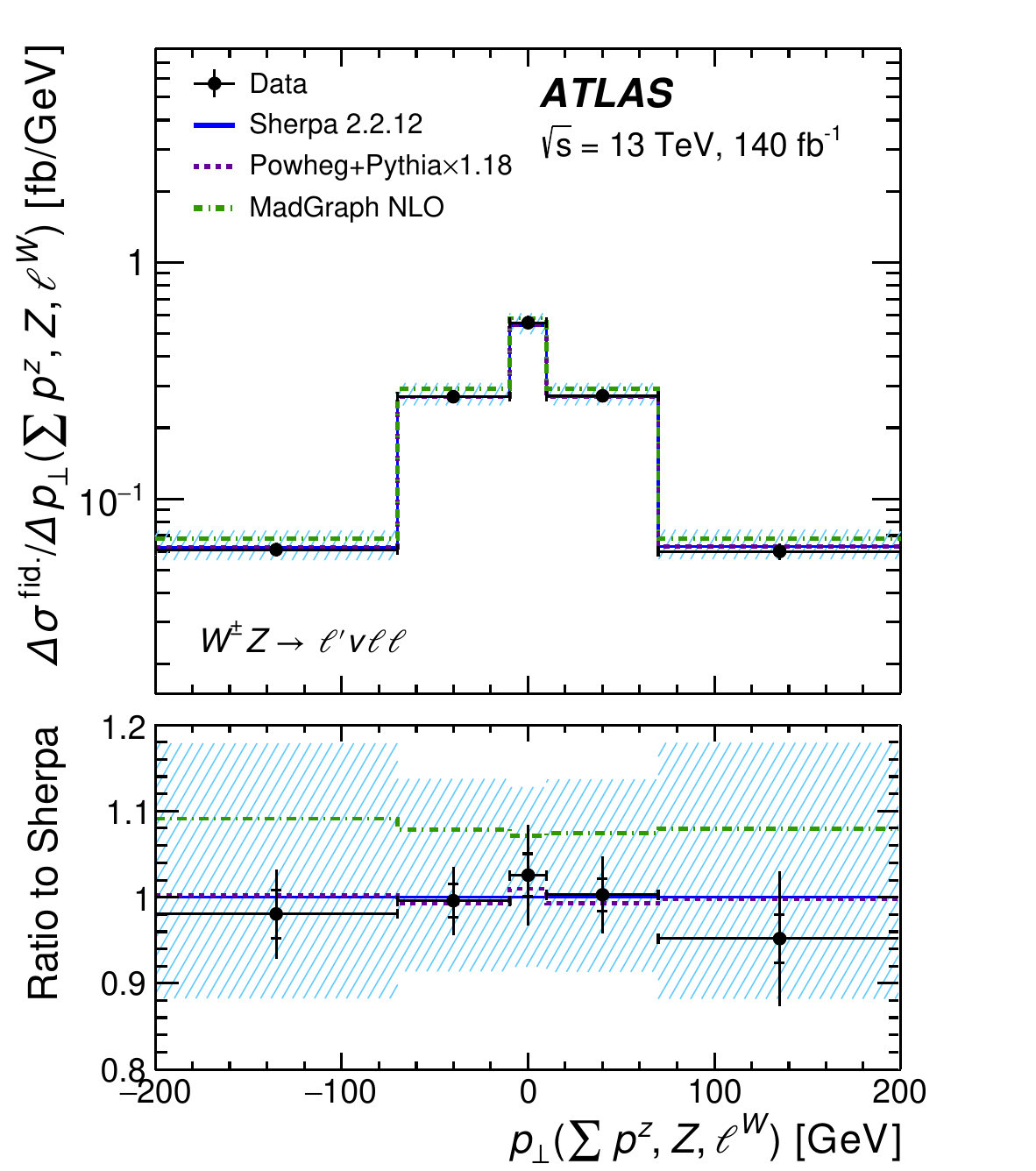}}
\subfloat[]{\includegraphics[width=0.4\textwidth]{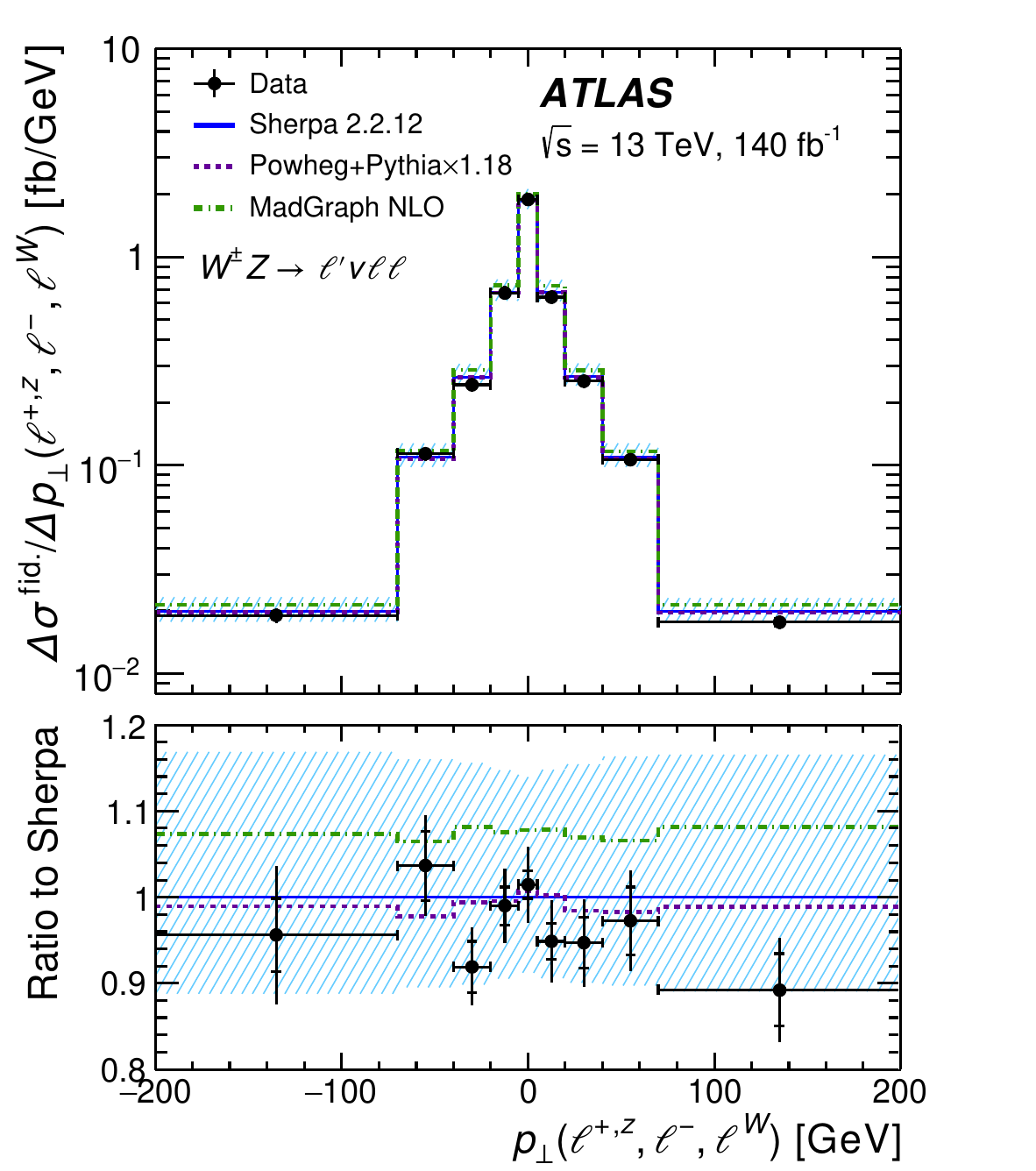}}
\caption{
The measured $W^{\pm}Z$ differential cross-section in the fiducial phase space as a function of (a) \phistarw, (b) \phistarz, (c) \TPzZWLep and (d) \TPZLeppzZLepmWLep.
The inner and outer error bars on the data points represent the statistical and total uncertainties, respectively.
The measurements are compared with the predictions from \sherpa (blue line), \powhegpythia (dashed-violet line)  and \MGPy8 (dotted-dashed green line).
The shaded band shows how the QCD scale uncertainties affect the \sherpa prediction.
}
\label{fig:DiffXSection:phiStar_and_TP}
\end{figure}

The exclusive multiplicity of jets above a \pt threshold of $25$~\GeV\ unfolded at particle level is presented in Figure~\ref{fig:unfoldResultNjetMjj}(a).
The measurements are compared with predictions from \SHERPA, \MGPy8 and \powhegpythia.
The \SHERPA prediction provide a better description of the ratio of $0$-jet to $1$-jet event cross-sections than \MGPy8  or \powhegpythia.
The \MGPy8 prediction, which models up to two partons at NLO, tends to overestimate the cross-section of events with two or three jets, while \SHERPA, which models only up to one parton at NLO, provides a slightly better description of data in these bins.
Finally, the measured $W^{\pm}Z$ differential cross-section as a function of the invariant mass, $m_{jj}$, of the two leading  jets with $\pT > 25$~\GeV\ is presented in Figure~\ref{fig:unfoldResultNjetMjj}(b).
All three MC predictions have difficulties in describing the data.
A similar observation was made in the production of \wzjj events in Ref.~\cite{ATLAS:2024ini}.

\begin{figure}[t]
\centering
\subfloat[]{\includegraphics[width=0.4\textwidth]{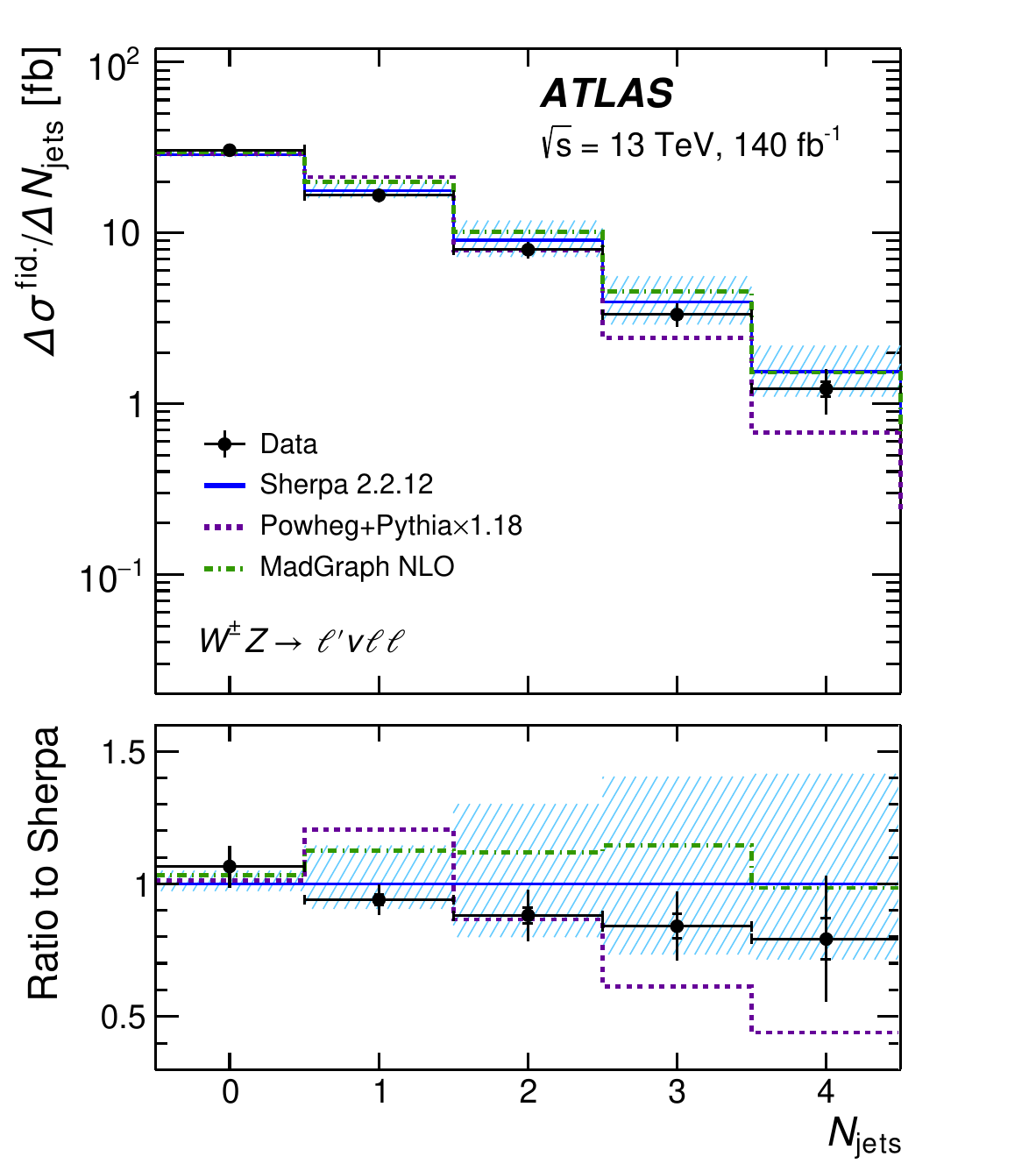}}
\subfloat[]{\includegraphics[width=0.4\textwidth]{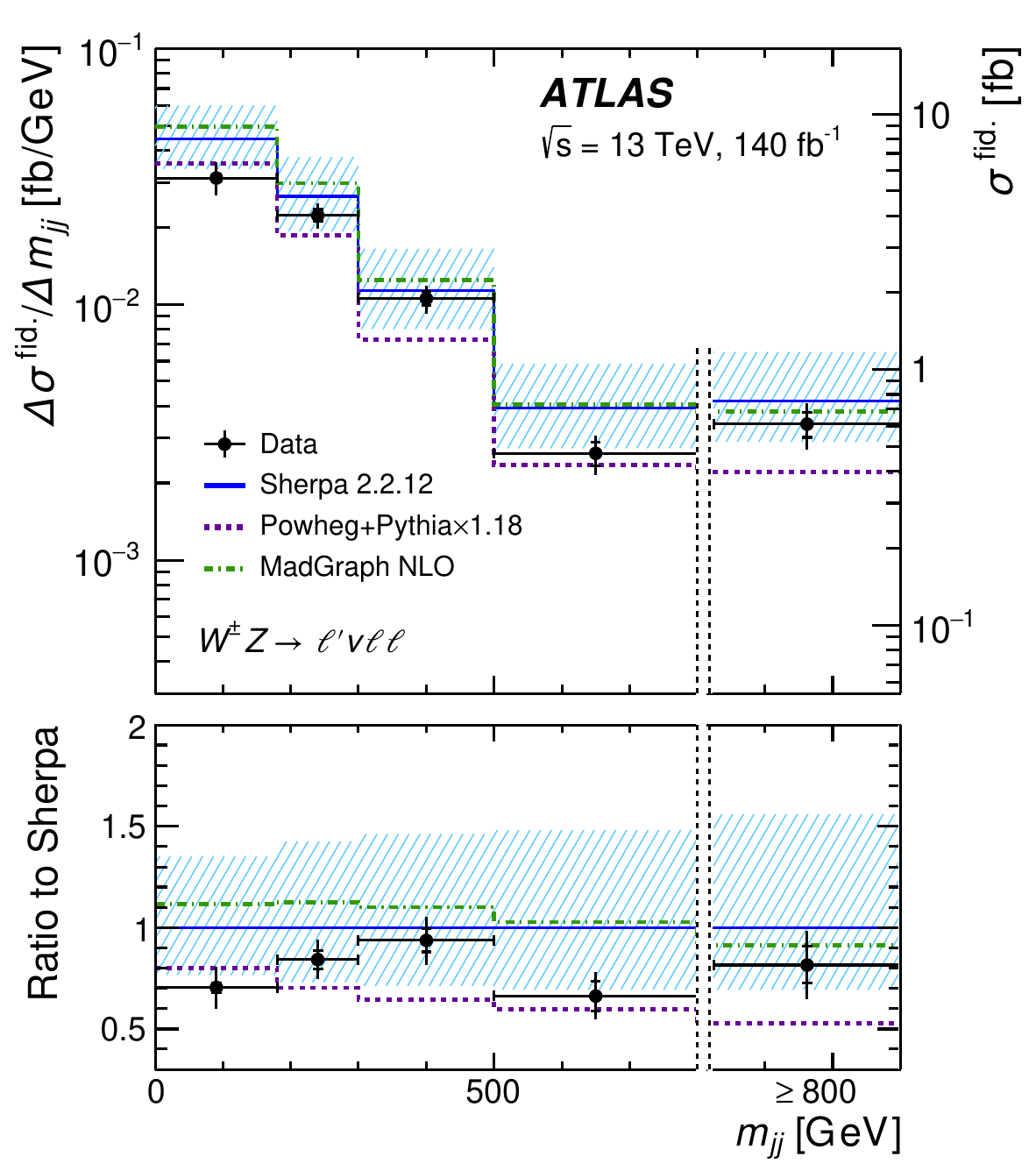}}
\caption{The measured $W^{\pm}Z$ differential cross-section in the fiducial phase space as a function of (a) the exclusive multiplicity of jets with $p_\textrm{T} > 25$~\GeV\  and of (b) the invariant mass of the two leading jets with $p_\textrm{T} > 25$~\GeV.
The inner and outer error bars on the data points represent the statistical and total uncertainties, respectively.
The measurements are compared with the predictions from \sherpa (blue line), \powhegpythia (dashed-violet line)  and \MGPy8 (dotted-dashed green line).
The shaded band shows how the QCD scale uncertainties affect the \sherpa prediction.
In (b) the right vertical axis refers to the last cross-section point, separated from the others by vertical dashed lines,
as this last bin is integrated up to the maximum value reached in the phase space and the cross-section is not divided by the bin width.
}
\label{fig:unfoldResultNjetMjj}
\end{figure}


\clearpage
\section{Constraints on anomalous interactions}
\label{sec:EFT}

\subsection{SMEFT parameterisation}\label{sec:EFTparam}


The measured detector-level distributions are used to constrain extensions to the SM that produce anomalous interactions.
The anomalous interactions are introduced using an effective field theory (EFT) for which the Lagrangian is given by:

\begin{equation} \label{eq:EFT_Lagrangian}
\mathcal{L}=\mathcal{L}_{\mathrm{SM}}+\mathcal{L}_{\mathrm{EFT}}=\mathcal{L}_{\mathrm{SM}}+\sum_i \frac{c_i^{(6)}}{\Lambda_i^2} \mathcal{O}_i^{(6)}+\sum_j \frac{c_j^{(8)}}{\Lambda_j^4} \mathcal{O}_j^{(8)}+\ldots\,,
\end{equation}

where $\mathcal{L}_{\mathrm{SM}}$  contains the SM Lagrangian, extended with the addition of an effective Lagrangian $\mathcal{L}_{\mathrm{EFT}}$. To comply with lepton and baryon number conservation, only operators of even dimensions are considered to construct $\mathcal{L}_{\mathrm{EFT}}$.
The terms $\mathcal{O}_{i}^{(6)}$ and $\mathcal{O}_{j}^{(8)}$ are dimension-6 and dimension-8 operators, respectively, involving only SM fields. The corresponding Wilson coefficients are denoted by ${c_i^{(6)}}/{\Lambda_i^2}$ and ${c_j^{(8)}}/{\Lambda_j^4}$.
The analysis  focuses on dimension-6 operators in the Warsaw basis~\cite{Grzadkowski:2010es}, since anomalous effects from operators of higher dimensionality are more strongly suppressed by inverse powers of the new physics scale $\Lambda$.

Theoretical predictions are constructed  using the Lagrangian in Eq.~\eqref{eq:EFT_Lagrangian}.
The amplitude for the \wz process is split into an SM part and a dimension-6 part which contains the anomalous interactions.
The  squared amplitude which enters in the cross-section is:
\begin{equation}\label{eq:matrix_element}
\begin{aligned}
|\mathcal{M}|^2 &= |\mathcal{M}_{\mathrm{SM}}|^2 + \sum_i \frac{c_i}{\Lambda_i^2} 2\Re \left\{ \mathcal{M}_{\mathrm{SM}}^* \mathcal{M}_{i,6}\right\} + \sum_i \frac{c_i^2}{\Lambda_i^4} |\mathcal{M}_{i,6}|^2 + \sum_{i\neq j} \frac{c_i c_j}{\Lambda_i^2 \Lambda_j^2}2\Re \left\{\mathcal{M}_{i,6}^* \mathcal{M}_{j,6}\right\}\,.
\end{aligned}
\end{equation}
The first term in Eq.~\eqref{eq:matrix_element} is the pure SM contribution.
The second term, called the \emph{linear term}, contains interferences between the SM and each considered SMEFT operator $i$, and is of order $\Lambda^{-2}$.
The third term, of order $\Lambda^{-4}$, stands for the pure contributions of all single dimension-6 operators and is called the \emph{quadratic term}.
Finally, the fourth term represents interference terms between two different dimension-6 operators and are referenced as the \emph{cross-term}.

Constraints are placed on the Wilson coefficients of three CP-conserving bosonic operators, \oW, \oHWB, \oHD and of two CP-violating bosonic operators, \oWtilde and \oHWBtilde, as well as on the Wilson coefficients of operators, denoted \oHjone, \oHjthree,  describing fermion-boson interactions.
The \cW/$\Lambda^2$ Wilson coefficient can only be measured in processes affected by modifications of the gauge-boson self couplings.
Diboson measurements are therefore of particular interest to constrain it.
Its effect increases rapidly with the centre-of-mass energy.
However,  in this regime, the SM amplitude and the anomalous amplitude are dominated by different helicity configurations, such that their interference is suppressed, reducing the impact of the operator on the measured cross-section~\cite{Azatov:2016sqh,Panico:2017frx,Azatov:2017kzw,Falkowski:2016cxu}.
For CP-odd operators the interference with SM is expected to give no contribution to CP-even observables like \mtwz.
The reduced sensitivity  to the interference poses a problem for the truncation of the series at dimension-6 operators because contributions that depend on  $\Lambda^{-4}$ which are introduced by the interference between SM and dimension-8 operators, which are expected to be subdominant in the EFT expansion, cannot be safely neglected.
Specific angular observables can be used to revive the interference~\cite{Panico:2017frx, Degrande:2021zpv} and are exploited for CP-odd operators.
For a more robust interpretation of the measurements, limits are therefore determined using parameterisations that exclude or include the pure dimension-6 quadratic contributions, and are studied as a function of an upper energy cut-off on the SMEFT contributions~\cite{Brivio:2022pyi}.

The pure SM contribution to the \wz differential cross-sections in Eq.~\eqref{eq:matrix_element} is taken to be the prediction from \textsc{Sherpa\,2.2.12} described in Section~\ref{sec:samples}.
The contributions arising from the interference and pure dimension-6 terms are generated at leading-order (LO) in perturbative QCD using \MGNLO[2.6.5], with the interactions from dimension-6 operators provided by the \textsc{SMEFTSim 3.0} package~\cite{Brivio:2017btx, Brivio:2020onw}.
A dynamic QCD scale is chosen, equal to the averaged sum of the two gauge-boson transverse masses and the \texttt{mW} EW scheme is used.
The \texttt{top} flavour scheme is chosen, implying that quarks from the third generation are described by independent fields with respect to the lighter quarks. The CKM matrix is considered diagonal and the only fermion assumed to be massive is the top quark.
To account for the real part of NLO QCD corrections, these events are merged with events generated at \LO in QCD using \MGNLO[2.6.5], containing three leptons, one neutrino and one jet in the final state.
The merging procedure is done at the PS level using \PYTHIA[8.240] in the CKKW-L scheme~\cite{Lonnblad:2001iq}.
\PYTHIA[8.240] is also used for the simulation of the hadronisation and underlying event.
The \NNPDF[3.0nlo] PDF set is used for the parton process generation.
These MC sets are referred to as \mg01j samples in the following.
They were passed through the ATLAS detector simulation and event reconstruction to be used in the determination of limits based on detector-level data.

For CP-even operators, the modelling of these samples is validated at particle-level against the predictions from \SMEFTatNLO~\cite{Degrande:2020evl} interfaced with \MGNLO[2.9.9] which includes calculations up to next-to-leading-order in QCD for both SM and SMEFT contributions.
The \SMEFTatNLO simulation was used to verify that the difference between \NLO and pure \LO prediction, or between \NLO and  0,1j@LO predictions, is not the same for the pure SM term and for the sum of the linear and quadratic EFT terms.
The QCD corrections affecting the pure SM term are found to be larger than those related to the linear or quadratic terms.
Therefore, any QCD correction derived from the pure SM simulation cannot be applied to the EFT contributions, in order to define a pseudo-NLO EFT MC simulation.
Half of the difference between SMEFT \madgraph MC simulations at \LO and at 0,1j@LO accuracy is used to define a modelling uncertainty affecting MC simulations of the linear and quadratic terms, accounting for missing higher-order QCD corrections.
Using the \SMEFTatNLO simulation, it is verified that this modelling uncertainty effectively covers the differences in shape and normalisation between the \NLO and 0,1j@LO MC simulations for the sum of the linear and quadratic terms.

For CP-odd operators, such a validation procedure with \SMEFTatNLO was not possible due to the lack of EFT models including CP-odd operators in  \SMEFTatNLO at that time.
The model uncertainty affecting the simulation of CP-odd operators is thus defined in the same way as explained for CP-even operators.


\subsection{Limits on CP-conserving operators}\label{sec:EFT_CPeven}

The effects of CP-conserving EFT operators are primarily expected to manifest themselves at high energies.
A large range of kinematic observables was tested to enhance the sensitivity to CP-conserving EFT contributions in inclusive \wz production, encompassing variables related to the energy of the \wz system, angular distributions of the final-state leptons, polarisation-related observables, as well as various multivariate combinations of them.
Among all options investigated, the \mtwz observable is found to provide the best sensitivity to effects of CP-even EFT operators.
Effects arising from higher-order QCD or EW corrections are also minimised in differential distributions of this observable.
The distribution of observed events as a function of \mtwz, as represented in Figure~\ref{fig:CPeven:fit_control_dist}, is used to define an extended binned likelihood function consisting of a product of Poisson probability terms over bins of the \mtwz distribution~\cite{Cranmer:1456844}.
The bin boundaries are optimised to obtain the best expected limits.
Experimental and theory uncertainties, as discussed in Section~\ref{sec:systematics}, are included as Gaussian-constrained nuisance parameters.
Confidence intervals for a given $c_i$ are determined using a profile-likelihood-ratio test statistic~\cite{Feldman:1997qc}, which is assumed to be $\chi^2$-distributed according to Wilks’ theorem~\cite{Wilks:1938dza}.
The validity of the $\chi^2$-approximation is confirmed using pseudo-experiments.
For each individual $c_i$, or pair $(c_i, c_j)$ for two-dimensional limits, a separate fit is performed, setting other coefficients to zero.
Possible effects of the probed SMEFT operator on the background predictions are not considered.
While detector-level distributions are expected to provide the highest sensitivity to BSM effects, the unfolded differential distribution of \mtwz was also used to derive limits.
For all operators better or comparable limits are obtained using the detector-level distribution.

\begin{figure}[tb]
\centering
\includegraphics[width=.45\textwidth]{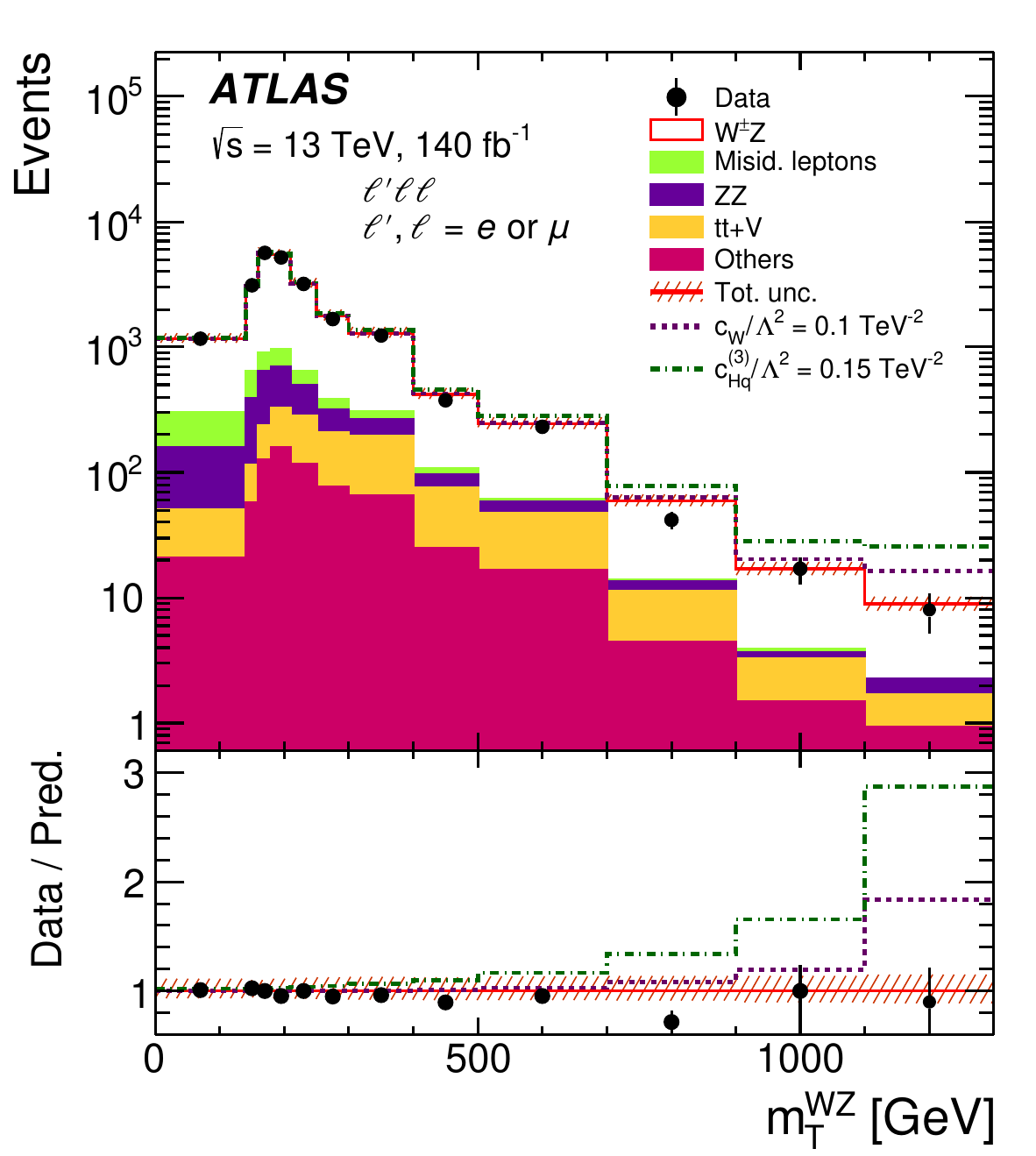}
\caption{Detector-level distribution of \mtwz used to obtain limits on all CP-even EFT Wilson coefficients.
The points correspond to the data with the error bars representing the statistical uncertainties, and the histograms correspond to the predictions of the various SM processes.
The sum of the background processes with misidentified leptons is labelled ``Misid. leptons''.
The \sherpa MC prediction is used for the \wz\ signal contribution.
The open red histogram shows the total prediction; the shaded band is the total uncertainty of this prediction.
The lower panel shows the ratio of the data points to the open red histogram with their respective uncertainties.
For illustration purposes, the predicted contributions of the \oW and \oHqthree operators with values of the corresponding Wilson coefficients of 0.1~\TeV$^{-2}$ and 0.15~\TeV$^{-2}$, respectively, are also represented.
The parameterisation including the pure dimension-6 contribution is used.
}
\label{fig:CPeven:fit_control_dist}
\end{figure}

Expected and observed 95\% confidence level (CL) intervals on CP-conserving Wilson coefficients are summarised in Table~\ref{tab:CPevent:expected_limits}.
Limits are derived considering either only linear contributions, arising from the interference between SM and
BSM amplitudes, or also including quadratic terms, i.e. including effects from pure BSM amplitudes.
Considering cross-terms between SMEFT operators, two-dimensional limits at $95$\% CL on two pairs of operators, (\oW,\oHqthree) and  (\oW,\oHWB), are also obtained and presented in Figure~\ref{fig:CPeven:2Dlimits}.

\begin{table}[tb]
\caption{Expected and observed 95\% CL intervals on CP-even Wilson coefficients.
Limits are presented using parameterisations excluding (``lin.'') and including (``lin.+quad.'') the pure dimension-6 contributions, respectively.
Only one Wilson coefficient is permitted to vary from zero at a time.
}
\label{tab:CPevent:expected_limits}
\centering
\begin{tabular}{lcccc}
\toprule
& \multicolumn{2}{c}{ Expected [\TeV$^{-2}$] } & \multicolumn{2}{c}{ Observed [\TeV$^{-2}$] } \\
& 95\% CL (lin.) & 95\% CL (lin.+quad.) & 95\% CL (lin.) & 95\% CL (lin.+quad.) \\
\midrule
\cW/$\Lambda^2$          & [-0.668, 0.733]  & [-0.103, 0.095] 	      	   & [-1.150, 0.181] &  [-0.093, 0.079] \\
\cHWB/$\Lambda^2$     &  [-0.326, 0.412]  &  [-0.326, 0.413] 		     & [-0.470, 1.440] &  [-0.458, 1.520] \\
\cHD/$\Lambda^2$         &  [-8.107, 9.096]  &  [-4.494, 3.976] 	  	   & [-18.0, 1.7] &   [-5.840, 1.680] \\
\cHjone/$\Lambda^2$    &  [-1.994, 2.122]   &  [-2.923, 1.615] 				& [-0.367, 3.980] &  [-0.503, 2.550] \\
\cHjthree/$\Lambda^2$  &  [-0.097, 0.106]  &  [-0.145, 0.074] 			 & [-0.146, 0.0275] & [-0.165, 0.015]  \\
\bottomrule
\end{tabular}

\end{table}
\begin{figure}[tb]
\centering
\subfloat[]{\includegraphics[width=.45\textwidth]{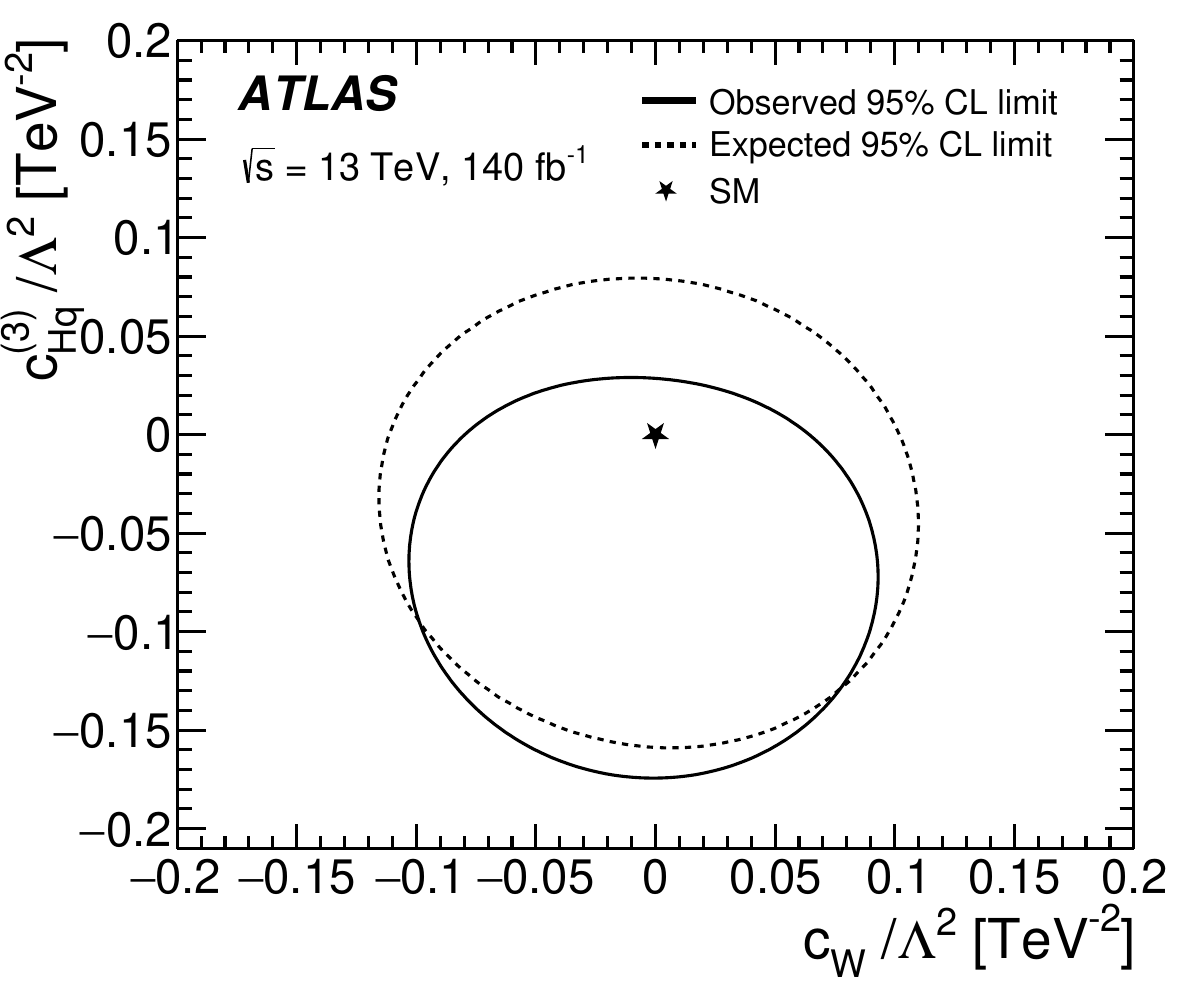}}
\subfloat[]{\includegraphics[width=.45\textwidth]{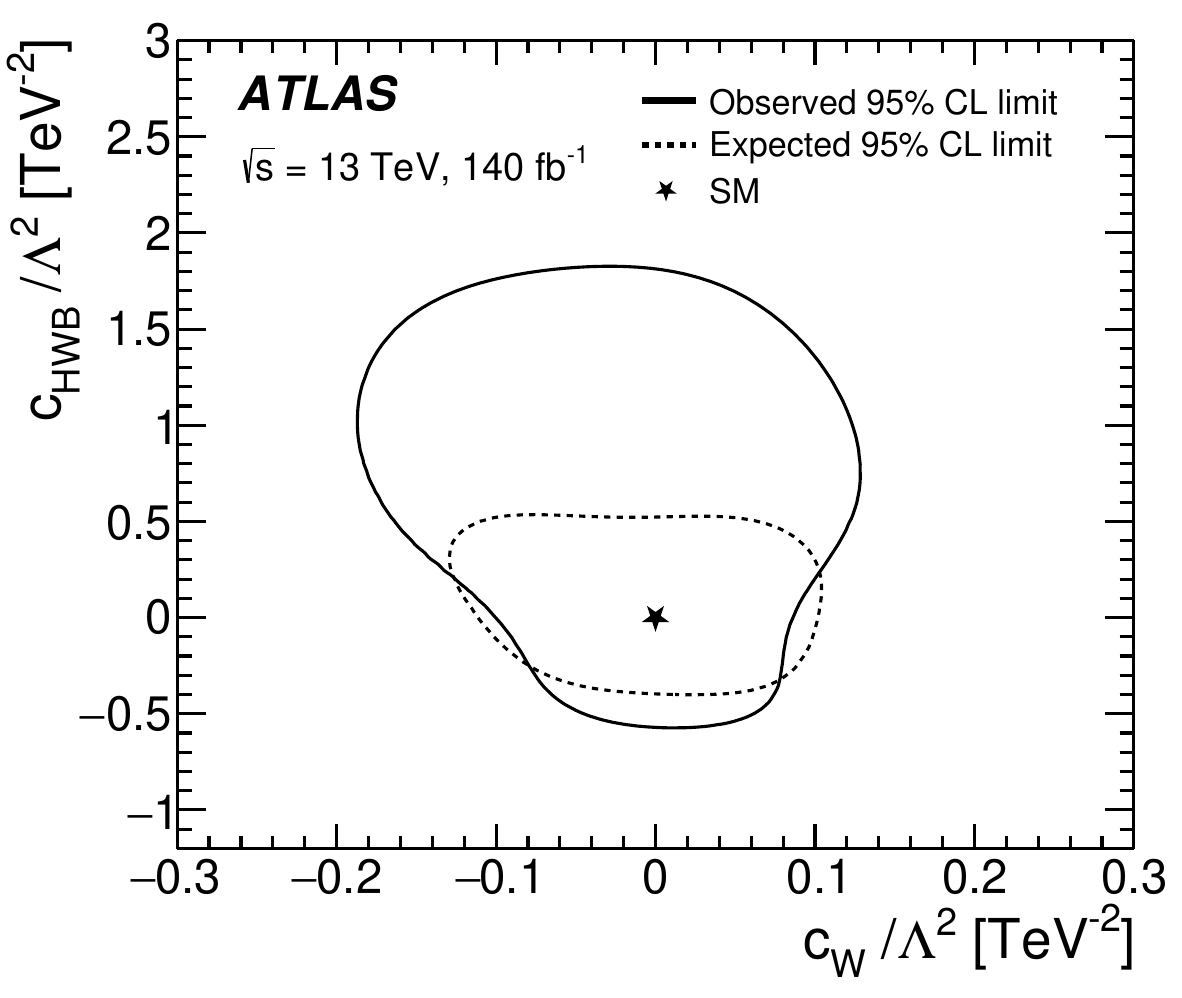}}
\caption{Two-dimensional expected (dashed line) and observed (solid line) 95\% CL intervals on Wilson coefficients corresponding to the pairs of (a) (\oW,\oHqthree) and (b) (\oW,\oHWB) operators.
The parameterisation including the pure dimension-6 contribution is used.}
\label{fig:CPeven:2Dlimits}
\end{figure}
All observed limits are compatible with the $c_i = 0$ assumption at 95\% CL, indicating that no significant deviations from the SM are observed.
Limits on \cHWB/$\Lambda^2$, \cHjthree/$\Lambda^2$, \cHjone/$\Lambda^2$, and, to a lower extent, on \cHD/$\Lambda^2$, are driven by interference effects, as the expected sensitivity is not strongly affected by the inclusion of the quadratic term.
Interference effects manifest at intermediate \mtwz values, between 500~\GeV and 1~\TeV.
The deficit of observed data events in the region from 700 to 900~\GeV induces, therefore, asymmetric observed limits for these Wilson coefficients.
The limit on \cW/$\Lambda^2$ is predominantly determined by the quadratic contributions.
The corresponding confidence intervals are therefore only valid for BSM models where contributions from dimension-8 operators are small with respect to the quadratic effects in dimension six.
The range of validity of the EFT approach must be assessed, since arbitrary growth of dimension-6 contributions at high energies leads to nonphysical results and to violations of unitarity in the underlying quantum field theory.
Following recommendations from the LHC EFT Working Group~\cite{Brivio:2022pyi}, confidence intervals are also derived as a function of an upper energy cut-off on the SMEFT contributions.
The diboson invariant mass \mwz, which represents to a good approximation the total energy of the interaction per event, is used to remove the SMEFT events at particle-level predicted above a certain threshold.
Confidence intervals were derived for cut-off values of 1~\TeV, 1.5~\TeV and above.
Only confidence intervals on  \cW/$\Lambda^2$ and \cHjthree/$\Lambda^2$ show a significant dependence on the cut-off value.
The extracted limits on these two Wilson coefficients are presented in Table~\ref{tab:CPevent:clipping}.
They are found to have only a mild dependence on the applied cut-off  after a value of 1.5~\TeV, degrading when lower cut-off values are applied.
This result provides reassurance that the nominal limits, obtained without any cut-off procedure, do not suffer from issues related to the region of validity of the EFT model.

\begin{table}[tb]
\caption{Evolution of the expected  95\% CL intervals on CP-even \cW/$\Lambda^2$ and \cHjthree/$\Lambda^2$ Wilson coefficients as a function of an upper energy cut-off on \mwz in the SMEFT contribution. Limits are presented considering both the linear and quadratic contributions. Only one Wilson coefficient is permitted to vary from zero at a time.}
\label{tab:CPevent:clipping}
\centering
\begin{tabular}{lccc}
\toprule
& \multicolumn{3}{c}{ Cut-off value on \mwz[\GeV]} \\
\midrule
& 1000 & 1500 & no cut-off\\
\midrule
\cW/$\Lambda^2$~ [\TeV$^{-2}$]   & [-0.236, 0.208]      & [-0.143, 0.130]  & [-0.103, 0.095] 	     \\
\cHjthree/$\Lambda^2$  [\TeV$^{-2}$]  &  [-0.286, 0.119]      &  [-0.195, 0.086]  &  [-0.145, 0.074] 	 \\
\bottomrule
\end{tabular}

\end{table}

The sensitivity to \cHjthree/$\Lambda^2$ and \cHjone/$\Lambda^2$ is nearly one order of magnitude worse than that achieved in direct Higgs boson production measurements~\cite{ATLAS:2020jwz,ATLAS:2020fcp,ATLAS:2022tnm} or in global fits combining Higgs, top-quark, and electroweak boson production~\cite{Celada:2024mcf,CMS:2025ugn}, where electroweak precision data is included~\cite{ALEPH:2005ab}.
A similar sensitivity as seen in $W^+W^-$ production~\cite{ATLAS:2025dhf} is obtained for these Wilson coefficients.
The limits extracted on the anomalous triple-gauge-coupling Wilson coefficient \cW surpass in sensitivity those obtained in global fits~\cite{Celada:2024mcf,CMS:2025ugn}, $H\to\gamma\gamma$ production~\cite{ATLAS:2022tnm}, and in other diboson processes such as \ww~\cite{ATLAS:2025dhf} by roughly a factor of two, showing a similar level of sensitivity than measurements performed by the CMS Collaboration in $W\gamma$~\cite{CMS:2021cxr} and \wz production~\cite{CMS:2021icx}.

%
%
%
%
%
%
%
%
%

%
%
%

%
%
%

%
%
%
%
%


\clearpage
\subsection{Limits on CP-violating operators}\label{sec:EFT_CPodd}

For CP-odd operators, the interference between the SM amplitude and the dimension-6 amplitude is also CP-odd.
Therefore, interference effects cancel out entirely for CP-even observables, such as inclusive cross-sections or transverse momentum distributions.
Interference effects however induce asymmetries in appropriately constructed CP-odd observables~\cite{Degrande:2021zpv}.
This asymmetry is a powerful tool to constrain the Wilson coefficients via a fit.
It can be enhanced using multivariate techniques to construct a CP-odd observable resulting from the combination of different high-level variables.
\begin{table}[tb]
\caption{Input variables used for the three BDTs trained to separate effects of CP-odd EFT operators from the SM.
The variables \mtw, \mtwz, \ayzlw, \dPhiWlZ, \phisw, \phisz and the triple products $p_{\perp}$ from three momenta  are described in sections~\ref{sec:RecoResults} and~\ref{sec:DiffCrossSections}.
The variables \ctw and \ctz are the cosines of the decay angle of the charged lepton in the $W$ or $Z$ rest frame relative to the $W$ or $Z$ direction in the \wz centre-of-mass frame~\cite{ATLAS:2022oge}.
The variable \ctv is the cosine of the scattering angle of the $W$ or $Z$  boson in the $WZ$ centre-of-mass frame, calculated relative to the beam axis~\cite{ATLAS:2022oge}.
The variable  $\coschi$ is the cosine of the angle between the decay planes of both bosons taken in the $WZ$ rest frame.
The variable \dPhillss is the difference in azimuthal angle between the two decay leptons of same sign from the $W$ and $Z$ bosons.
}
\label{tab:CPodd:BDTvariables}
\centering
\begin{tabular}{lcc}
\toprule
Variable & \Sp or \Sn & \Szero \\
\midrule
$\ptvtwo/\ptvone$ & \checkmark &  \\
$\mtwz$ & \checkmark &  \checkmark \\
$\mtw$ &  & \checkmark \\
$\ptwz$ & \checkmark &  \\
$\etmiss$ & \checkmark &  \\
$\TPzZWLep$ & \checkmark & \checkmark \\
$\TPWZLeppZWLep$ & \checkmark & \checkmark \\
$\TPWZLepmWLep$ & \checkmark & \checkmark \\
$\TPZLepmzWLepZLepp$ &  & \checkmark \\
$\TPZLeppzZLepmWLep$ &  & \checkmark \\
$\phisz$ & \checkmark & \checkmark \\
$\phisw$ &  & \checkmark \\
$\ctv$ & \checkmark & \checkmark \\
$\ctw$ & \checkmark & \checkmark \\
$\ctz$ &  & \checkmark \\
$\coschi$ &  & \checkmark \\
$\dPhillss$ & \checkmark &  \\
$\dPhiWlZ$ &  & \checkmark \\
$\ayzlw$ & \checkmark & \checkmark \\
\bottomrule
\end{tabular}
\end{table}

The interference between SM and CP-violating EFT operators integrated over the total phase space brings a zero contribution to the total cross-section of the process.
This means that the interference term can be modelled either as a positive or a negative contribution to the differential cross-section, depending on the region of the phase space.
This is translated in the MC as events generated with a negative weight.
Exploiting this sign difference of MC event weights, a multivariate approach is developed that combines angular and energy-dependent observables to construct an optimised CP-sensitive observable.
Following ideas from Refs.~\cite{Gritsan:2020pib,Bhardwaj:2021ujv,Hall:2022bme,}, MC event samples of the interference between the \oHWtilB or \oWtil EFT operators and the SM are used to train three different boosted decision trees (BDT), as implemented in the TMVA package~\cite{Hocker:2007ht}.
Events from the EFT MC simulation are separated into two sets of positively and negatively-weighted events.
A first BDT is trained to separate events generated with a positive weight from events generated with a negative weight.
It combines single angular observables sensitive to CP-violation  and outputs a score, called \Szero, which enhances the asymmetry specific to CP-violation already present in each input observable.
Two other BDTs are trained to discriminate EFT events from pure SMs \wz production.
One BDT, with an output score \Sp, separates positively-weighted EFT MC events from SM events, while the second BDT, with an output score \Sn, is used for negatively-weighted EFT events.
Input variables used in the training of each of the three BDTs are detailed in Table~\ref{tab:CPodd:BDTvariables}.
The same input variables are used for  \Sp and \Sn BDTs.
Detector-level distributions of \Szero for SM \wz events and for interference contributions of the \oHWtilB and \oWtil EFT operators are compared in Figure~\ref{fig:CPodd:S0_vs_Scomb}.
The \Szero distribution of SM \wz events is symmetric, with values peaking around zero, while the distribution of the interference contributions of EFT CP-odd operators is asymmetric around zero.
This behaviour is further enhanced by combining the three BDT scores to form a combined observable \Scomb as

\begin{equation*}
\Scomb = \frac{\Sp + 1}{2} \times \frac{\Sn + 1}{2} \times \Szero \, .
\label{eq:Scomb_definition}
\end{equation*}

As observed in Figure~\ref{fig:CPodd:S0_vs_Scomb}, the asymmetry between SM and EFT interference events is enhanced by a factor of about eight in \Scomb compared with \Szero.
A specific training of the BDTs is used for each of the two different \oHWtilB and \oWtil operators, as the kinematic properties of their interference term distributions are different.

\begin{figure}[tb]
\centering
\subfloat[]{\includegraphics[width=.45\textwidth]{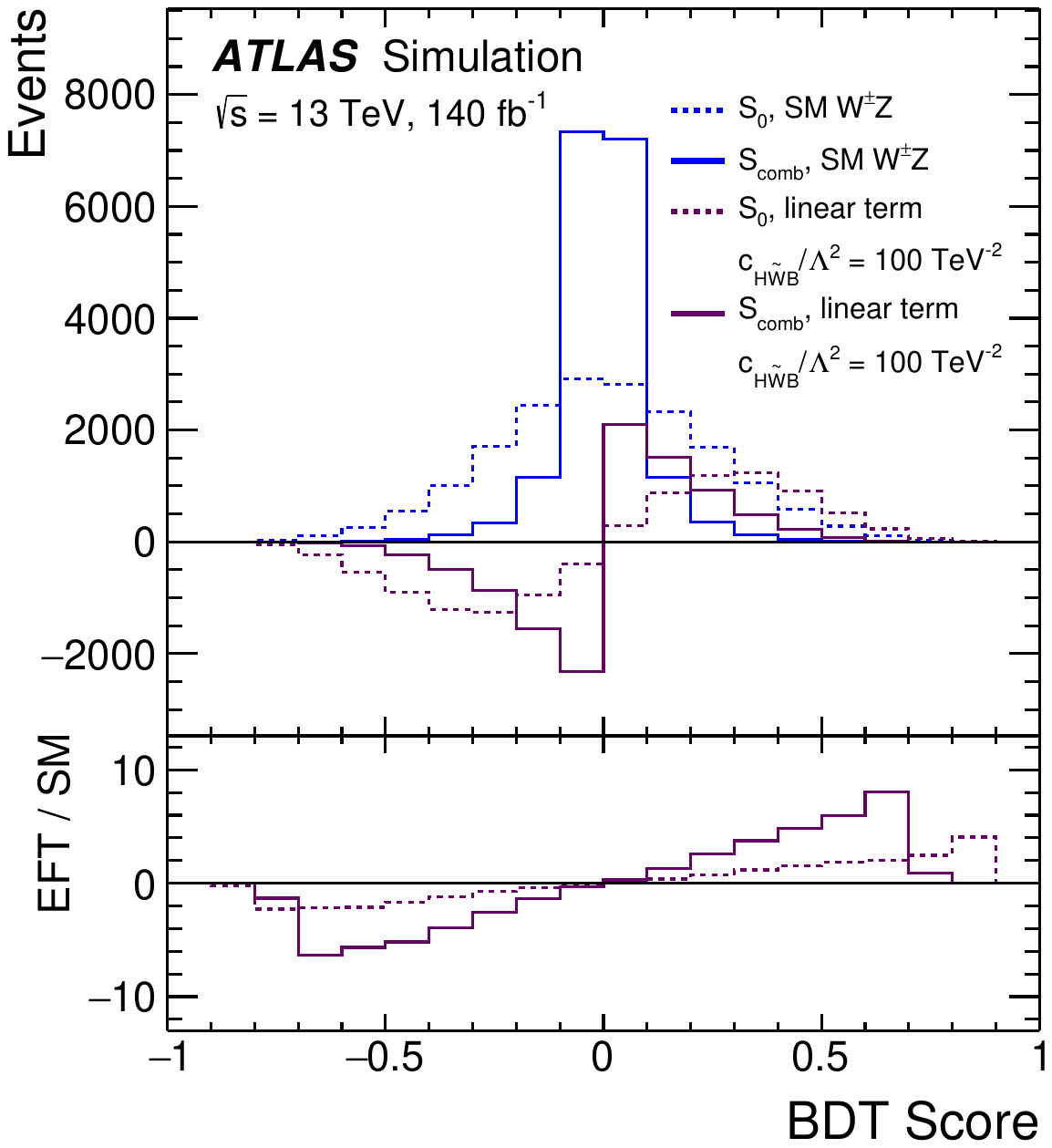}}
\subfloat[]{\includegraphics[width=.45\textwidth]{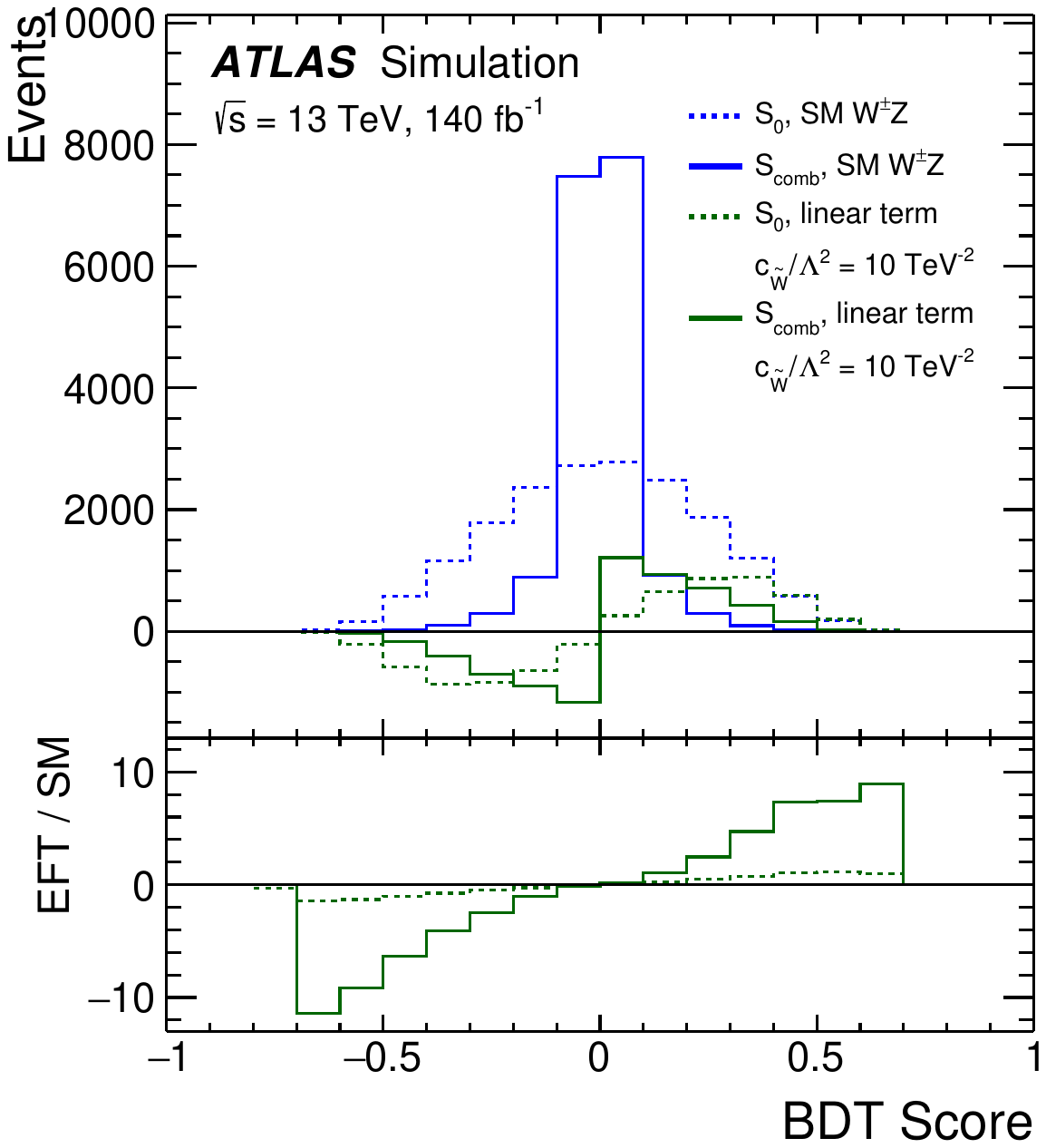}}
\caption{Comparison of the detector-level \Szero (dashed line) and \Scomb (solid line) BDT score distributions for SM and interference events from the (a) \oHWtilB  and (b) \oWtil operators. The lower panels in each figure show the ratio of the EFT contribution to the SM. For better visibility, large values of the Wilson coefficients are used to represent the interference distributions of \oHWtilB and \oWtil operators.}
\label{fig:CPodd:S0_vs_Scomb}
\end{figure}

As non-zero EFT contributions also enhance the \wz production at large diboson invariant masses, further sensitivity to EFT effects is gained by separating events in two different categories, with \mtwz smaller or greater then 600~\GeV.
Detector-level events distributed according to \Scomb and \mtwz are thus used to look for EFT CP-odd effects.

Nine bins in the \Scomb observable ($[-1,-0.4,-0.2,-0.1,-0.02,0.02,0.1,0.2,0.4,1]$) are used.
The bin boundaries are optimised to obtain the best expected limits.
The nine bins in \Scomb for the two categories in \mtwz are arranged in a one-dimensional histogram of 18   statistically independent bins, as represented in Figure~\ref{fig:CPodd:fit_control_dists}.
The distribution is used to define an extended likelihood function.
Similarly to Section~\ref{sec:EFT_CPeven}, a profile-likelihood-ratio test statistic is constructed to estimate the confidence intervals for a given $c_i$.
\begin{figure}[tb]
\centering
\subfloat[]{\includegraphics[width=.45\textwidth]{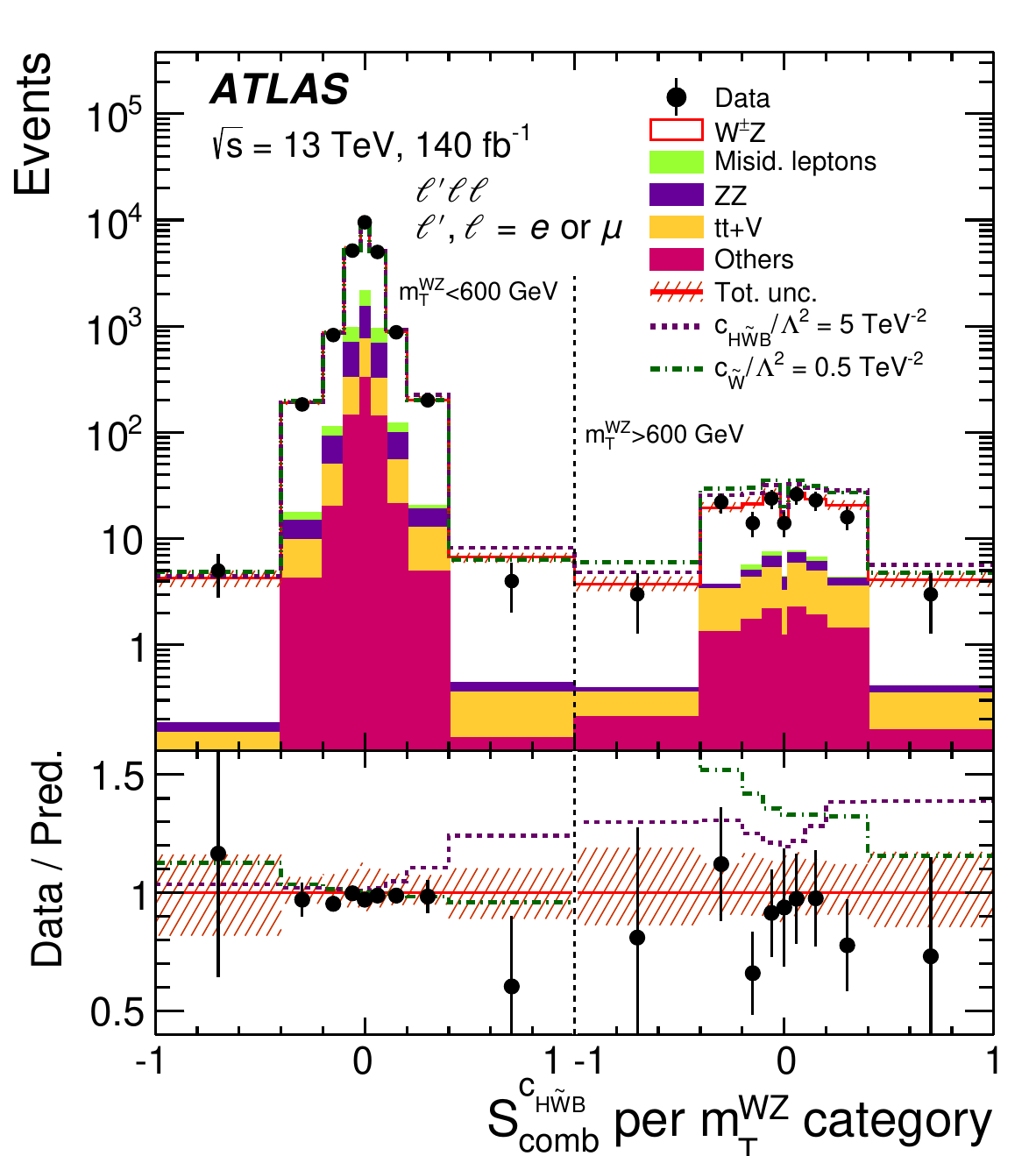}}
\subfloat[]{\includegraphics[width=.45\textwidth]{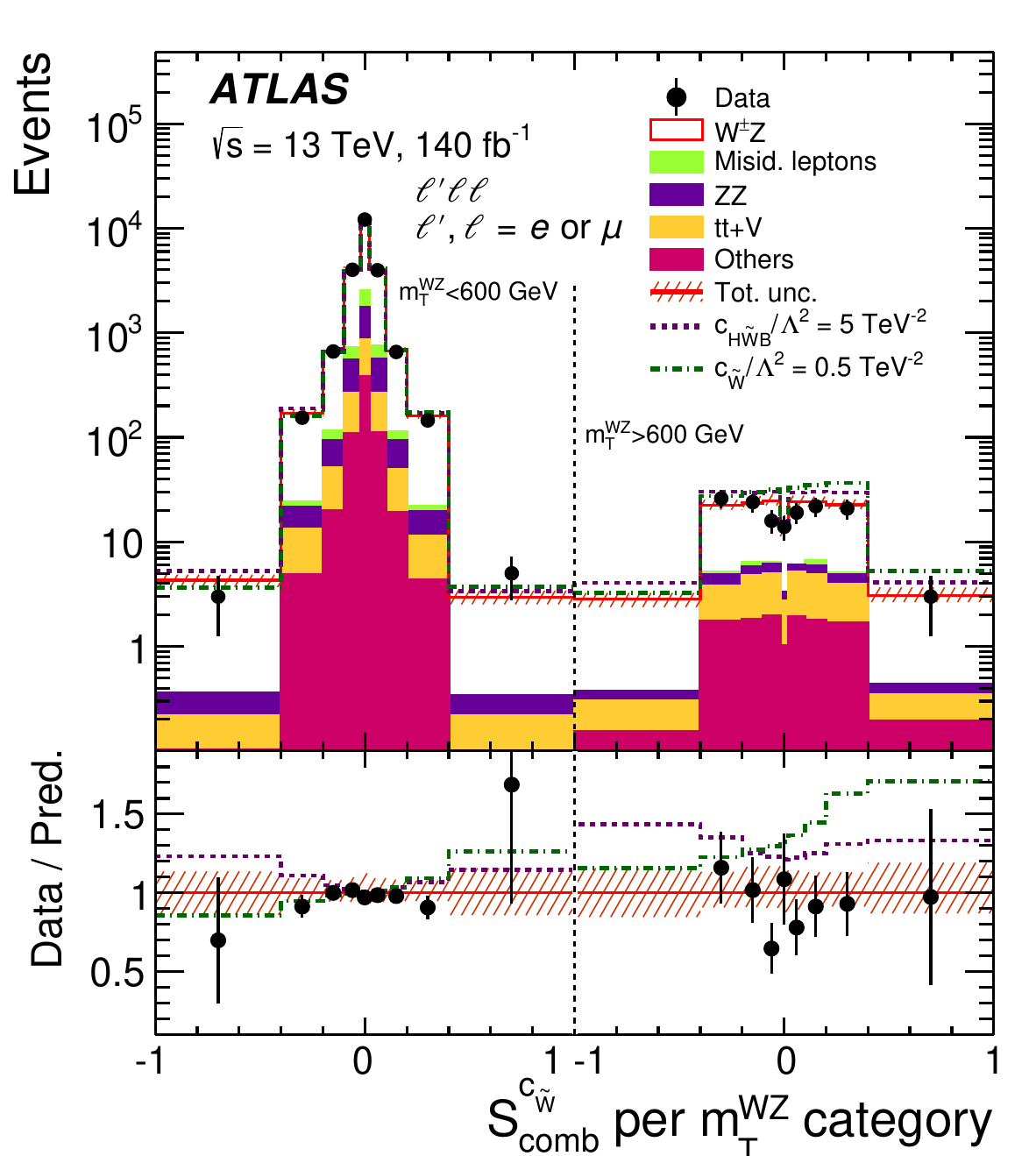}}
\caption{Detector-level one-dimensional distributions of the \Scomb observable in two \mtwz categories used to obtain limits on (a) \cHWtilB  and (b) \cWtil  EFT Wilson coefficients.
The points correspond to the data with the error bars representing the statistical uncertainties, and the histograms correspond to the predictions of the various SM processes.
The sum of the background processes with misidentified leptons is labelled ``Misid. leptons''.
The \sherpa MC prediction is used for the \wz\ signal contribution.
The open red histogram shows the total prediction; the shaded band is the total uncertainty of this prediction.
The lower panels in each figure show the ratio of the data points to the open red histogram with their respective uncertainties.
For illustration purposes, contribution of the \oHWtilB and \oWtil operators with values of the corresponding Wilson coefficients of 5~\TeV$^{-2}$ and 0.5~\TeV$^{-2}$, respectively, are also represented.
The parameterisation including the pure dimension-6 contribution is used.
}
\label{fig:CPodd:fit_control_dists}
\end{figure}

No deviation with respect to the SM predictions is observed and the expected and observed $95$\% CL lower and upper limits on the given Wilson coefficients are presented in Table~\ref{tab:CPodd:expected_limits}.
The use of the \Scomb observable improves the expected limits by factors of 2.5 and 2.9 for \cHWtilB/$\Lambda^2$ and \cWtil/$\Lambda^2$, respectively, compared to using a single observable like \mtwz or a triple product.
For the \cWtil/$\Lambda^2$ coefficient, the \Scomb observable brings an important sensitivity to the linear term alone that is not possible to reach using a single observable.
This makes the results less sensitive to missing higher-order operators in the EFT expansion.
The \cHWtilB/$\Lambda^2$ coefficient is itself completely constrained by the contribution from the linear term.

Limits obtained on \cWtil/$\Lambda^2$ are comparable to those derived from VBF $Z$ measurements~\cite{ATLAS:2020nzk}.
The limit on \cHWtilB is only slightly weaker by a factor of 1.5 to 2 than what is obtained in VBF Z or Higgs measurements~\cite{ATLAS:2023mqy,ATLAS:2023pwa,ATLAS:2025hki}.
However, the limit derived from $H \to WW^*$ measurements is completely dominated by effects of the quadratic term, while in VBF $Z$ production and in the present result the limit purely results from effects of the linear term.

Confidence intervals on CP-conserving and CP-violating operators are summarised together in Figure~\ref{fig:CPodd:c_limits}.
Results for CP-conserving and CP-violating operators are also presented in Figure~\ref{fig:CPodd:Lambda_limits} in terms of upper limits on the scale of new physics $\Lambda$ assuming different input values of $c_i = 0.01, 1.0, (4\pi)^2$ and symmetrising the confidence intervals.

\begin{table}[tb]
\caption{Expected and observed 95\% CL intervals for CP-odd Wilson coefficients.
Limits are presented using parameterisations excluding (``lin.'') and including (``lin.+quad.'') the pure dimension-6 contributions, respectively.
Only one Wilson coefficient is constrained at a time.
}
\label{tab:CPodd:expected_limits}
\centering
\begin{tabular}{lcccc}
\toprule
& \multicolumn{2}{c}{ Expected [\TeV$^{-2}$] } & \multicolumn{2}{c}{ Observed [\TeV$^{-2}$] } \\
& 95\% CL (lin.) & 95\% CL (lin.+quad.) & 95\% CL (lin.) & 95\% CL (lin.+quad.) \\
\midrule
\cHWtilB/$\Lambda^2$ & [-1.463, 1.456] & [-1.458, 1.459] 	& [-1.625, 1.332]  &  [-1.505, 1.263]  \\
\cWtil/$\Lambda^2$   & [-0.162, 0.162] & [-0.127, 0.128] 	& [-0.186, 0.139]  &   [-0.109, 0.093]  \\
\bottomrule
\end{tabular}

\end{table}
\begin{figure}[tb]
\centering
\includegraphics[width=.55\textwidth]{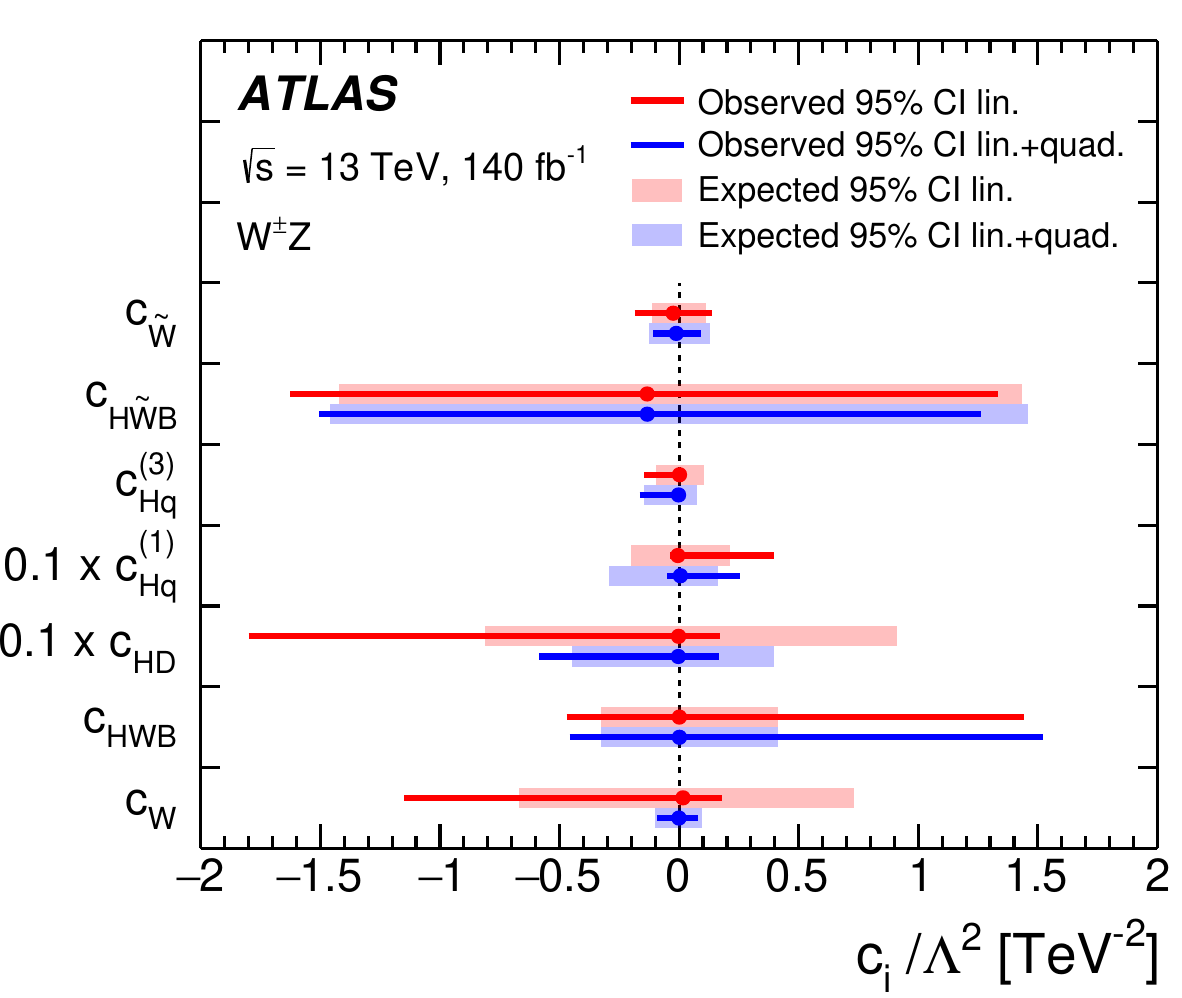}
\caption{The observed and expected values of the Wilson coefficients from CP-even and CP-odd operators obtained with only one Wilson coefficient left floating at a time in the fit while all others are set to zero.
The horizontal bands (bars) represent the expected (observed) intervals at 95\% CL using the parameterisations excluding (``lin.'') and including (``lin.+quad.'') the pure dimension-6 contributions, respectively.
For the observed intervals, the point represent the best-fit values.
The values of \cHD/$\Lambda^2$ and \cHjone/$\Lambda^2$ are scaled by a factor of 0.1.
}
\label{fig:CPodd:c_limits}
\end{figure}
\begin{figure}[tb]
\centering
\includegraphics[width=.55\textwidth]{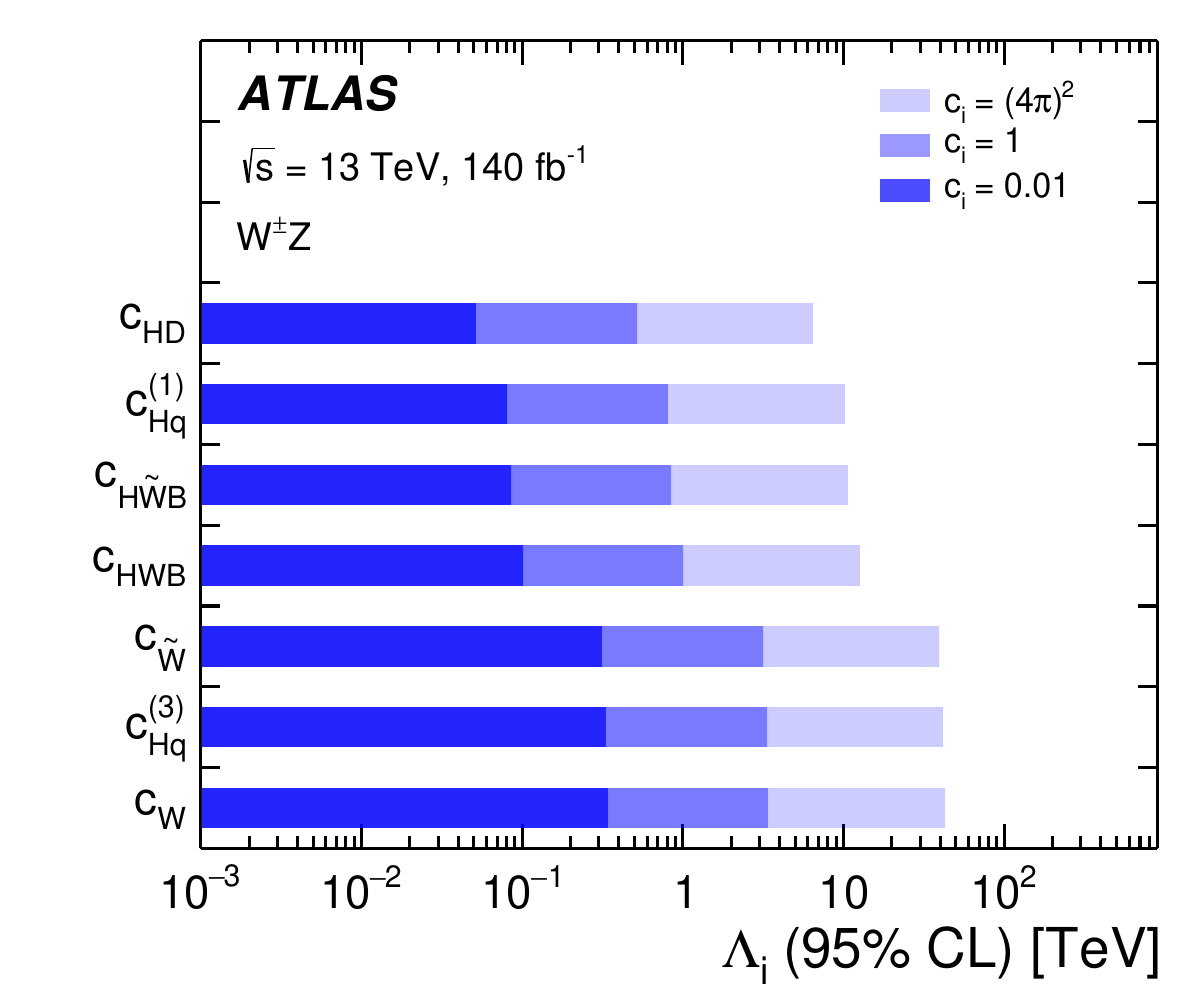}
\caption{Observed exclusion limits at 95\% CL on the scale of new physics $\Lambda$ for different values of the anomalous coupling strength $c_i$.
The parameterisation including the pure dimension-6 contribution is used.
}
\label{fig:CPodd:Lambda_limits}
\end{figure}
%


\newpage
\clearpage

\FloatBarrier

\section{Conclusion}
\label{sec:conclusion}
%

Measurements of \wz\ production cross-sections in $\sqrt{s} = 13$~\TeV\ $pp$ collisions at the LHC are presented.
The data analysed were collected with the ATLAS detector from $2015$ to $2018$ and correspond to an integrated luminosity of $140~\ifb$.
The measurements use leptonic decay modes of the gauge bosons into electrons or muons and are performed
in a fiducial phase space closely matching the detector acceptance.
The measured inclusive cross-section in the fiducial region for leptonic decay modes (electrons or muons) is  measured with a relative precision of 4\% and yields
$\sigma_{W^\pm Z \rightarrow \ell^{'} \nu \ell \ell}^{\textrm{fid.}} =  60.7  \, \pm~0.5~\textrm{(stat.)} \, \pm~2.3~\textrm{(syst.)} \, \pm~0.6~\textrm{(lumi.)}$~fb,
in agreement with the Standard Model \matrix prediction of $61.4 \pm 1.3~\textrm{(scale)}$~fb at \nnloxew.
The ratio of the cross-sections for $W^+Z$ and $W^-Z$ production is also measured.
Differential \wz\ production cross-sections are measured as functions of several kinematic variables and compared with SM predictions at NNLO QCD and NLO EW precision from the \matrix calculation and from the \SHERPA and \powhegpythia MC event generators.
The differential cross-section distributions are fairly well described by the theory predictions, except at low transverse momenta of the bosons where resummed calculations would be needed to describe data.
Measurements from the tail of the transverse momentum of the harder boson, \ptvone, tend to favour the multiplicative \nnloxew scheme for the combination of \NNLO QCD corrections with \NLO EW corrections instead of the additive one.

New observables sensitive to CP-violation effects are also measured for the first time in \wz production.
An effective field theory interpretation of the measurements is performed, considering a set of both CP-conserving and CP-violating dimension-6 operators impacting the $W^{\pm}Z$ production.
Limits are extracted on CP-conserving Wilson coefficients using the transverse mass of the $W^{\pm}Z$ system.
For CP-violating coefficients a novel machine learning approach is used to design an observable with enhanced sensitivity to CP-violation effects, allowing a significant improvement in sensitivity compared to the use of a single kinematic variable.
The limit obtained on the CP-even bosonic coupling operator is comparable to previously existing limits from single measurements.
Limits reached on CP-odd operators have the same sensitivity as the best existing limits derived from other final states, and therefore bring new complementary information.

%
%


%
%
%
%
%
\section*{Acknowledgements}
%

%

%
%

%
%

We thank CERN for the very successful operation of the LHC and its injectors, as well as the support staff at
CERN and at our institutions worldwide without whom ATLAS could not be operated efficiently.

The crucial computing support from all WLCG partners is acknowledged gratefully, in particular from CERN, the ATLAS Tier-1 facilities at TRIUMF/SFU (Canada), NDGF (Denmark, Norway, Sweden), CC-IN2P3 (France), KIT/GridKA (Germany), INFN-CNAF (Italy), NL-T1 (Netherlands), PIC (Spain), RAL (UK) and BNL (USA), the Tier-2 facilities worldwide and large non-WLCG resource providers. Major contributors of computing resources are listed in Ref.~\cite{ATL-SOFT-PUB-2025-001}.

We gratefully acknowledge the support of ANPCyT, Argentina; YerPhI, Armenia; ARC, Australia; BMWFW and FWF, Austria; ANAS, Azerbaijan; CNPq and FAPESP, Brazil; NSERC, NRC and CFI, Canada; CERN; ANID, Chile; CAS, MOST and NSFC, China; Minciencias, Colombia; MEYS CR, Czech Republic; DNRF and DNSRC, Denmark; IN2P3-CNRS and CEA-DRF/IRFU, France; SRNSFG, Georgia; BMFTR, HGF and MPG, Germany; GSRI, Greece; RGC and Hong Kong SAR, China; ICHEP and Academy of Sciences and Humanities, Israel; INFN, Italy; MEXT and JSPS, Japan; CNRST, Morocco; NWO, Netherlands; RCN, Norway; MNiSW, Poland; FCT, Portugal; MNE/IFA, Romania; MSTDI, Serbia; MSSR, Slovakia; ARIS and MVZI, Slovenia; DSI/NRF, South Africa; MICIU/AEI, Spain; SRC and Wallenberg Foundation, Sweden; SERI, SNSF and Cantons of Bern and Geneva, Switzerland; NSTC, Taipei; TENMAK, T\"urkiye; STFC/UKRI, United Kingdom; DOE and NSF, United States of America.

Individual groups and members have received support from BCKDF, CANARIE, CRC and DRAC, Canada; CERN-CZ, FORTE and PRIMUS, Czech Republic; COST, ERC, ERDF, Horizon 2020, ICSC-NextGenerationEU and Marie Sk{\l}odowska-Curie Actions, European Union; Investissements d'Avenir Labex, Investissements d'Avenir Idex and ANR, France; DFG and AvH Foundation, Germany; Herakleitos, Thales and Aristeia programmes co-financed by EU-ESF and the Greek NSRF, Greece; BSF-NSF and MINERVA, Israel; NCN and NAWA, Poland; La Caixa Banking Foundation, CERCA Programme Generalitat de Catalunya and PROMETEO and GenT Programmes Generalitat Valenciana, Spain; G\"{o}ran Gustafssons Stiftelse, Sweden; The Royal Society and Leverhulme Trust, United Kingdom.

In addition, individual members wish to acknowledge support from CERN: European Organization for Nuclear Research (CERN DOCT); Chile: Agencia Nacional de Investigaci\'on y Desarrollo (FONDECYT 1230812, FONDECYT 1240864, Fondecyt 3240661); China: Chinese Ministry of Science and Technology (MOST-2023YFA1605700, MOST-2023YFA1609300), National Natural Science Foundation of China (NSFC - 12175119, NSFC 12275265); Czech Republic: Czech Science Foundation (GACR - 24-11373S), Ministry of Education Youth and Sports (ERC-CZ-LL2327, FORTE CZ.02.01.01/00/22\_008/0004632), PRIMUS Research Programme (PRIMUS/21/SCI/017); EU: H2020 European Research Council (ERC - 101002463); European Union: European Research Council (BARD No. 101116429, ERC - 948254, ERC 101089007), European Regional Development Fund (SMASH COFUND 101081355, SLO ERDF), Horizon 2020 Framework Programme (MUCCA - CHIST-ERA-19-XAI-00), European Union, Future Artificial Intelligence Research (FAIR-NextGenerationEU PE00000013), Italian Center for High Performance Computing, Big Data and Quantum Computing (ICSC, NextGenerationEU); France: Agence Nationale de la Recherche (ANR-21-CE31-0022, ANR-22-EDIR-0002); Germany: Baden-Württemberg Stiftung (BW Stiftung-Postdoc Eliteprogramme), Deutsche Forschungsgemeinschaft (DFG - 469666862, DFG - CR 312/5-2); China: Research Grants Council (GRF); Italy: Istituto Nazionale di Fisica Nucleare (ICSC, NextGenerationEU), Ministero dell'Università e della Ricerca (NextGenEU 153D23001490006 M4C2.1.1, NextGenEU I53D23000820006 M4C2.1.1, NextGenEU I53D23001490006 M4C2.1.1, SOE2024\_0000023); Japan: Japan Society for the Promotion of Science (JSPS KAKENHI JP22H01227, JSPS KAKENHI JP22H04944, JSPS KAKENHI JP22KK0227, JSPS KAKENHI JP24K23939, JSPS KAKENHI JP24KK0251, JSPS KAKENHI JP25H00650, JSPS KAKENHI JP25H01291, JSPS KAKENHI JP25K01023); Norway: Research Council of Norway (RCN-314472); Poland: Ministry of Science and Higher Education (IDUB AGH, POB8, D4 no 9722), Polish National Science Centre (NCN 2021/42/E/ST2/00350, NCN OPUS 2023/51/B/ST2/02507, NCN OPUS nr 2022/47/B/ST2/03059, NCN UMO-2019/34/E/ST2/00393, UMO-2022/47/O/ST2/00148, UMO-2023/49/B/ST2/04085, UMO-2023/51/B/ST2/00920, UMO-2024/53/N/ST2/00869); Portugal: Foundation for Science and Technology (FCT); Spain: Ministry of Science and Innovation (MCIN \& NextGenEU PCI2022-135018-2, MICIN \& FEDER PID2021-125273NB, RYC2019-028510-I, RYC2020-030254-I, RYC2021-031273-I, RYC2022-038164-I); Sweden: Carl Trygger Foundation (Carl Trygger Foundation CTS 22:2312), Swedish Research Council (Swedish Research Council 2023-04654, VR 2021-03651, VR 2022-03845, VR 2022-04683, VR 2023-03403, VR 2024-05451), Knut and Alice Wallenberg Foundation (KAW 2018.0458, KAW 2022.0358, KAW 2023.0366); Switzerland: Swiss National Science Foundation (SNSF - PCEFP2\_194658); United Kingdom: Royal Society (NIF-R1-231091); United States of America: U.S. Department of Energy (ECA DE-AC02-76SF00515), Neubauer Family Foundation.

%
%


%
%
%
%
%
%

%
%

%
%
%
\printbibliography
\clearpage
\input{atlas_authlist}
%

%
%

%

%
%
%
%

\end{document}

%% file: atlas_authlist.tex
 
\begin{flushleft}
\hypersetup{urlcolor=black}
{\Large The ATLAS Collaboration}

\bigskip

\AtlasOrcid[0000-0002-6665-4934]{G.~Aad}$^\textrm{\scriptsize 104}$,
\AtlasOrcid[0000-0001-7616-1554]{E.~Aakvaag}$^\textrm{\scriptsize 17}$,
\AtlasOrcid[0000-0002-5888-2734]{B.~Abbott}$^\textrm{\scriptsize 123}$,
\AtlasOrcid[0000-0002-0287-5869]{S.~Abdelhameed}$^\textrm{\scriptsize 119a}$,
\AtlasOrcid[0000-0002-1002-1652]{K.~Abeling}$^\textrm{\scriptsize 55}$,
\AtlasOrcid[0000-0001-5763-2760]{N.J.~Abicht}$^\textrm{\scriptsize 49}$,
\AtlasOrcid[0000-0002-8496-9294]{S.H.~Abidi}$^\textrm{\scriptsize 30}$,
\AtlasOrcid[0009-0003-6578-220X]{M.~Aboelela}$^\textrm{\scriptsize 45}$,
\AtlasOrcid[0000-0002-9987-2292]{A.~Aboulhorma}$^\textrm{\scriptsize 36e}$,
\AtlasOrcid[0000-0001-5329-6640]{H.~Abramowicz}$^\textrm{\scriptsize 157}$
% ...
\bigskip
\\

% ...
$^{17}$Department for Physics and Technology, University of Bergen, Bergen; Norway.\\
% ...
$^{30}$Physics Department, Brookhaven National Laboratory, Upton NY; United States of America.\\
% ...
$^{36}$$^{(a)}$Facult\'e des Sciences Ain Chock, Universit\'e Hassan II de Casablanca;$^{(b)}$Facult\'{e} des Sciences, Universit\'{e} Ibn-Tofail, K\'{e}nitra;$^{(c)}$Facult\'e des Sciences Semlalia, Universit\'e Cadi Ayyad, LPHEA-Marrakech;$^{(d)}$LPMR, Facult\'e des Sciences, Universit\'e Mohamed Premier, Oujda;$^{(e)}$Facult\'e des sciences, Universit\'e Mohammed V, Rabat;$^{(f)}$Institute of Applied Physics, Mohammed VI Polytechnic University, Ben Guerir; Morocco.\\
% ...
$^{45}$Physics Department, Southern Methodist University, Dallas TX; United States of America.\\
% ...
$^{49}$Fakult\"{a}t Physik , Technische Universit{\"a}t Dortmund, Dortmund; Germany.\\
% ...
$^{55}$II. Physikalisches Institut, Georg-August-Universit\"{a}t G\"ottingen, G\"ottingen; Germany.\\
% ...
$^{104}$CPPM, Aix-Marseille Universit\'e, CNRS/IN2P3, Marseille; France.\\
% ...
$^{119}$$^{(a)}$New York University Abu Dhabi, Abu Dhabi;$^{(b)}$United Arab Emirates University, Al Ain; United Arab Emirates.\\
% ...
$^{123}$Homer L. Dodge Department of Physics and Astronomy, University of Oklahoma, Norman OK; United States of America.\\
% ...
$^{157}$Raymond and Beverly Sackler School of Physics and Astronomy, Tel Aviv University, Tel Aviv; Israel.\\
% ...
$^{a}$ Also at Affiliated with an institute formerly covered by a cooperation agreement with CERN.\\
$^{b}$ Also at An-Najah National University, Nablus; Palestine.\\
$^{c}$ Also at Borough of Manhattan Community College, City University of New York, New York NY; United States of America.\\
$^{d}$ Also at Center for Interdisciplinary Research and Innovation (CIRI-AUTH), Thessaloniki; Greece.\\
$^{e}$ Also at Centre of Physics of the Universities of Minho and Porto (CF-UM-UP); Portugal.\\
$^{f}$ Also at CERN, Geneva; Switzerland.\\
% ...

\end{flushleft}


%% file: ANA-STDM-2019-26-PAPER.bib
@Article{	  Aaboud:2016lpj,
  author	= "{ATLAS Collaboration}",
  title		= "{Measurement of the cross-section for producing a W boson
		  in association with a single top quark in pp collisions at
		  $ \sqrt{s}=13 $ TeV with ATLAS}",
  collaboration	= "ATLAS",
  journal	= "JHEP",
  volume	= "01",
  year		= "2018",
  pages		= "063",
  doi		= "10.1007/JHEP01(2018)063",
  eprint	= "1612.07231",
  archiveprefix	= "arXiv",
  primaryclass	= "hep-ex",
  reportnumber	= "CERN-EP-2016-238",
  slaccitation	= "%%CITATION = ARXIV:1612.07231;%%"
}

@Article{	  Aaboud:2016mmw,
  author	= "{ATLAS Collaboration}",
  title		= "{Measurement of the Inelastic Proton-Proton Cross Section
		  at $\sqrt{s} = 13$ TeV with the ATLAS Detector at the LHC}",
  collaboration	= "ATLAS",
  journal	= "Phys. Rev. Lett.",
  volume	= "117",
  year		= "2016",
  number	= "18",
  pages		= "182002",
  doi		= "10.1103/PhysRevLett.117.182002",
  eprint	= "1606.02625",
  archiveprefix	= "arXiv",
  primaryclass	= "hep-ex",
  reportnumber	= "CERN-EP-2016-140",
  slaccitation	= "%%CITATION = ARXIV:1606.02625;%%"
}

@Article{	  Aaboud:2016yus,
  author	= "{ATLAS Collaboration}",
  title		= "{Measurement of the $W^{\pm}Z$ boson pair-production cross
		  section in $pp$ collisions at $\sqrt{s}=13$ TeV with the
		  ATLAS Detector}",
  collaboration	= "ATLAS",
  journal	= "Phys. Lett. B",
  volume	= "762",
  year		= "2016",
  pages		= "1-22",
  doi		= "10.1016/j.physletb.2016.08.052",
  eprint	= "1606.04017",
  archiveprefix	= "arXiv",
  primaryclass	= "hep-ex",
  reportnumber	= "CERN-EP-2016-108",
  slaccitation	= "%%CITATION = ARXIV:1606.04017;%%"
}

@Article{	  Aaboud:2017rwm,
  author	= "{ATLAS Collaboration}",
  title		= "{$ZZ \to \ell^{+}\ell^{-}\ell^{\prime +}\ell^{\prime -}$
		  cross-section measurements and search for anomalous triple
		  gauge couplings in 13 TeV $pp$ collisions with the ATLAS
		  detector}",
  collaboration	= "ATLAS",
  journal	= "Phys. Rev. D",
  volume	= "97",
  year		= "2018",
  number	= "3",
  pages		= "032005",
  doi		= "10.1103/PhysRevD.97.032005",
  eprint	= "1709.07703",
  archiveprefix	= "arXiv",
  primaryclass	= "hep-ex",
  reportnumber	= "CERN-EP-2017-163",
  slaccitation	= "%%CITATION = ARXIV:1709.07703;%%"
}

@Article{	  Aad:2010ah,
  author	= "{ATLAS Collaboration}",
  title		= "{The ATLAS Simulation Infrastructure}",
  collaboration	= "ATLAS",
  journal	= "Eur. Phys. J. C",
  volume	= "70",
  year		= "2010",
  pages		= "823",
  doi		= "10.1140/epjc/s10052-010-1429-9",
  eprint	= "1005.4568",
  archiveprefix	= "arXiv",
  primaryclass	= "physics.ins-det",
  slaccitation	= "%%CITATION = ARXIV:1005.4568;%%"
}

@Article{	  Aad:2012twa,
  author	= "{ATLAS Collaboration}",
  title		= "{Measurement of $WZ$ production in proton-proton
		  collisions at $\sqrt{s}=7$ \TeV{} with the ATLAS detector}",
  collaboration	= "ATLAS Collaboration",
  journal	= "Eur. Phys. J. C",
  volume	= "72",
  pages		= "2173",
  doi		= "10.1140/epjc/s10052-012-2173-0",
  year		= "2012",
  eprint	= "1208.1390",
  archiveprefix	= "arXiv",
  primaryclass	= "hep-ex",
  reportnumber	= "CERN-PH-EP-2012-179",
  slaccitation	= "%%CITATION = ARXIV:1208.1390;%%"
}

@Article{	  Aad:2014pda,
  author	= "{ATLAS Collaboration}",
  title		= "{Search for supersymmetry at $\sqrt{s}$=8 TeV in final
		  states with jets and two same-sign leptons or three leptons
		  with the ATLAS detector}",
  collaboration	= "ATLAS",
  journal	= "JHEP",
  volume	= "06",
  year		= "2014",
  pages		= "035",
  doi		= "10.1007/JHEP06(2014)035",
  eprint	= "1404.2500",
  archiveprefix	= "arXiv",
  primaryclass	= "hep-ex",
  reportnumber	= "CERN-PH-EP-2014-044",
  slaccitation	= "%%CITATION = ARXIV:1404.2500;%%"
}

@Article{	  Aad:2015ina,
  author	= "{ATLAS Collaboration}",
  title		= "{Performance of pile-up mitigation techniques for jets in
		  $pp$ collisions at $\sqrt{s}=8$ TeV using the ATLAS
		  detector}",
  collaboration	= "ATLAS",
  journal	= "Eur. Phys. J. C",
  volume	= "76",
  year		= "2016",
  number	= "11",
  pages		= "581",
  doi		= "10.1140/epjc/s10052-016-4395-z",
  eprint	= "1510.03823",
  archiveprefix	= "arXiv",
  primaryclass	= "hep-ex",
  reportnumber	= "CERN-PH-EP-2015-206",
  slaccitation	= "%%CITATION = ARXIV:1510.03823;%%"
}

@Article{	  Aad:2020wog,
  author	= "{ATLAS Collaboration}",
  collaboration	= "ATLAS",
  title		= "{Observation of the associated production of a top quark
		  and a $Z$ boson in $pp$ collisions at $\sqrt{s} = 13$ TeV
		  with the ATLAS detector}",
  eprint	= "2002.07546",
  archiveprefix	= "arXiv",
  primaryclass	= "hep-ex",
  reportnumber	= "CERN-EP-2019-273",
  doi		= "10.1007/JHEP07(2020)124",
  journal	= "JHEP",
  volume	= "07",
  pages		= "124",
  year		= "2020"
}

@Article{	  Aaron:2009bp,
  author	= "{H1 Collaboration}",
  title		= "{Measurement of the Inclusive ep Scattering Cross Section
		  at Low $Q^2$ and x at HERA}",
  collaboration	= "H1",
  journal	= "Eur.~Phys.~J.~C",
  volume	= "63",
  pages		= "625",
  doi		= "10.1140/epjc/s10052-009-1128-6",
  year		= "2009",
  eprint	= "0904.0929",
  archiveprefix	= "arXiv",
  primaryclass	= "hep-ex",
  reportnumber	= "DESY-08-171",
  slaccitation	= "%%CITATION = ARXIV:0904.0929;%%"
}

@Article{	  Aaron:2009wt,
  author	= "{H1 and ZEUS Collaborations}",
  collaboration	= "H1 and ZEUS",
  title		= "{Combined measurement and QCD analysis of the inclusive $e^\pm p$
		  scattering cross sections at HERA}",
  journal	= "JHEP",
  volume	= "01",
  year		= "2010",
  pages		= "109",
  eprint	= "0911.0884",
  archiveprefix	= "arXiv",
  primaryclass	= "hep-ex",
  doi		= "10.1007/JHEP01(2010)109",
  slaccitation	= "%%CITATION = 0911.0884;%%"
}

@Article{	  Accomando:2005xp,
  author	= "Accomando, E. and Kaiser, A.",
  title		= "{Electroweak corrections and anomalous triple gauge-boson
		  couplings in $W^{+} W^{-}$ and $W^\pm Z$ production at the
		  LHC}",
  journal	= "Phys. Rev. D",
  volume	= "73",
  year		= "2006",
  pages		= "093006",
  doi		= "10.1103/PhysRevD.73.093006",
  eprint	= "hep-ph/0511088",
  archiveprefix	= "arXiv",
  primaryclass	= "hep-ph",
  reportnumber	= "PSI-PR-05-10, ZU-TH-19-05, DFTT-36-05",
  slaccitation	= "%%CITATION = HEP-PH/0511088;%%"
}

@Article{	  Agostinelli:2002hh,
  author	= "Agostinelli, S. and others",
  collaboration	= "GEANT4",
  title		= "{\textsc{Geant4}--a simulation toolkit}",
  journal	= "Nucl. Instrum. Meth. A",
  volume	= "506",
  year		= "2003",
  pages		= "250-303",
  doi		= "10.1016/S0168-9002(03)01368-8",
  slaccitation	= "%%CITATION = NUIMA,A506,250;%%"
}

@Article{	  Alioli:2010xd,
  author	= "Alioli, Simone and Nason, Paolo and Oleari, Carlo and Re,
		  Emanuele",
  title		= "{A general framework for implementing NLO calculations in
		  shower Monte Carlo programs: the POWHEG BOX}",
  journal	= "JHEP",
  volume	= "06",
  year		= "2010",
  pages		= "043",
  doi		= "10.1007/JHEP06(2010)043",
  eprint	= "1002.2581",
  archiveprefix	= "arXiv",
  primaryclass	= "hep-ph",
  reportnumber	= "DESY-10-018, SFB-CPP-10-22, IPPP-10-11, DCPT-10-22",
  slaccitation	= "%%CITATION = ARXIV:1002.2581;%%"
}

@Article{	  antikt,
  author	= "Cacciari, Matteo and Salam, Gavin P. and Soyez, Gregory",
  title		= "{The anti-$k_t$ jet clustering algorithm}",
  journal	= "JHEP",
  volume	= "04",
  year		= "2008",
  pages		= "063",
  doi		= "10.1088/1126-6708/2008/04/063",
  eprint	= "0802.1189",
  archiveprefix	= "arXiv",
  primaryclass	= "hep-ph",
  reportnumber	= "LPTHE-07-03",
  slaccitation	= "%%CITATION = ARXIV:0802.1189;%%"
}

@Booklet{	  ATL-PHYS-PUB-2016-002,
  author	= "{ATLAS Collaboration}",
  title		= "{Multi-Boson Simulation for 13 \TeV{} ATLAS Analyses}",
  howpublished	= "{ATL-PHYS-PUB-2016-002}",
  url		= "http://cds.cern.ch/record/2119986",
  year		= "2016"
}

@TechReport{	  ATL-PHYS-PUB-2016-005,
  title		= "{Modelling of the $t\bar{t}H$ and $t\bar{t}V$ $(V=W,Z)$
		  processes for $\sqrt{s}=13$ TeV ATLAS analyses}",
  institution	= "CERN",
  collaboration	= "ATLAS Collaboration",
  address	= "Geneva",
  number	= "ATL-PHYS-PUB-2016-005",
  month		= "Jan",
  year		= "2016",
  reportnumber	= "ATL-PHYS-PUB-2016-005",
  url		= "https://cds.cern.ch/record/2120826"
}

@TechReport{	  ATL-PHYS-PUB-2016-017,
  author	= "{ATLAS Collaboration}",
  title		= "{The Pythia 8 A3 tune description of ATLAS minimum bias
		  and inelastic measurements incorporating the
		  Donnachie-Landshoff diffractive model}",
  institution	= "CERN",
  collaboration	= "ATLAS Collaboration",
  address	= "Geneva",
  reportnumber	= "ATL-PHYS-PUB-2016-017",
  month		= "Aug",
  year		= "2016",
  url		= "https://cds.cern.ch/record/2206965"
}

@Article{	  ATLAS:2017ghe,
  author	= "{ATLAS Collaboration}",
  collaboration	= "ATLAS",
  title		= "{Jet reconstruction and performance using particle flow
		  with the ATLAS Detector}",
  eprint	= "1703.10485",
  archiveprefix	= "arXiv",
  primaryclass	= "hep-ex",
  reportnumber	= "CERN-EP-2017-024",
  doi		= "10.1140/epjc/s10052-017-5031-2",
  journal	= "Eur. Phys. J. C",
  volume	= "77",
  number	= "7",
  pages		= "466",
  year		= "2017"
}

@Article{	  ATLAS:2019dpa,
  author	= "{ATLAS Collaboration}",
  collaboration	= "ATLAS",
  title		= "{Performance of electron and photon triggers in ATLAS
		  during LHC Run 2}",
  eprint	= "1909.00761",
  archiveprefix	= "arXiv",
  primaryclass	= "hep-ex",
  reportnumber	= "CERN-EP-2019-169",
  doi		= "10.1140/epjc/s10052-019-7500-2",
  journal	= "Eur. Phys. J. C",
  volume	= "80",
  number	= "1",
  pages		= "47",
  year		= "2020"
}

@Article{	  ATLAS:2019qmc,
  author	= "{ATLAS Collaboration}",
  collaboration	= "ATLAS",
  title		= "{Electron and photon performance measurements with the
		  ATLAS detector using the 2015\textendash{}2017 LHC
		  proton-proton collision data}",
  eprint	= "1908.00005",
  archiveprefix	= "arXiv",
  primaryclass	= "hep-ex",
  reportnumber	= "CERN-EP-2019-145",
  doi		= "10.1088/1748-0221/14/12/P12006",
  journal	= "JINST",
  volume	= "14",
  number	= "12",
  pages		= "P12006",
  year		= "2019"
}

@Article{	  ATLAS:2020auj,
  author	= "{ATLAS Collaboration}",
  collaboration	= "ATLAS",
  title		= "{Muon reconstruction and identification efficiency in
		  ATLAS using the full Run 2 $pp$ collision data set at
		  $\sqrt{s}=13$ TeV}",
  eprint	= "2012.00578",
  archiveprefix	= "arXiv",
  primaryclass	= "hep-ex",
  reportnumber	= "CERN-EP-2020-199",
  doi		= "10.1140/epjc/s10052-021-09233-2",
  journal	= "Eur. Phys. J. C",
  volume	= "81",
  number	= "7",
  pages		= "578",
  year		= "2021"
}

@Article{	  ATLAS:2020gty,
  author	= "{ATLAS Collaboration}",
  collaboration	= "ATLAS",
  title		= "{Performance of the ATLAS muon triggers in Run 2}",
  eprint	= "2004.13447",
  archiveprefix	= "arXiv",
  primaryclass	= "physics.ins-det",
  reportnumber	= "CERN-EP-2020-031",
  doi		= "10.1088/1748-0221/15/09/p09015",
  journal	= "JINST",
  volume	= "15",
  number	= "09",
  pages		= "P09015",
  year		= "2020"
}

@Article{	  ATLAS:2021fzm,
  author	= "{ATLAS Collaboration}",
  collaboration	= "ATLAS",
  title		= "{Measurements of the inclusive and differential production
		  cross sections of a top-quark\textendash{}antiquark pair in
		  association with a Z~boson at $\sqrt{s} = 13$~TeV with the
		  ATLAS detector}",
  eprint	= "2103.12603",
  archiveprefix	= "arXiv",
  primaryclass	= "hep-ex",
  reportnumber	= "CERN-EP-2021-01",
  doi		= "10.1140/epjc/s10052-021-09439-4",
  journal	= "Eur. Phys. J. C",
  volume	= "81",
  number	= "8",
  pages		= "737",
  year		= "2021"
}

@Article{	  ATLAS:2022swp,
  author	= "{ATLAS Collaboration}",
  collaboration	= "ATLAS",
  title		= "{Tools for estimating fake/non-prompt lepton backgrounds
		  with the ATLAS detector at the LHC}",
  eprint	= "2211.16178",
  archiveprefix	= "arXiv",
  primaryclass	= "hep-ex",
  reportnumber	= "CERN-EP-2022-214",
  doi		= "10.1088/1748-0221/18/11/T11004",
  journal	= "JINST",
  volume	= "18",
  number	= "11",
  pages		= "T11004",
  year		= "2023"
}

@Article{	  ATLASWZ_8TeV,
  author	= "{ATLAS Collaboration}",
  title		= "{Measurements of $W^\pm Z$ production cross sections in
		  $pp$ collisions at $\sqrt{s} = 8$ TeV with the ATLAS
		  detector and limits on anomalous gauge boson
		  self-couplings}",
  collaboration	= "ATLAS",
  journal	= "Phys. Rev. D",
  volume	= "93",
  year		= "2016",
  pages		= "092004",
  doi		= "10.1103/PhysRevD.93.092004",
  eprint	= "1603.02151",
  archiveprefix	= "arXiv",
  primaryclass	= "hep-ex",
  reportnumber	= "CERN-EP-2016-017",
  slaccitation	= "%%CITATION = ARXIV:1603.02151;%%"
}

@Article{	  Azatov:2017kzw,
  author	= "Azatov, A. and Elias-Miró, J. and Reyimuaji, Y. and
		  Venturini, E.",
  title		= "{Novel measurements of anomalous triple gauge couplings
		  for the LHC}",
  journal	= "JHEP",
  volume	= "10",
  year		= "2017",
  pages		= "027",
  doi		= "10.1007/JHEP10(2017)027",
  eprint	= "1707.08060",
  archiveprefix	= "arXiv",
  primaryclass	= "hep-ph",
  reportnumber	= "SISSA-35-2017-FISI",
  slaccitation	= "%%CITATION = ARXIV:1707.08060;%%"
}

@Article{	  AZNLO:2014,
  author	= "{ATLAS Collaboration}",
  title		= "{Measurement of the $Z/\gamma^{\ast}$ boson transverse
		  momentum distribution in pp collisions at $\sqrt{s} = 7$
		  \TeV{} with the ATLAS detector}",
  journal	= "JHEP",
  volume	= "09",
  pages		= "145",
  doi		= "10.1007/JHEP09(2014)145",
  year		= "2014",
  eprint	= "1406.3660",
  archiveprefix	= "arXiv",
  primaryclass	= "hep-ex",
  reportnumber	= "CERN-PH-EP-2014-075",
  slaccitation	= "%%CITATION = ARXIV:1406.3660;%%"
}

@Article{	  Ball:2012cx,
  author	= "{NNPDF Collaboration}",
  title		= "{Parton distributions with LHC data}",
  journal	= "Nucl. Phys. B",
  volume	= "867",
  year		= "2013",
  pages		= "244",
  doi		= "10.1016/j.nuclphysb.2012.10.003",
  eprint	= "1207.1303",
  archiveprefix	= "arXiv",
  primaryclass	= "hep-ph",
  reportnumber	= "EDINBURGH-2012-08, IFUM-FT-997, FR-PHENO-2012-014,
		  RWTH-TTK-12-25, CERN-PH-TH-2012-037, SFB-CPP-12-47\,
		  --CERN-PH-TH-2012-037",
  slaccitation	= "%%CITATION = ARXIV:1207.1303;%%"
}

@Article{	  Ball:2014uwa,
  author	= "{NNPDF Collaboration}",
  title		= "{Parton distributions for the LHC Run II}",
  collaboration	= "NNPDF",
  journal	= "JHEP",
  volume	= "04",
  year		= "2015",
  pages		= "040",
  doi		= "10.1007/JHEP04(2015)040",
  eprint	= "1410.8849",
  archiveprefix	= "arXiv",
  primaryclass	= "hep-ph",
  reportnumber	= "EDINBURGH-2014-15, IFUM-1034-FT,
		  CERN-PH-TH-2013-253,OUTP-14-11P, CAVENDISH-HEP-14-11",
  slaccitation	= "%%CITATION = ARXIV:1410.8849;%%"
}

@Article{	  Baur:1994aj,
  author	= "Baur, U. and Han, Tao and Ohnemus, J.",
  title		= "{$W Z$ production at hadron colliders: Effects of
		  nonstandard $W W Z$ couplings and QCD corrections}",
  journal	= "Phys. Rev. D",
  volume	= "51",
  pages		= "3381",
  doi		= "10.1103/PhysRevD.51.3381",
  year		= "1995",
  eprint	= "hep-ph/9410266",
  archiveprefix	= "arXiv",
  primaryclass	= "hep-ph",
  reportnumber	= "FSU-HEP-941010, UCD-94-22",
  slaccitation	= "%%CITATION = HEP-PH/9410266;%%"
}

@Article{	  Biedermann:2017oae,
  author	= "Biedermann, Benedikt and Denner, Ansgar and Hofer, Lars",
  title		= "{Next-to-leading-order electroweak corrections to the
		  production of three charged leptons plus missing energy at
		  the LHC}",
  journal	= "JHEP",
  volume	= "10",
  year		= "2017",
  pages		= "043",
  doi		= "10.1007/JHEP10(2017)043",
  eprint	= "1708.06938",
  archiveprefix	= "arXiv",
  primaryclass	= "hep-ph",
  reportnumber	= "ICCUB-17-016",
  slaccitation	= "%%CITATION = ARXIV:1708.06938;%%"
}

@Article{	  Bierlich:2019rhm,
  author	= "Bierlich, Christian and others",
  title		= "{Robust Independent Validation of Experiment and Theory:
		  Rivet version 3}",
  eprint	= "1912.05451",
  archiveprefix	= "arXiv",
  primaryclass	= "hep-ph",
  reportnumber	= "MCnet-19-26",
  doi		= "10.21468/SciPostPhys.8.2.026",
  journal	= "SciPost Phys.",
  volume	= "8",
  pages		= "026",
  year		= "2020"
}

@Article{	  Buchmuller:1985jz,
  author	= "Buchmüller, W. and Wyler, D.",
  title		= "{Effective lagrangian analysis of new interactions and
		  flavor conservation}",
  journal	= "Nucl. Phys. B",
  volume	= "268",
  year		= "1986",
  pages		= "621",
  doi		= "10.1016/0550-3213(86)90262-2",
  reportnumber	= "CERN-TH-4254/85",
  slaccitation	= "%%CITATION = NUPHA,B268,621;%%"
}

@Article{	  Butterworth:2015oua,
  author	= "Butterworth, Jon and others",
  title		= "{PDF4LHC recommendations for LHC Run II}",
  journal	= "J. Phys. G",
  volume	= "43",
  year		= "2016",
  pages		= "023001",
  doi		= "10.1088/0954-3899/43/2/023001",
  eprint	= "1510.03865",
  archiveprefix	= "arXiv",
  primaryclass	= "hep-ph",
  reportnumber	= "OUTP-15-17P, SMU-HEP-15-12, TIF-UNIMI-2015-14,
		  LCTS-2015-27, CERN-PH-TH-2015-249",
  slaccitation	= "%%CITATION = ARXIV:1510.03865;%%"
}

@Article{	  Cascioli:2011va,
  author	= "Cascioli, Fabio and Maierh{\"o}fer, Philipp and Pozzorini,
		  Stefano",
  title		= "{Scattering Amplitudes with Open Loops}",
  journal	= "Phys. Rev. Lett.",
  volume	= "108",
  year		= "2012",
  pages		= "111601",
  doi		= "10.1103/PhysRevLett.108.111601",
  eprint	= "1111.5206",
  archiveprefix	= "arXiv",
  primaryclass	= "hep-ph",
  reportnumber	= "ZU-TH-23-11, LPN11-66",
  slaccitation	= "%%CITATION = ARXIV:1111.5206;%%"
}

@Article{	  Catani:2007vq,
  author	= "Catani, Stefano and Grazzini, Massimiliano",
  title		= "{Next-to-Next-to-Leading-Order Subtraction Formalism in
		  Hadron Collisions and its Application to Higgs-Boson
		  Production at the Large Hadron Collider}",
  journal	= "Phys. Rev. Lett.",
  volume	= "98",
  year		= "2007",
  pages		= "222002",
  doi		= "10.1103/PhysRevLett.98.222002",
  eprint	= "hep-ph/0703012",
  archiveprefix	= "arXiv",
  slaccitation	= "%%CITATION = HEP-PH/0703012;%%"
}

@Article{	  Catani:2012qa,
  author	= "Catani, Stefano and Cieri, Leandro and de Florian, Daniel
		  and Ferrera, Giancarlo and Grazzini, Massimiliano",
  title		= "{Vector-boson production at hadron colliders:
		  hard-collinear coefficients at the NNLO}",
  journal	= "Eur. Phys. J. C",
  volume	= "72",
  year		= "2012",
  pages		= "2195",
  doi		= "10.1140/epjc/s10052-012-2195-7",
  eprint	= "1209.0158",
  archiveprefix	= "arXiv",
  primaryclass	= "hep-ph",
  reportnumber	= "ZU-TH-16-12",
  slaccitation	= "%%CITATION = ARXIV:1209.0158;%%"
}

@Article{	  CDF:2012WZ,
  author	= "{CDF Collaboration}",
  title		= "{Measurement of the $WZ$ Cross Section and Triple Gauge
		  Couplings in $p \bar{p}$ Collisions at $\sqrt{s} = 1.96$
		  \TeV{}}",
  xcollaboration= "CDF",
  journal	= "Phys. Rev. D",
  volume	= "86",
  year		= "2012",
  pages		= "031104",
  doi		= "10.1103/PhysRevD.86.031104",
  eprint	= "1202.6629",
  archiveprefix	= "arXiv",
  primaryclass	= "hep-ex",
  reportnumber	= "FERMILAB-PUB-12-054-E",
  slaccitation	= "%%CITATION = ARXIV:1202.6629;%%"
}

@Article{	  CMS:2019efc,
  author	= "{CMS Collaboration}",
  collaboration	= "CMS",
  title		= "{Measurements of the pp $\to$ WZ inclusive and
		  differential production cross sections and constraints on
		  charged anomalous triple gauge couplings at $\sqrt{s} =$ 13
		  TeV}",
  eprint	= "1901.03428",
  archiveprefix	= "arXiv",
  primaryclass	= "hep-ex",
  reportnumber	= "CMS-SMP-18-002, CERN-EP-2018-322",
  doi		= "10.1007/JHEP04(2019)122",
  journal	= "JHEP",
  volume	= "04",
  pages		= "122",
  year		= "2019"
}

@article{CMS:2021icx,
    author = "{CMS Collaboration}",
    collaboration = "CMS",
    title = "{Measurement of the inclusive and differential WZ production cross sections, polarization angles, and triple gauge couplings in pp collisions at $ \sqrt{s} $ = 13 TeV}",
    eprint = "2110.11231",
    archivePrefix = "arXiv",
    primaryClass = "hep-ex",
    reportNumber = "CMS-SMP-20-014, CERN-EP-2021-163",
    doi = "10.1007/JHEP07(2022)032",
    journal = "JHEP",
    volume = "07",
    pages = "032",
    year = "2022"
}

@Article{	  CT10,
  author	= "Lai, Hung-Liang and Guzzi, Marco and Huston, Joey and Li,
		  Zhao and Nadolsky, Pavel M. and Pumplin, Jon and Yuan, C.
		  -P.",
  title		= "{New parton distributions for collider physics}",
  journal	= "Phys. Rev. D",
  volume	= "82",
  year		= "2010",
  pages		= "074024",
  doi		= "10.1103/PhysRevD.82.074024",
  eprint	= "1007.2241",
  archiveprefix	= "arXiv",
  primaryclass	= "hep-ph",
  reportnumber	= "MSUHEP-100707, SMU-HEP-10-10",
  slaccitation	= "%%CITATION = ARXIV:1007.2241;%%"
}

@Article{	  D0:2012WZ,
  author	= "{D0 Collaboration}",
  title		= "{A measurement of the $WZ$ and $ZZ$ production cross
		  sections using leptonic final states in 8.6 fb$^{-1}$ of
		  $p\bar{p}$ collisions}",
  xcollaboration= "D0",
  journal	= "Phys. Rev. D",
  volume	= "85",
  year		= "2012",
  pages		= "112005",
  doi		= "10.1103/PhysRevD.85.112005",
  eprint	= "1201.5652",
  archiveprefix	= "arXiv",
  primaryclass	= "hep-ex",
  reportnumber	= "FERMILAB-PUB-12-022-E",
  slaccitation	= "%%CITATION = ARXIV:1201.5652;%%"
}

@Article{	  DAgostini:1994zf,
  author	= "D'Agostini, G.",
  title		= "{A Multidimensional unfolding method based on Bayes'
		  theorem}",
  journal	= "Nucl. Instrum. Meth. A",
  volume	= "362",
  year		= "1995",
  pages		= "487",
  doi		= "10.1016/0168-9002(95)00274-X",
  reportnumber	= "DESY-94-099",
  slaccitation	= "%%CITATION = NUIMA,A362,487;%%"
}

@Article{	  DAPR-2021-01,
  author	= "{ATLAS Collaboration}",
  collaboration	= "ATLAS",
  title		= "{Luminosity determination in $pp$ collisions at
		  $\sqrt{s}=13$ TeV using the ATLAS detector at the LHC}",
  eprint	= "2212.09379",
  archiveprefix	= "arXiv",
  primaryclass	= "hep-ex",
  reportnumber	= "CERN-EP-2022-281",
  doi		= "10.1140/epjc/s10052-023-11747-w",
  journal	= "Eur. Phys. J. C",
  volume	= "83",
  number	= "10",
  pages		= "982",
  year		= "2023"
}

@Article{	  Degrande:2012wf,
  author	= "Degrande, Celine and Greiner, Nicolas and Kilian, Wolfgang
		  and Mattelaer, Olivier and Mebane, Harrison and Stelzer,
		  Tim and Willenbrock, Scott and Zhang, Cen",
  title		= "{Effective Field Theory: A Modern Approach to Anomalous
		  Couplings}",
  journal	= "Annals Phys.",
  volume	= "335",
  year		= "2013",
  pages		= "21",
  doi		= "10.1016/j.aop.2013.04.016",
  eprint	= "1205.4231",
  archiveprefix	= "arXiv",
  primaryclass	= "hep-ph",
  reportnumber	= "MPP-2011-149, SI-HEP-2011-17, CP3-12-25",
  slaccitation	= "%%CITATION = ARXIV:1205.4231;%%"
}

@article{Degrande:2020evl,
    author = "Degrande, C\'eline and Durieux, Gauthier and Maltoni, Fabio and Mimasu, Ken and Vryonidou, Eleni and Zhang, Cen",
    title = "{Automated one-loop computations in the standard model effective field theory}",
    eprint = "2008.11743",
    archivePrefix = "arXiv",
    primaryClass = "hep-ph",
    reportNumber = "CERN-TH-2020-140, CP3-20-42",
    doi = "10.1103/PhysRevD.103.096024",
    journal = "Phys. Rev. D",
    volume = "103",
    number = "9",
    pages = "096024",
    year = "2021"
}

@Article{	  Denner:2016kdg,
  author	= "Denner, Ansgar and Dittmaier, Stefan and Hofer, Lars",
  title		= "{Collier: A fortran-based complex one-loop library in
		  extended regularizations}",
  journal	= "Comput. Phys. Commun.",
  volume	= "212",
  year		= "2017",
  pages		= "220-238",
  doi		= "10.1016/j.cpc.2016.10.013",
  eprint	= "1604.06792",
  archiveprefix	= "arXiv",
  primaryclass	= "hep-ph",
  reportnumber	= "FR-PHENO-2016-003, ICCUB-16-016",
  slaccitation	= "%%CITATION = ARXIV:1604.06792;%%"
}

@Article{	  EFT,
  author	= "Degrande, Celine and Greiner, Nicolas and Kilian, Wolfgang
		  and Mattelaer, Olivier and Mebane, Harrison and Stelzer,
		  Tim and Willenbrock, Scott and Zhang, Cen",
  title		= "{Effective Field Theory: A Modern Approach to Anomalous
		  Couplings}",
  journal	= "Annals Phys.",
  volume	= "335",
  year		= "2013",
  pages		= "21-32",
  doi		= "10.1016/j.aop.2013.04.016",
  eprint	= "1205.4231",
  archiveprefix	= "arXiv",
  primaryclass	= "hep-ph",
  reportnumber	= "MPP-2011-149, SI-HEP-2011-17, CP3-12-25",
  slaccitation	= "%%CITATION = ARXIV:1205.4231;%%"
}

@Article{	  Fastjet,
  author	= "Cacciari, Matteo and Salam, Gavin P. and Soyez, Gregory",
  title		= "{FastJet user manual}",
  journal	= "Eur. Phys. J. C",
  volume	= "72",
  year		= "2012",
  pages		= "1896",
  doi		= "10.1140/epjc/s10052-012-1896-2",
  eprint	= "1111.6097",
  archiveprefix	= "arXiv",
  primaryclass	= "hep-ph",
  reportnumber	= "CERN-PH-TH-2011-297",
  slaccitation	= "%%CITATION = ARXIV:1111.6097;%%"
}

@Article{	  Frederix:2012ps,
  author	= "Frederix, Rikkert and Frixione, Stefano",
  title		= "{Merging meets matching in MC@NLO}",
  eprint	= "1209.6215",
  archiveprefix	= "arXiv",
  primaryclass	= "hep-ph",
  reportnumber	= "CERN-PH-TH-2012-247, ZU-TH-21-12",
  doi		= "10.1007/JHEP12(2012)061",
  journal	= "JHEP",
  volume	= "12",
  pages		= "061",
  year		= "2012"
}

@Article{	  Gehrmann:2015ora,
  author	= "Gehrmann, Thomas and von Manteuffel, Andreas and Tancredi,
		  Lorenzo",
  title		= "{The two-loop helicity amplitudes for
		  $q\overline{q}^{\prime} \rightarrow {V}_1{V}_2 \rightarrow
		  4 $ leptons}",
  journal	= "JHEP",
  volume	= "09",
  year		= "2015",
  pages		= "128",
  doi		= "10.1007/JHEP09(2015)128",
  eprint	= "1503.04812",
  archiveprefix	= "arXiv",
  primaryclass	= "hep-ph",
  reportnumber	= "ZU-TH-03-15, MITP-15-011, TTP15-011",
  slaccitation	= "%%CITATION = ARXIV:1503.04812;%%"
}

@Article{	  Glazov:2005rn,
  author	= "Glazov, A.",
  title		= "{Averaging of DIS cross section data}",
  journal	= "AIP~Conf.~Proc.",
  volume	= "792",
  pages		= "237",
  doi		= "10.1063/1.2122026",
  year		= "2005",
  slaccitation	= "%%CITATION = APCPC,792,237;%%"
}

@Article{	  Gleisberg:2008fv,
  author	= "Gleisberg, Tanju and H{\"o}che, Stefan",
  title		= "{Comix, a new matrix element generator}",
  journal	= "JHEP",
  volume	= "12",
  year		= "2008",
  pages		= "039",
  doi		= "10.1088/1126-6708/2008/12/039",
  eprint	= "0808.3674",
  archiveprefix	= "arXiv",
  primaryclass	= "hep-ph",
  reportnumber	= "SLAC-PUB-13232, IPPP-08-31, DCPT-08-62, MCNET-08-08",
  slaccitation	= "%%CITATION = ARXIV:0808.3674;%%"
}

@article{Grazzini:2016swo,
    author = "Grazzini, Massimiliano and Kallweit, Stefan and Rathlev, Dirk and Wiesemann, Marius",
    title = "{$W^{\pm}Z$ production at hadron colliders in NNLO QCD}",
    eprint = "1604.08576",
    archivePrefix = "arXiv",
    primaryClass = "hep-ph",
    reportNumber = "ZU-TH-15-16, MITP-16-037, NSF-KITP-16-046, DESY-16-074",
    doi = "10.1016/j.physletb.2016.08.017",
    journal = "Phys. Lett. B",
    volume = "761",
    pages = "179--183",
    year = "2016"
}

@Article{	  Grazzini:2017ckn,
  author	= "{M.~Grazzini, S.~Kallweit, D.~Rathlev and M.~Wiesemann}",
  title		= "{$W^\pm Z$ production at the LHC: fiducial cross sections
		  and distributions in NNLO QCD}",
  journal	= "JHEP",
  volume	= "05",
  year		= "2017",
  pages		= "139",
  doi		= "10.1007/JHEP05(2017)139",
  eprint	= "1703.09065",
  archiveprefix	= "arXiv",
  primaryclass	= "hep-ph",
  reportnumber	= "ZU-TH-06-17, CERN-TH-2017-065",
  slaccitation	= "%%CITATION = ARXIV:1703.09065;%%"
}

@Article{	  Hirschi:2015iia,
  author	= "Hirschi, Valentin and Mattelaer, Olivier",
  title		= "{Automated event generation for loop-induced processes}",
  eprint	= "1507.00020",
  archiveprefix	= "arXiv",
  primaryclass	= "hep-ph",
  reportnumber	= "IPPP-15-35, DCPT-15-70, MCNET-15-14",
  doi		= "10.1007/JHEP10(2015)146",
  journal	= "JHEP",
  volume	= "10",
  pages		= "146",
  year		= "2015"
}

@Article{	  Hocker:2007ht,
  author	= "Hoecker, Andreas and Speckmayer, Peter and Stelzer, Joerg
		  and Therhaag, Jan and von Toerne, Eckhard and Voss, Helge",
  title		= "{TMVA - Toolkit for Multivariate Data Analysis}",
  journal	= "PoS ACAT",
  volume	= "",
  year		= "2007",
  pages		= "040",
  eprint	= "physics/0703039",
  archiveprefix	= "arXiv",
  slaccitation	= "%%CITATION = PHYSICS/0703039;%%"
}

@Article{	  Hoeche:2012yf,
  author	= "H{\"o}che, Stefan and Krauss, Frank and Sch{\"o}nherr,
		  Marek and Siegert, Frank",
  title		= "{QCD matrix elements + parton showers. The NLO case}",
  journal	= "JHEP",
  volume	= "04",
  year		= "2013",
  pages		= "027",
  doi		= "10.1007/JHEP04(2013)027",
  eprint	= "1207.5030",
  archiveprefix	= "arXiv",
  primaryclass	= "hep-ph",
  reportnumber	= "SLAC-PUB-15191, IPPP-12-52, DCPT-12-104, LPN12-081,
		  FR-PHENO-2012-017, MCNET-12-09, --FR-PHENO-2012-017",
  slaccitation	= "%%CITATION = ARXIV:1207.5030;%%"
}

@Article{	  Khachatryan:2016poo,
  author	= "{CMS Collaboration}",
  title		= "{Measurement of the WZ production cross section in pp
		  collisions at $\sqrt{s} = 7$ and 8 $\,\text{TeV}$ and
		  search for anomalous triple gauge couplings at $\sqrt{s} =
		  8\,\text{TeV} $}",
  collaboration	= "CMS",
  journal	= "Eur. Phys. J. C",
  volume	= "77",
  year		= "2017",
  number	= "4",
  pages		= "236",
  doi		= "10.1140/epjc/s10052-017-4730-z",
  eprint	= "1609.05721",
  archiveprefix	= "arXiv",
  primaryclass	= "hep-ex",
  reportnumber	= "CMS-SMP-14-014, CERN-EP-2016-205",
  slaccitation	= "%%CITATION = ARXIV:1609.05721;%%"
}

@Article{	  Lonnblad:2001iq,
  author	= "L{\"o}nnblad, Leif",
  title		= "{Correcting the Colour-Dipole Cascade Model with Fixed
		  Order Matrix Elements}",
  eprint	= "hep-ph/0112284",
  archiveprefix	= "arXiv",
  reportnumber	= "LU-TP-01-38",
  doi		= "10.1088/1126-6708/2002/05/046",
  journal	= "JHEP",
  volume	= "05",
  pages		= "046",
  year		= "2002"
}

@Article{	  LUCID2,
  author	= {G. Avoni and others},
  title		= {The new LUCID-2 detector for luminosity measurement and
		  monitoring in ATLAS},
  journal	= {JINST},
  volume	= {13},
  number	= {07},
  pages		= {P07017},
  doi		= "10.1088/1748-0221/13/07/P07017",
  year		= {2018}
}

@Article{	  Maguire:2017ypu,
  author	= "Maguire, Eamonn and Heinrich, Lukas and Watt, Graeme",
  editor	= "Mount, Richard and Tull, Craig",
  title		= "{HEPData: a repository for high energy physics data}",
  eprint	= "1704.05473",
  archiveprefix	= "arXiv",
  primaryclass	= "hep-ex",
  reportnumber	= "IPPP-17-31",
  doi		= "10.1088/1742-6596/898/10/102006",
  journal	= "J. Phys. Conf. Ser.",
  volume	= "898",
  number	= "10",
  pages		= "102006",
  year		= "2017"
}

@Article{	  Malaescu:2011yg,
  author	= "Malaescu, Bogdan",
  title		= "{An Iterative, Dynamically Stabilized(IDS) Method of Data
		  Unfolding}",
  journal	= "{Proceedings of the PHYSTAT 2011 Workshop, CERN, Geneva,
		  Switzerland}",
  xvolume	= "362",
  year		= "2011",
  pages		= "271",
  eprint	= "1106.3107",
  archiveprefix	= "arXiv",
  primaryclass	= "physics.data-an",
  reportnumber	= "CERN-2011-00"
}

@Article{	  Melia:2011tj,
  author	= "Melia, Tom and Nason, Paolo and R{\"o}ntsch, Raoul and
		  Zanderighi, Giulia",
  title		= "{$W^{+}W^{-}$, $WZ$ and $ZZ$ production in the POWHEG
		  BOX}",
  journal	= "JHEP",
  volume	= "11",
  year		= "2011",
  pages		= "078",
  doi		= "10.1007/JHEP11(2011)078",
  eprint	= "1107.5051",
  archiveprefix	= "arXiv",
  primaryclass	= "hep-ph",
  slaccitation	= "%%CITATION = ARXIV:1107.5051;%%"
}

@Article{	  Panico:2017frx,
  author	= "Panico, Giuliano and Riva, Francesco and Wulzer, Andrea",
  title		= "{Diboson interference resurrection}",
  journal	= "Phys. Lett. B",
  volume	= "776",
  year		= "2018",
  pages		= "473-480",
  doi		= "10.1016/j.physletb.2017.11.068",
  eprint	= "1708.07823",
  archiveprefix	= "arXiv",
  primaryclass	= "hep-ph",
  reportnumber	= "CERN-TH-2017-185",
  slaccitation	= "%%CITATION = ARXIV:1708.07823;%%"
}

@Article{	  PDF4LHC,
  author	= "Butterworth, Jon and others",
  title		= "{PDF4LHC recommendations for LHC Run II}",
  journal	= "J. Phys. G",
  volume	= "43",
  year		= "2016",
  pages		= "023001",
  doi		= "10.1088/0954-3899/43/2/023001",
  eprint	= "1510.03865",
  archiveprefix	= "arXiv",
  primaryclass	= "hep-ph",
  reportnumber	= "OUTP-15-17P, SMU-HEP-15-12, TIF-UNIMI-2015-14,
		  LCTS-2015-27, CERN-PH-TH-2015-249",
  slaccitation	= "%%CITATION = ARXIV:1510.03865;%%"
}

@Article{	  PDG,
  author	= "Beringer, J. and others",
  title		= "{The Review of Particle Physics}",
  journal	= "Phys. Rev. D",
  volume	= "86",
  year		= "2012"
}

@Article{	  PhysRevD.98.030001,
  author	= "{Particle Data Group}",
  title		= "{Review of Particle Physics}",
  collaboration	= "Particle Data Group",
  journal	= "Phys. Rev. D",
  volume	= "98",
  pages		= "030001",
  numpages	= "1898",
  year		= "2018",
  month		= "Aug",
  publisher	= "American Physical Society",
  doi		= "10.1103/PhysRevD.98.030001"
}

@Article{	  powheg,
  author	= "Frixione, Stefano and Nason, Paolo and Oleari, Carlo",
  title		= "{Matching NLO QCD computations with Parton Shower
		  simulations: the POWHEG method}",
  journal	= "JHEP",
  volume	= "11",
  year		= "2007",
  pages		= "070",
  doi		= "10.1088/1126-6708/2007/11/070",
  eprint	= "0709.2092",
  archiveprefix	= "arXiv",
  primaryclass	= "hep-ph",
  reportnumber	= "BICOCCA-FT-07-9, GEF-TH-21-2007",
  slaccitation	= "%%CITATION = ARXIV:0709.2092;%%"
}

@Article{	  powheg1:2004,
  author	= "Nason, Paolo",
  title		= "{A New method for combining NLO QCD with shower Monte
		  Carlo algorithms}",
  journal	= "JHEP",
  volume	= "11",
  year		= "2004",
  pages		= "040",
  doi		= "10.1088/1126-6708/2004/11/040",
  eprint	= "hep-ph/0409146",
  archiveprefix	= "arXiv",
  primaryclass	= "hep-ph",
  reportnumber	= "BICOCCA-FT-04-11",
  slaccitation	= "%%CITATION = HEP-PH/0409146;%%"
}

@Article{	  Pumplin:2002vw,
  author	= "Pumplin, J. and Stump, D. R. and Huston, J. and Lai, H. L.
		  and Nadolsky, Pavel M. and Tung, W. K.",
  title		= "{New Generation of Parton Distributions with Uncertainties
		  from Global QCD Analysis}",
  journal	= "JHEP",
  volume	= "07",
  year		= "2002",
  pages		= "012",
  doi		= "10.1088/1126-6708/2002/07/012",
  eprint	= "hep-ph/0201195",
  archiveprefix	= "arXiv",
  primaryclass	= "hep-ph",
  reportnumber	= "MSU-HEP-011101",
  slaccitation	= "%%CITATION = HEP-PH/0201195;%%"
}

@Article{	  RooUnfold,
  author	= "Adye, T.",
  title		= "{Unfolding algorithms and tests using RooUnfold}",
  journal	= "{Proceedings of the PHYSTAT 2011 Workshop, CERN, Geneva,
		  Switzerland}",
  xvolume	= "362",
  year		= "2011",
  pages		= "313",
  eprint	= "1105.1160",
  archiveprefix	= "arXiv",
  primaryclass	= "physics.data-an",
  reportnumber	= "CERN-2011-00"
}

@Article{	  Sapeta:2012,
  author	= "Campanario, Francisco and Sapeta, Sebastian",
  title		= "{WZ production beyond NLO for high-pT observables}",
  journal	= "Phys. Lett. B",
  volume	= "718",
  year		= "2012",
  pages		= "100",
  doi		= "10.1016/j.physletb.2012.10.013",
  eprint	= "1209.4595",
  archiveprefix	= "arXiv",
  primaryclass	= "hep-ph",
  reportnumber	= "FTUV-12-0920, KA-TP-36-2012, LPN12-102, SFB-CPP-12-70,
		  IPPP-12-70, DCPT-12-140, --DCPT-12-140, --DCPT-12-140",
  slaccitation	= "%%CITATION = ARXIV:1209.4595;%%"
}

@Article{	  Schumann:2007mg,
  author	= "Schumann, Steffen and Krauss, Frank",
  title		= "{A Parton shower algorithm based on Catani-Seymour dipole
		  factorisation}",
  journal	= "JHEP",
  volume	= "03",
  year		= "2008",
  pages		= "038",
  doi		= "10.1088/1126-6708/2008/03/038",
  eprint	= "0709.1027",
  archiveprefix	= "arXiv",
  primaryclass	= "hep-ph",
  reportnumber	= "DCPT-07-86, IPPP-07-43",
  slaccitation	= "%%CITATION = ARXIV:0709.1027;%%"
}

@Article{	  Sjostrand:2007gs,
  author	= "Sj{\"o}strand, Torbjorn and Mrenna, Stephen and Skands,
		  Peter Z.",
  title		= "{A Brief Introduction to PYTHIA 8.1}",
  journal	= "Comput. Phys. Commun.",
  volume	= "178",
  year		= "2008",
  pages		= "852",
  doi		= "10.1016/j.cpc.2008.01.036",
  eprint	= "0710.3820",
  archiveprefix	= "arXiv",
  primaryclass	= "hep-ph",
  reportnumber	= "CERN-LCGAPP-2007-04, LU-TP-07-28,
		  FERMILAB-PUB-07-512-CD-T",
  slaccitation	= "%%CITATION = ARXIV:0710.3820;%%"
}

@Article{	  Sjostrand:2014zea,
 author	= "Sj{\"o}strand, Torbjorn and Ask, Stefan and Christiansen,
		  Jesper R. and Corke, Richard and Desai, Nishita and Ilten,
		  Philip and Mrenna, Stephen and Prestel, Stefan and
		  Rasmussen, Christine O. and Skands, Peter Z.",
  title		= "{An introduction to \textsc{PYTHIA} 8.2}",
  journal	= "Comput. Phys. Commun.",
  volume	= "191",
  year		= "2015",
  pages		= "159-177",
  doi		= "10.1016/j.cpc.2015.01.024",
  eprint	= "1410.3012",
  archiveprefix	= "arXiv",
  primaryclass	= "hep-ph",
  reportnumber	= "LU-TP-14-36, MCNET-14-22, CERN-PH-TH-2014-190,
		  FERMILAB-PUB-14-316-CD, DESY-14-178, SLAC-PUB-16122",
  slaccitation	= "%%CITATION = ARXIV:1410.3012;%%"
}

@TechReport{	  TruthWS,
  author	= "{ATLAS Collaboration}",
  title		= "{Proposal for truth particle observable definitions in
		  physics measurements}",
  institution	= "CERN",
  address	= "Geneva",
  number	= "ATL-PHYS-PUB-2015-013",
  month		= "Jun",
  year		= "2015",
  reportnumber	= "ATL-PHYS-PUB-2015-013",
  url		= "http://cds.cern.ch/record/2022743"
}

@Article{	  WZPol36fb,
  author	= "{ATLAS Collaboration}",
  title		= "{Measurement of $W^{\pm}Z$ production cross sections and
		  gauge boson polarisation in $pp$ collisions at $\sqrt{s} =
		  13$ TeV with the ATLAS detector}",
  eprint	= "1902.05759",
  archiveprefix	= "arXiv",
  primaryclass	= "hep-ex",
  reportnumber	= "CERN-EP-2018-327",
  doi		= "10.1140/epjc/s10052-019-7027-6",
  journal	= "Eur. Phys. J. C",
  volume	= "79",
  number	= "6",
  pages		= "535",
  year		= "2019"
}

@Article{	  WZVBS36fb,
  author	= "{ATLAS Collaboration}",
  title		= "{Observation of electroweak $W^{\pm}Z$ boson pair
		  production in association with two jets in $pp$ collisions
		  at $\sqrt{s} =$ 13 TeV with the ATLAS detector}",
  eprint	= "1812.09740",
  archiveprefix	= "arXiv",
  primaryclass	= "hep-ex",
  reportnumber	= "CERN-EP-2018-286",
  doi		= "10.1016/j.physletb.2019.05.012",
  journal	= "Phys. Lett. B",
  volume	= "793",
  pages		= "469",
  year		= "2019"
}

@article{Grazzini:2019jkl,
    author = "Grazzini, Massimiliano and Kallweit, S. and Lindert, Jonas M. and Pozzorini, Stefano and Wiesemann, Marius",
    title = "{NNLO QCD + NLO EW with Matrix+OpenLoops: precise predictions for vector-boson pair production}",
    eprint = "1912.00068",
    archivePrefix = "arXiv",
    primaryClass = "hep-ph",
    reportNumber = "MPP-2019-175, ZU-TH 49/19",
    doi = "10.1007/JHEP02(2020)087",
    journal = "JHEP",
    volume = "02",
    pages = "087",
    year = "2020",
    addendum = "{(We thank M.~Grazzini, S.~Kallweit, J.M.~Lindert, S.~Pozzorini and
		  M.~Wiesemann for useful discussions)}"
}

@article{ATLAS:2020nzk,
    author = "{ATLAS Collaboration}",
    collaboration = "ATLAS",
    title = "{Differential cross-section measurements for the electroweak production of dijets in association with a $Z$ boson in proton\textendash{}proton collisions at ATLAS}",
    eprint = "2006.15458",
    archivePrefix = "arXiv",
    primaryClass = "hep-ex",
    reportNumber = "CERN-EP-2020-045",
    doi = "10.1140/epjc/s10052-020-08734-w",
    journal = "Eur. Phys. J. C",
    volume = "81",
    number = "2",
    pages = "163",
    year = "2021"
}

@article{ATLAS:2024wfv,
    author = "{ATLAS Collaboration}",
    collaboration = "ATLAS",
    title = "{Differential cross-section measurements of Higgs boson production in the H \textrightarrow{} \ensuremath{\tau}$^{+}$\ensuremath{\tau}$^{-}$ decay channel in pp collisions at $ \sqrt{s} $ = 13 TeV with the ATLAS detector}",
    eprint = "2407.16320",
    archivePrefix = "arXiv",
    primaryClass = "hep-ex",
    reportNumber = "CERN-EP-2024-198",
    doi = "10.1007/JHEP03(2025)010",
    journal = "JHEP",
    volume = "03",
    pages = "010",
    year = "2025"
}

@article{ATLAS:2023mqy,
    author = "{ATLAS Collaboration}",
    collaboration = "ATLAS",
    title = "{Test of CP-invariance of the Higgs boson in vector-boson fusion production and in its decay into four leptons}",
    eprint = "2304.09612",
    archivePrefix = "arXiv",
    primaryClass = "hep-ex",
    reportNumber = "CERN-EP-2023-030",
    doi = "10.1007/JHEP05(2024)105",
    journal = "JHEP",
    volume = "05",
    pages = "105",
    year = "2024"
}

@article{ATLAS:2022tan,
    author = "{ATLAS Collaboration}",
    collaboration = "ATLAS",
    title = "{Test of CP Invariance in Higgs Boson Vector-Boson-Fusion Production Using the H\textrightarrow{}\ensuremath{\gamma}\ensuremath{\gamma} Channel with the ATLAS Detector}",
    eprint = "2208.02338",
    archivePrefix = "arXiv",
    primaryClass = "hep-ex",
    reportNumber = "CERN-EP-2022-134",
    doi = "10.1103/PhysRevLett.131.061802",
    journal = "Phys. Rev. Lett.",
    volume = "131",
    number = "6",
    pages = "061802",
    year = "2023"
}

@article{Degrande:2021zpv,
    author = "Degrande, C\'eline and Touch\`eque, Julien",
    title = "{A reduced basis for CP violation in SMEFT at colliders and its application to diboson production}",
    eprint = "2110.02993",
    archivePrefix = "arXiv",
    primaryClass = "hep-ph",
    doi = "10.1007/JHEP04(2022)032",
    journal = "JHEP",
    volume = "04",
    pages = "032",
    year = "2022"
}

@article{ATLAS:2022oge,
	author = "{ATLAS Collaboration}",
	collaboration = "ATLAS",
	title = "{Observation of gauge boson joint-polarisation states in $W^\pm Z$ production from pp collisions at $\sqrt{s}=$13 TeV with the ATLAS detector}",
	eprint = "2211.09435",
	archivePrefix = "arXiv",
	primaryClass = "hep-ex",
	reportNumber = "CERN-EP-2022-202",
	doi = "10.1016/j.physletb.2023.137895",
	journal = "Phys. Lett. B",
	volume = "843",
	pages = "137895",
	year = "2023"
}

@article{CMS:2024ild,
    author = "{CMS Collaboration}",
    collaboration = "CMS",
    title = "{Measurement of the inclusive WZ production cross section in pp collisions at $ \sqrt{\textrm{s}} $ = 13.6 TeV}",
    eprint = "2412.02477",
    archivePrefix = "arXiv",
    primaryClass = "hep-ex",
    reportNumber = "CMS-SMP-24-005, CERN-EP-2024-293",
    doi = "10.1007/JHEP04(2025)115",
    journal = "JHEP",
    volume = "04",
    pages = "115",
    year = "2025"
}

@article{Grzadkowski:2010es,
    author = "Grzadkowski, B. and Iskrzyński, M. and Misiak, M. and Rosiek, J.",
    title = "{Dimension-Six Terms in the Standard Model Lagrangian}",
    eprint = "1008.4884",
    archivePrefix = "arXiv",
    primaryClass = "hep-ph",
    reportNumber = "IFT-9-2010, TTP10-35",
    doi = "10.1007/JHEP10(2010)085",
    journal = "JHEP",
    volume = "10",
    pages = "085",
    year = "2010"
}

@article{Celada:2024mcf,
    author = "Celada, Eugenia and Giani, Tommaso and ter Hoeve, Jaco and Mantani, Luca and Rojo, Juan and Rossia, Alejo N. and Thomas, Marion O. A. and Vryonidou, Eleni",
    title = "{Mapping the SMEFT at high-energy colliders: from LEP and the (HL-)LHC to the FCC-ee}",
    eprint = "2404.12809",
    archivePrefix = "arXiv",
    primaryClass = "hep-ph",
    doi = "10.1007/JHEP09(2024)091",
    journal = "JHEP",
    volume = "09",
    pages = "091",
    year = "2024"
}

@article{Brivio:2017btx,
    author = "Brivio, Ilaria and Jiang, Yun and Trott, Michael",
    title = "{The SMEFTsim package, theory and tools}",
    eprint = "1709.06492",
    archivePrefix = "arXiv",
    primaryClass = "hep-ph",
    doi = "10.1007/JHEP12(2017)070",
    journal = "JHEP",
    volume = "12",
    pages = "070",
    year = "2017"
}

@article{Brivio:2020onw,
    author = "Brivio, Ilaria",
    title = "{SMEFTsim 3.0 \textemdash{} a practical guide}",
    eprint = "2012.11343",
    archivePrefix = "arXiv",
    primaryClass = "hep-ph",
    doi = "10.1007/JHEP04(2021)073",
    journal = "JHEP",
    volume = "04",
    pages = "073",
    year = "2021"
}

@article{Sakharov:1967dj,
    author = "Sakharov, A. D.",
    title = "{Violation of CP Invariance, C asymmetry, and baryon asymmetry of the universe}",
    doi = "10.1070/PU1991v034n05ABEH002497",
    journal = "Pisma Zh. Eksp. Teor. Fiz.",
    volume = "5",
    pages = "32--35",
    year = "1967"
}

@article{CMS:2021pqj,
    author = "{CMS Collaboration}",
    collaboration = "CMS",
    title = "{Measurements of the electroweak diboson production cross sections in proton-proton collisions at $\sqrt{s} =$ 5.02 TeV using leptonic decays}",
    eprint = "2107.01137",
    archivePrefix = "arXiv",
    primaryClass = "hep-ex",
    reportNumber = "CMS-SMP-20-012, CERN-EP-2021-111",
    doi = "10.1103/PhysRevLett.127.191801",
    journal = "Phys. Rev. Lett.",
    volume = "127",
    number = "19",
    pages = "191801",
    year = "2021"
}

@article{Kallweit:2020gva,
    author = "Kallweit, Stefan and Re, Emanuele and Rottoli, Luca and Wiesemann, Marius",
    title = "{Accurate single- and double-differential resummation of colour-singlet processes with MATRIX+RADISH: W$^{+}$W$^{-}$ production at the LHC}",
    eprint = "2004.07720",
    archivePrefix = "arXiv",
    primaryClass = "hep-ph",
    reportNumber = "MPP-2020-53, LAPTH-012/20",
    doi = "10.1007/JHEP12(2020)147",
    journal = "JHEP",
    volume = "12",
    pages = "147",
    year = "2020"
}

@article{Grazzini:2015wpa,
    author = "Grazzini, Massimiliano and Kallweit, Stefan and Rathlev, Dirk and Wiesemann, Marius",
    title = "{Transverse-momentum resummation for vector-boson pair production at NNLL+NNLO}",
    eprint = "1507.02565",
    archivePrefix = "arXiv",
    primaryClass = "hep-ph",
    reportNumber = "ZU-TH-20-15, MITP-15-051",
    doi = "10.1007/JHEP08(2015)154",
    journal = "JHEP",
    volume = "08",
    pages = "154",
    year = "2015"
}

@article{Brivio:2022pyi,
    author = "Brivio, Ilaria and others",
    title = "{Truncation, validity, uncertainties}",
    eprint = "2201.04974",
    archivePrefix = "arXiv",
    primaryClass = "hep-ph",
    reportNumber = "CERN-LHCEFTWG-2021-002, CERN-LPCC-2022-01",
    month = "1",
    year = "2022"
}

@article{Cranmer:1456844,
    author = "{ROOT Collaboration}",
    title = "{HistFactory: A tool for creating statistical models for use with RooFit and RooStats}",
    journal = "CERN-OPEN-2012-016",
    year = "2012",
    doi = "10.17181/CERN-OPEN-2012-016",
}

@article{ATLAS:2023dxj,
    author = "{ATLAS Collaboration}",
    collaboration = "ATLAS",
    title = "{Electron and photon efficiencies in LHC Run 2 with the ATLAS experiment}",
    eprint = "2308.13362",
    archivePrefix = "arXiv",
    primaryClass = "hep-ex",
    reportNumber = "CERN-EP-2023-182",
    doi = "10.1007/JHEP05(2024)162",
    journal = "JHEP",
    volume = "05",
    pages = "162",
    year = "2024"
}

@article{ATLAS:2023mnw,
    author = "{ATLAS Collaboration}",
    collaboration = "ATLAS",
    title = "{Electron and photon energy calibration with the ATLAS detector using LHC Run~2 data}",
    eprint = "2309.05471",
    archivePrefix = "arXiv",
    primaryClass = "hep-ex",
    reportNumber = "CERN-EP-2023-128",
    doi = "10.1088/1748-0221/19/02/P02009",
    journal = "JINST",
    volume = "19",
    number = "02",
    pages = "P02009",
    year = "2024"
}

@article{Leung:1984ni,
    author = "Leung, Chung Ngoc and Love, S. T. and Rao, S.",
    title = "{Low-energy manifestations of a new interactions scale: Operator analysis}",
    reportNumber = "FERMILAB-PUB-84-074-T",
    doi = "10.1007/BF01588041",
    journal = "Z. Phys. C",
    volume = "31",
    pages = "433",
    year = "1986"
}

@article{Brivio:2017vri,
    author = "Brivio, Ilaria and Trott, Michael",
    title = "{The Standard Model as an Effective Field Theory}",
    eprint = "1706.08945",
    archivePrefix = "arXiv",
    primaryClass = "hep-ph",
    doi = "10.1016/j.physrep.2018.11.002",
    journal = "Phys. Rept.",
    volume = "793",
    pages = "1",
    year = "2019"
}

@article{Isidori:2023pyp,
    author = "Isidori, Gino and Wilsch, Felix and Wyler, Daniel",
    title = "{The standard model effective field theory at work}",
    eprint = "2303.16922",
    archivePrefix = "arXiv",
    primaryClass = "hep-ph",
    reportNumber = "ZU-TH 14/23",
    doi = "10.1103/RevModPhys.96.015006",
    journal = "Rev. Mod. Phys.",
    volume = "96",
    number = "1",
    pages = "015006",
    year = "2024"
}

@article{ATLAS:2020jwz,
    author = "{ATLAS Collaboration}",
    collaboration = "ATLAS",
    title = "{Measurement of the associated production of a Higgs boson decaying into $b$-quarks with a vector boson at high transverse momentum in $pp$ collisions at $\sqrt{s} = 13$ TeV with the ATLAS detector}",
    eprint = "2008.02508",
    archivePrefix = "arXiv",
    primaryClass = "hep-ex",
    reportNumber = "CERN-EP-2020-093",
    doi = "10.1016/j.physletb.2021.136204",
    journal = "Phys. Lett. B",
    volume = "816",
    pages = "136204",
    year = "2021"
}

@article{ATLAS:2020fcp,
    author = "{ATLAS Collaboration}",
    collaboration = "ATLAS",
    title = "{Measurements of $WH$ and $ZH$ production in the $H \rightarrow b\bar{b}$ decay channel in $pp$ collisions at 13 TeV with the ATLAS detector}",
    eprint = "2007.02873",
    archivePrefix = "arXiv",
    primaryClass = "hep-ex",
    reportNumber = "CERN-EP-2020-087",
    doi = "10.1140/epjc/s10052-020-08677-2",
    journal = "Eur. Phys. J. C",
    volume = "81",
    number = "2",
    pages = "178",
    year = "2021"
}

@article{ATLAS:2022tnm,
    author = "{ATLAS Collaboration}",
    collaboration = "ATLAS",
    title = "{Measurement of the properties of Higgs boson production at $\sqrt{s} = 13$ TeV in the $H\to\gamma\gamma$ channel using $139$ fb$^{-1}$ of $pp$ collision data with the ATLAS experiment}",
    eprint = "2207.00348",
    archivePrefix = "arXiv",
    primaryClass = "hep-ex",
    reportNumber = "CERN-EP-2022-094",
    doi = "10.1007/JHEP07(2023)088",
    journal = "JHEP",
    volume = "07",
    pages = "088",
    year = "2023"
}

@article{ALEPH:2005ab,
    author = "Schael, S. and others",
    collaboration = "ALEPH, DELPHI, L3, OPAL, SLD, LEP Electroweak Working Group, SLD Electroweak Group, SLD Heavy Flavour Group",
    title = "{Precision electroweak measurements on the $Z$ resonance}",
    eprint = "hep-ex/0509008",
    archivePrefix = "arXiv",
    reportNumber = "SLAC-R-774",
    doi = "10.1016/j.physrep.2005.12.006",
    journal = "Phys. Rept.",
    volume = "427",
    pages = "257--454",
    year = "2006"
}

@article{ATLAS:2025dhf,
    author = "{ATLAS Collaboration}",
    collaboration = "ATLAS",
    title = "{Measurements of $W^+W^-$ production cross-sections in pp collisions at $\sqrt{s}=13$ TeV with the ATLAS detector}",
    eprint = "2505.11310",
    archivePrefix = "arXiv",
    primaryClass = "hep-ex",
    reportNumber = "CERN-EP-2025-091",
    doi = "10.1007/JHEP08(2025)142",
    journal = "JHEP",
    volume = "08",
    pages = "142",
    year = "2025"
}

@article{CMS:2025ugn,
    author = "{CMS Collaboration}",
    collaboration = "CMS",
    title = "{Combined effective field theory interpretation of Higgs boson, electroweak vector boson, top quark, and multi-jet measurements}",
    eprint = "2504.02958",
    archivePrefix = "arXiv",
    primaryClass = "hep-ex",
    reportNumber = "CMS-SMP-24-003, CERN-EP-2025-035",
    month = "4",
    year = "2025"
}

@article{CMS:2021cxr,
    author = "{CMS Collaboration}",
    collaboration = "CMS",
    title = "{Measurement of W$^\pm\gamma$ differential cross sections in proton-proton collisions at $\sqrt{s}$ = 13 TeV and effective field theory constraints}",
    eprint = "2111.13948",
    archivePrefix = "arXiv",
    primaryClass = "hep-ex",
    reportNumber = "CMS-SMP-20-005, CERN-EP-2021-219",
    doi = "10.1103/PhysRevD.105.052003",
    journal = "Phys. Rev. D",
    volume = "105",
    number = "5",
    pages = "052003",
    year = "2022"
}

@article{Azatov:2016sqh,
    author = "Azatov, Aleksandr and Contino, Roberto and Machado, Camila S. and Riva, Francesco",
    title = "{Helicity selection rules and noninterference for BSM amplitudes}",
    eprint = "1607.05236",
    archivePrefix = "arXiv",
    primaryClass = "hep-ph",
    reportNumber = "CERN-TH-2016-165",
    doi = "10.1103/PhysRevD.95.065014",
    journal = "Phys. Rev. D",
    volume = "95",
    number = "6",
    pages = "065014",
    year = "2017"
}

@article{Falkowski:2016cxu,
    author = "Falkowski, Adam and González-Alonso, Martin and Greljo, Admir and Marzocca, David and Son, Minho",
    title = "{Anomalous Triple Gauge Couplings in the Effective Field Theory Approach at the LHC}",
    eprint = "1609.06312",
    archivePrefix = "arXiv",
    primaryClass = "hep-ph",
    reportNumber = "ZU-TH-34-16",
    doi = "10.1007/JHEP02(2017)115",
    journal = "JHEP",
    volume = "02",
    pages = "115",
    year = "2017"
}

@article{Feldman:1997qc,
    author = "Feldman, Gary J. and Cousins, Robert D.",
    title = "{Unified approach to the classical statistical analysis of small signals}",
    eprint = "physics/9711021",
    archivePrefix = "arXiv",
    reportNumber = "HUTP-97-A096",
    doi = "10.1103/PhysRevD.57.3873",
    journal = "Phys. Rev. D",
    volume = "57",
    pages = "3873",
    year = "1998"
}

@article{Wilks:1938dza,
    author = "Wilks, S. S.",
    title = "{The Large-Sample Distribution of the Likelihood Ratio for Testing Composite Hypotheses}",
    doi = "10.1214/aoms/1177732360",
    journal = "Annals Math. Statist.",
    volume = "9",
    number = "1",
    pages = "60",
    year = "1938"
}

@article{ATLAS:2023pwa,
    author = "{ATLAS Collaboration}",
    collaboration = "ATLAS",
    title = "{Integrated and differential fiducial cross-section measurements for the vector boson fusion production of the Higgs boson in the $H \rightarrow WW^{\ast}\rightarrow e\nu\mu\nu$ decay channel at 13 $\text{TeV}$ with the ATLAS detector}",
    eprint = "2304.03053",
    archivePrefix = "arXiv",
    primaryClass = "hep-ex",
    reportNumber = "CERN-EP-2023-025",
    doi = "10.1103/PhysRevD.108.072003",
    journal = "Phys. Rev. D",
    volume = "108",
    number = "7",
    pages = "072003",
    year = "2023"
}

@article{Bertone:2017bme,
    author = "Bertone, Valerio and Carrazza, Stefano and Hartland, Nathan P. and Rojo, Juan",
    collaboration = "NNPDF",
    title = "{Illuminating the photon content of the proton within a global PDF analysis}",
    eprint = "1712.07053",
    archivePrefix = "arXiv",
    primaryClass = "hep-ph",
    reportNumber = "Nikhef/2017-064, CERN-TH-2017-235, NIKHEF-2017-064",
    doi = "10.21468/SciPostPhys.5.1.008",
    journal = "SciPost Phys.",
    volume = "5",
    number = "1",
    pages = "008",
    year = "2018"
}

@article{Xie:2021equ,
    author = "Xie, Keping and Hobbs, T. J. and Hou, Tie-Jiun and Schmidt, Carl and Yan, Mengshi and Yuan, C. -P.",
    collaboration = "CTEQ-TEA",
    title = "{Photon PDF within the CT18 global analysis}",
    eprint = "2106.10299",
    archivePrefix = "arXiv",
    primaryClass = "hep-ph",
    reportNumber = "MSUHEP-21-013, PITT-PACC-2112, FERMILAB-PUB-21-370-QIS-SCD-T, SMU-HEP-21-06, SMU-HEP-21-06,
  FERMILAB-PUB-21-370-QIS-SCD-T",
    doi = "10.1103/PhysRevD.105.054006",
    journal = "Phys. Rev. D",
    volume = "105",
    number = "5",
    pages = "054006",
    year = "2022"
}

@article{Hou:2019efy,
    author = "Hou, Tie-Jiun and others",
    title = "{New CTEQ global analysis of quantum chromodynamics with high-precision data from the LHC}",
    eprint = "1912.10053",
    archivePrefix = "arXiv",
    primaryClass = "hep-ph",
    reportNumber = "MSUHEP-19-025, PITT-PACC-1911, SMU-HEP-19-03",
    doi = "10.1103/PhysRevD.103.014013",
    journal = "Phys. Rev. D",
    volume = "103",
    number = "1",
    pages = "014013",
    year = "2021"
}

@article{Bailey:2020ooq,
    author = "Bailey, S. and Cridge, T. and Harland-Lang, L. A. and Martin, A. D. and Thorne, R. S.",
    title = "{Parton distributions from LHC, HERA, Tevatron and fixed target data: MSHT20 PDFs}",
    eprint = "2012.04684",
    archivePrefix = "arXiv",
    primaryClass = "hep-ph",
    reportNumber = "IPPP/20/58",
    doi = "10.1140/epjc/s10052-021-09057-0",
    journal = "Eur. Phys. J. C",
    volume = "81",
    number = "4",
    pages = "341",
    year = "2021"
}

@article{Cridge:2021pxm,
    author = "Cridge, T. and Harland-Lang, L. A. and Martin, A. D. and Thorne, R. S.",
    title = "{QED parton distribution functions in the MSHT20 fit}",
    eprint = "2111.05357",
    archivePrefix = "arXiv",
    primaryClass = "hep-ph",
    reportNumber = "IPPP/21/44",
    doi = "10.1140/epjc/s10052-022-10028-2",
    journal = "Eur. Phys. J. C",
    volume = "82",
    number = "1",
    pages = "90",
    year = "2022"
}

@article{Bhardwaj:2021ujv,
    author = "Bhardwaj, Akanksha and Englert, Christoph and Hankache, Robert and Pilkington, Andrew D.",
    title = "{Machine-enhanced CP-asymmetries in the Higgs sector}",
    eprint = "2112.05052",
    archivePrefix = "arXiv",
    primaryClass = "hep-ph",
    doi = "10.1016/j.physletb.2022.137246",
    journal = "Phys. Lett. B",
    volume = "832",
    pages = "137246",
    year = "2022"
}

@article{Hall:2022bme,
    author = "Hall, Noah Clarke and Criddle, Isaac and Crossland, Archie and Englert, Christoph and Forbes, Patrick and Hankache, Robert and Pilkington, Andrew D.",
    title = "{Machine-enhanced CP-asymmetries in the electroweak sector}",
    eprint = "2209.05143",
    archivePrefix = "arXiv",
    primaryClass = "hep-ph",
    doi = "10.1103/PhysRevD.107.016008",
    journal = "Phys. Rev. D",
    volume = "107",
    number = "1",
    pages = "016008",
    year = "2023"
}

@article{Gritsan:2020pib,
    author = "Gritsan, Andrei V. and Roskes, Jeffrey and Sarica, Ulascan and Schulze, Markus and Xiao, Meng and Zhou, Yaofu",
    title = "{New features in the JHU generator framework: constraining Higgs boson properties from on-shell and off-shell production}",
    eprint = "2002.09888",
    archivePrefix = "arXiv",
    primaryClass = "hep-ph",
    reportNumber = "HU-EP-20/02",
    doi = "10.1103/PhysRevD.102.056022",
    journal = "Phys. Rev. D",
    volume = "102",
    number = "5",
    pages = "056022",
    year = "2020"
}

@article{ATLAS:2024ini,
    author = "{ATLAS Collaboration}",
    collaboration = "ATLAS",
    title = "{Measurements of electroweak W$^{\pm}$Z boson pair production in association with two jets in pp collisions at $ \sqrt{s} $ = 13 TeV with the ATLAS detector}",
    eprint = "2403.15296",
    archivePrefix = "arXiv",
    primaryClass = "hep-ex",
    reportNumber = "CERN-EP-2024-082",
    doi = "10.1007/JHEP06(2024)192",
    journal = "JHEP",
    volume = "06",
    pages = "192",
    year = "2024"
}

@article{ATLAS:2025bts,
    author = "{ATLAS Collaboration}",
    collaboration = "ATLAS",
    title = "{Probing the Higgs boson CP properties in vector-boson fusion production in the $H\rightarrow \tau^+\tau^-$ channel with the ATLAS detector}",
    eprint = "2506.19395",
    archivePrefix = "arXiv",
    primaryClass = "hep-ex",
    reportNumber = "CERN-EP-2025-127",
    doi = "10.1007/JHEP10(2025)092",
    journal = "JHEP",
    volume = "10",
    pages = "092",
    year = "2025"
}

@article{ATLAS:2025hki,
    author = "{ATLAS Collaboration}",
    title = "{Measurements of Higgs boson production via gluon-gluon fusion and vector-boson fusion using $H\rightarrow WW^\ast \rightarrow \ell\nu\ell\nu$ decays in $pp$ collisions with the ATLAS detector and their effective field theory interpretations}",
    eprint = "2504.07686",
    archivePrefix = "arXiv",
    primaryClass = "hep-ex",
    reportNumber = "CERN-EP-2025-054",
    month = "4",
    year = "2025"
}

@Article{Evans:2008zzb,
      author         = "Evans, Lyndon and Bryant, Philip",
      title          = "{LHC Machine}",
      journal        = "JINST",
      volume         = "3",
      pages          = "S08001",
      doi            = "10.1088/1748-0221/3/08/S08001",
      year           = "2008",
      SLACcitation   = "%%CITATION = JINST,3,S08001;%%",
}

@article{Dittmaier:2009cr,
    author = "Dittmaier, Stefan and Huber, Max",
    title = "{Radiative corrections to the neutral-current Drell-Yan process in the Standard Model and its minimal supersymmetric extension}",
    eprint = "0911.2329",
    archivePrefix = "arXiv",
    primaryClass = "hep-ph",
    reportNumber = "MPP-2009-164",
    doi = "10.1007/JHEP01(2010)060",
    journal = "JHEP",
    volume = "01",
    pages = "060",
    year = "2010"
}


%% file: ATLAS-useful.bib
@Article{Lange:2001uf,
      author         = "Lange, D. J.",
      title          = "{The EvtGen particle decay simulation package}",
      booktitle      = "{Proceedings, 7th International Conference on B physics
                        at hadron machines (BEAUTY 2000)}",
      journal        = "Nucl. Instrum. Meth. A",
      volume         = "462",
      year           = "2001",
      pages          = "152",
      doi            = "10.1016/S0168-9002(01)00089-4",
      SLACcitation   = "%%CITATION = NUIMA,A462,152;%%"
}

@article{Bothmann:2019yzt,
      author = "Bothmann, Enrico and others",
      title = "{Event generation with Sherpa 2.2}",
      journal = "SciPost Phys.",
      volume = "7",
      year = "2019",
      number = "3",
      pages = "034",
      doi = "10.21468/SciPostPhys.7.3.034",
      reportNumber = "FERMILAB-PUB-19-218-T, SLAC-PUB-17433, IPPP/19/42, MCNET-19-11",
      eprint = "1905.09127",
      archivePrefix = "arXiv",
      primaryClass = "hep-ph",
}


%% file: ATLAS.bib
@Article{PERF-2007-01,
    author         = "{ATLAS Collaboration}",
    title          = "{The ATLAS Experiment at the CERN Large Hadron Collider}",
    journal        = "JINST",
    volume         = "3",
    year           = "2008",
    pages          = "S08003",
    doi            = "10.1088/1748-0221/3/08/S08003",
    primaryClass   = "hep-ex",
}

@Article{TRIG-2016-01,
    author         = "{ATLAS Collaboration}",
    title          = "{Performance of the ATLAS trigger system in 2015}",
    journal        = "Eur. Phys. J. C",
    volume         = "77",
    year           = "2017",
    pages          = "317",
    doi            = "10.1140/epjc/s10052-017-4852-3",
    reportNumber   = "CERN-EP-2016-241",
    eprint         = "1611.09661",
    archivePrefix  = "arXiv",
    primaryClass   = "hep-ex",
}

@Article{DAPR-2018-01,
    author         = "{ATLAS Collaboration}",
    title          = "{ATLAS data quality operations and performance for 2015--2018 data-taking}",
    journal        = "JINST",
    volume         = "15",
    year           = "2020",
    pages          = "P04003",
    doi            = "10.1088/1748-0221/15/04/P04003",
    reportNumber   = "CERN-EP-2019-207",
    eprint         = "1911.04632",
    archivePrefix  = "arXiv",
    primaryClass   = "physics.ins-det",
}

@Article{JETM-2018-05,
    author         = "{ATLAS Collaboration}",
    title          = "{Jet energy scale and resolution measured in proton--proton collisions at \(\sqrt{s} = 13\,\text{TeV}\) with the ATLAS detector}",
    journal        = "Eur. Phys. J. C",
    volume         = "81",
    year           = "2021",
    pages          = "689",
    doi            = "10.1140/epjc/s10052-021-09402-3",
    reportNumber   = "CERN-EP-2020-083",
    eprint         = "2007.02645",
    archivePrefix  = "arXiv",
    primaryClass   = "hep-ex",
}

@Article{STDM-2018-17,
    author         = "{ATLAS Collaboration}",
    title          = "{Precise measurements of \(W\)- and \(Z\)-boson transverse momentum spectra with the ATLAS detector using \(pp\) collisions at \(\sqrt{s} = 5.02\,\text{TeV}\) and \(13\,\text{TeV}\)}",
    journal        = "Eur. Phys. J. C",
    volume         = "84",
    year           = "2024",
    pages          = "1126",
    doi            = "10.1140/epjc/s10052-024-13414-0",
    reportNumber   = "CERN-EP-2024-080",
    eprint         = "2404.06204",
    archivePrefix  = "arXiv",
    primaryClass   = "hep-ex",
}

@Article{FTAG-2019-07,
    author         = "{ATLAS Collaboration}",
    title          = "{ATLAS flavour-tagging algorithms for the LHC Run~2 \(pp\) collision dataset}",
    journal        = "Eur. Phys. J. C",
    volume         = "83",
    year           = "2023",
    pages          = "681",
    doi            = "10.1140/epjc/s10052-023-11699-1",
    reportNumber   = "CERN-EP-2022-226",
    eprint         = "2211.16345",
    archivePrefix  = "arXiv",
    primaryClass   = "physics.data-an",
}

@Article{JETM-2020-03,
    author         = "{ATLAS Collaboration}",
    title          = "{The performance of missing transverse momentum reconstruction and its significance with the ATLAS detector using \(140\,\text{fb}^{-1}\) of \(\sqrt{s} = 13\,\text{TeV}\) \(pp\) collisions}",
    journal        = "Eur. Phys. J. C",
    volume         = "85",
    year           = "2025",
    pages          = "606",
    doi            = "10.1140/epjc/s10052-025-14062-8",
    reportNumber   = "CERN-EP-2024-023",
    eprint         = "2402.05858",
    archivePrefix  = "arXiv",
    primaryClass   = "hep-ex",
}

@Article{STDM-2020-01,
    author         = "{ATLAS Collaboration}",
    title          = "{Studies of the Energy Dependence of Diboson Polarization Fractions and the Radiation-Amplitude-Zero Effect in \(WZ\) Production with the ATLAS Detector}",
    journal        = "Phys. Rev. Lett.",
    volume         = "133",
    year           = "2024",
    pages          = "101802",
    doi            = "10.1103/PhysRevLett.133.101802",
    reportNumber   = "CERN-EP-2024-045",
    eprint         = "2402.16365",
    archivePrefix  = "arXiv",
    primaryClass   = "hep-ex",
    related        = "STDM-2020-01-err",
    relatedstring  = "Erratum:",
}

@Article{SOFT-2022-02,
    author         = "{ATLAS Collaboration}",
    title          = "{Software and computing for Run~3 of the ATLAS experiment at the LHC}",
    journal        = "Eur. Phys. J. C",
    volume         = "85",
    year           = "2025",
    pages          = "234",
    doi            = "10.1140/epjc/s10052-024-13701-w",
    reportNumber   = "CERN-EP-2024-100",
    eprint         = "2404.06335",
    archivePrefix  = "arXiv",
    primaryClass   = "hep-ex",
}

@Booklet{ATL-SOFT-PUB-2025-001,
    author         = "{ATLAS Collaboration}",
    title          = "{ATLAS Computing Acknowledgements}",
    howpublished   = "{ATL-SOFT-PUB-2025-001}",
    url            = "https://cds.cern.ch/record/2922210",
    year           = "2025",
}


%% file: PubNotes.bib
@Booklet{ATL-PHYS-PUB-2014-021,
    author         = "{ATLAS Collaboration}",
    title          = "{ATLAS Pythia~8 tunes to \(7~\text{TeV}\) data}",
    howpublished   = "{ATL-PHYS-PUB-2014-021}",
    url            = "https://cds.cern.ch/record/1966419",
    year           = "2014",
}

@Booklet{ATL-PHYS-PUB-2016-020,
    author         = "{ATLAS Collaboration}",
    title          = "{Studies on top-quark Monte Carlo modelling for Top2016}",
    howpublished   = "{ATL-PHYS-PUB-2016-020}",
    url            = "https://cds.cern.ch/record/2216168",
    year           = "2016",
}

@Booklet{ATL-PHYS-PUB-2019-026,
    author         = "{ATLAS Collaboration}",
    title          = "{Forward jet vertex tagging using the particle flow algorithm}",
    howpublished   = "{ATL-PHYS-PUB-2019-026}",
    url            = "https://cds.cern.ch/record/2683100",
    year           = "2019",
}
